\documentclass[trackchanges, twocolumn]{aastex701}
\usepackage{tabularx}
\usepackage{orcidlink}
\usepackage{multirow}
\usepackage{graphicx}
\usepackage{dcolumn}
\usepackage{bm}
\usepackage{bbding}
\usepackage{amsmath}
\usepackage{comment}

\begin{document}


\title{On the Astrophysical Origin of Binary Black Hole Subpopulations: A Tale of Three Channels?}

\author[0000-0002-7322-4748]{Anarya\, Ray}\email[show]{anarya.ray@northwestern.edu} \affiliation{Center for Interdisciplinary Exploration and Research in Astrophysics (CIERA), Northwestern University, 1800 Sherman Ave,
Evanston, IL 60201, USA}
\affiliation{NSF-Simons AI Institute for the Sky (SkAI), 172 E. Chestnut Street, Chicago, IL 60611, USA}

\author{Shirsha Mukherjee\, \orcidlink{0009-0007-0866-4599}}
\email{shirsha.mukherjee@ou.edu}  
\affiliation{University of Oklahoma}

\author{Michael Zevin\, \orcidlink{0000-0002-0147-0835}}
\email{mzevin@adlerplanetarium.org}
\affiliation{Adler Planetarium, 1300 South DuSable Lake Shore Drive, Chicago, IL 60605, USA}
\affiliation{Center for Interdisciplinary Exploration and Research in Astrophysics (CIERA), Northwestern University, 1800 Sherman Ave,
Evanston, IL 60201, USA}
\affiliation{NSF-Simons AI Institute for the Sky (SkAI), 172 E. Chestnut Street, Chicago, IL 60611, USA}
\author{Vicky Kalogera\, \orcidlink{0000-0001-9236-5469}}
\email{vicky@northwestern.edu}
\affiliation{Center for Interdisciplinary Exploration and Research in Astrophysics (CIERA), Northwestern University, 1800 Sherman Ave,
Evanston, IL 60201, USA}
\affiliation{NSF-Simons AI Institute for the Sky (SkAI), 172 E. Chestnut Street, Chicago, IL 60611, USA}
\affiliation{Department of Physics and Astronomy, Northwestern University, 2145 Sheridan Road, Evanston, IL 60208, USA}

\date{\today}

\begin{abstract}
There is increasing evidence for multiple binary black hole~(BBH) subpopulations in the cumulative gravitational wave catalog by the LIGO-Virgo-KAGRA Collaboration. The astrophysical interpretation of this complex underlying population is subject to theoretical uncertainties in treatments of binary stellar evolution, core collapse, and host environments. In this \textit{Letter}, using parametrized mixture models, we show that the BBH detection sample comprises three astrophysical subpopulations that are likely dominated by specific formation channels. In particular, we show that the $10M_{\odot}$ peak and the $35M_{\odot}$ feature in the BBH mass spectrum correspond to distinct mass-ratio, spin alignment, spin precession, and redshift evolution properties. We show that mass-based transitions reported in the distribution of BBH parameters naturally emerge from our inferred distributions without explicit modeling. Our results are consistent with the current observed population arising from specific relative abundances of isolated binary evolution, dynamical formation in globular clusters, and higher-generation BBH mergers. Under this interpretation, we constrain the relative underlying fraction of these channels to be $79.0^{+11.5}_{-10.9}\%$, $14.5^{+11.6}_{-8.0}\%$, and, $2.5^{+5.5}_{-1.8}\%$, respectively, and find these relative fractions to be evolving over cosmic time with more than $1\sigma$ confidence. Our interpretation relies on simple theoretical predictions that are mostly robust against uncertainties in BBH formation, with more definite conclusions expected in the near future.
\end{abstract}

\section{Introduction}


The fourth gravitational wave transient catalog~\citep[GWTC-4,][]{LIGOScientific:2025slb} by LIGO-Virgo-KAGRA~\citep[LVK,][]{LIGOScientific:2014pky,VIRGO:2014yos, KAGRA:2020agh} has revealed a complex underlying population of binary black hole~(BBH) mergers~\citep{LIGOScientific:2025pvj}. With over 150 detections~\citep{LIGOScientific:2025slb}, there is increasing evidence of multiple subpopulations~\citep{Banagiri:2025dmy, Ray:2025xti, Farah:2026jlc, Vijaykumar:2026zjy, Sridhar:2025kvi, Wang:2025nhf, Afroz:2024fzp, Afroz:2025ikg, Galaudage:2026opk}, indicating contributions from several formation channels~\citep[see, e.g.][]{Zevin:2020gbd, Cheng:2023ddt, Colloms:2025hib}. However, theoretical uncertainties in binary stellar evolution, BH formation through stellar collapse, and host environments often limit the scope of self-consistent astrophysical interpretation~\citep[see, e.g.][for reviews]{Mandel:2021smh, Breivik:2025edm}. Nevertheless, certain predictions remain robust against such uncertainties, which can be modeled in the growing population dataset to derive empirical constraints on key astrophysical parameters.

Broadly, theoretical models for BBH formation can be grouped into isolated binary evolution in galactic fields and dynamically interacting binaries in dense stellar environments, triple systems, or disks of active galactic nuclei~\citep[see, e.g.,][for reviews]{Mandel:2018hfr, Mapelli:2021taw, Mandel:2021smh}. Dynamical formation can be further categorized into first-generation ``1G+1G" mergers that only comprise BHs formed directly through stellar collapse, and hierarchical mergers that comprise remnants of previous mergers retained by the environment~\citep[see, e.g.,][for a review]{Gerosa:2021mno}. Despite the uncertainties in theoretical predictions, it is possible to disentangle these three subpopulations by modeling certain robust features that are largely independent of the underlying assumptions of binary stellar evolution, BH formation, and treatment of host environments. 

In particular, BHs formed directly through stellar collapse are expected to be slowly spinning at birth~\citep{Fuller:2019ckz, Ma:2019cpr, Fuller:2019sxi} as long as significant tidal interactions do not persist prior to core collapse~\citep{Bavera:2020uch, Zevin:2022wrw, Broekgaarden:2022nst} and exhibit a gap in the higher end of their mass-spectrum due to the pair instability process limiting the remnant mass or completely disrupting the high-mass progenitor in a pair instability supernova (PISN)~\citep{Heger:2001cd, Woosley:2007qp, Belczynski:2016jno, Spera:2017fyx, Farmer:2019jed, Farmer:2020xne, vanSon:2020zbk, Ziegler:2020klg, Hendriks:2023yrw}. On the other hand, remnants of previous mergers that retain a part of the original binary's angular momentum prior to merger typically have high spin magnitude values of $\sim0.67$~\citep{Berti:2008af, Hofmann:2016yih, Rodriguez:2019huv, Borchers:2025sid, Fishbach:2017dwv}, and masses that can pollute the PISN gap~\citep{OLeary:2005vqo, Antonini:2016gqe,Tagawa:2020qll, Mapelli:2021syv, Antonini:2022vib, Torniamenti:2024uxl, Vaccaro:2025ogk, Rodriguez:2019huv}. In addition, the strong recoil kicks received by merger remnants often lead to their expulsion from the dense environment~\citep{Fitchett:1983qzq,PortegiesZwart:1999nm,Favata:2004wz,Gonzalez:2006md,Lousto:2009ka,Gerosa:2018qay,Mahapatra:2021hme, Zevin:2022bfa}. Hence, the increasing rarity of higher-generation BHs that are available for forming binaries indicates a higher rate for hierarchical mergers between a second and a first generation companion~(from here on 2G+1G) than all other combinations, leading to a strong preference for mass asymmetry in this subpopulation~\citep{Zevin:2022bfa, Rodriguez:2019huv, Gerosa:2017kvu, Kimball:2020opk, Kimball:2020qyd}.

Informed by these predictions, previous investigations that rely on semi-parametric mixture models have constrained the branching fraction of hierarchical mergers in the current observational sample, which is robust against the uncertainties in binary evolution and stellar collapse~\citep{Farah:2026jlc,Vijaykumar:2026zjy,Wang:2025nhf, Plunkett:2026pxt, Tong:2025xir, Antonini:2024het, Antonini:2025ilj, Kimball:2020opk, Kimball:2020qyd, Mould:2022ccw}. From the maximum mass of the remaining ensemble, these studies have constrained the PISN cutoff and hence the carbon-oxygen reaction rate in massive stars, a persistent theoretical uncertainty in the treatments of massive star evolution and the onset of the pair instability process. They have further provided a self-consistent astrophysical explanation for previously reported correlations in the BBH population, namely the broadening of the effective spin distribution with redshift and mass ratio, by showing that the highly-spinning, asymmetric-mass hierarchical subpopulation has a much steeper redshift evolution than the rest of the ensemble~\citep{Vijaykumar:2026zjy,Farah:2026jlc}.

However, there is additional substructure in the slowly spinning subpopulation whose interpretation requires further investigation~\citep{Sridhar:2025kvi, Ray:2024hos, Banagiri:2025dmy, Sadiq:2021fin, Sadiq:2023zee}. This ensemble comprises $\sim80\%$ of all astrophysical BBH mergers in the local universe~\citep{Farah:2026jlc,Vijaykumar:2026zjy,Wang:2025nhf, Plunkett:2026pxt, Tong:2025xir} and exhibits two robust features in the mass spectrum near $10M_{\odot}$ and $35M_{\odot}$~\citep{LIGOScientific:2025pvj, Galaudage:2024meo, Tiwari:2020otp, Tiwari:2025oah}. Furthermore, it shows mass-based transitions in the distribution of BBH mass ratios and effective spins near $\sim 20M_{\odot}$ and $\sim 40M_{\odot}$~\citep{Banagiri:2025dmy, Sridhar:2025kvi}. Due to small spin magnitudes and preference for symmetric mass-ratio values, it is likely that this subpopulation comprises of systems where both components are first-generation BHs. In order to physically interpret the transition mass scales and isolate specific formation channels that might be responsible for various features in the mass distribution~\citep{Godfrey:2023oxb, Roy:2025ktr}, it is necessary to model distinct observational signatures of 1G+1G dynamical assembly and field evolution as BBH formation pathways that are also robust against persisting theoretical uncertainties.

\begin{table*}[htt]
\centering
\begin{tabular}{cccccc}
\hline
Comp.  & $p(m_1)$  & $p(q|m_1)$ &$p(\chi_{eff})$&$p(\chi_p)$ & r(z)\\
\hline
1 & \texttt{PLP} & $\mathcal{N}_{[\frac{m_{2,min}}{m_1},1]}^{\mu_{q}^1,\sigma_{q}^1} $ & $\mathcal{N}_{[-1,1]}^{\mu_{\chi_{eff}}^1,\sigma_{\chi_{eff}}^1}$ &$\mathcal{N}_{[0,1]}^{\mu_{\chi_{p}}^1,\sigma_{\chi_{p}}^1} $ & $(1+z)^{\kappa_1}$\\
2 & \texttt{BPL} & \texttt{SPL}$_{[\frac{m_{2,min}}{m_1},1]}^{\beta_q^2, \frac{\delta_m}{m_1}} $ & $\mathcal{N}_{[-1,1]}^{\mu_{\chi_{eff}}^2,\sigma_{\chi_{eff}}^2}$ &$\mathcal{N}_{[0,1]}^{\mu_{\chi_{p}}^2,\sigma_{\chi_{p}}^2} $ & $(1+z)^{\kappa_2}$ \\
3 & \texttt{BPL} & $\mathcal{N}_{[\frac{m_{2,min}}{m_1},1]}^{0.5,\sigma_{q}^3} $ & $\mathcal{U}(-w,w)$ &$\mathcal{N}_{[0,1]}^{\mu_{\chi_{p}}^3,\sigma_{\chi_{p}}^3} $ & $(1+z)^{\kappa_3}$\\
\hline
\end{tabular}
\caption{\label{tab:models} Parametrizations of each component. Here \texttt{SPL}$_{a,b}^{\beta, \delta}$ denotes a single smoothed powerlaw with index $\beta$, and smoothing parameter $\delta$, truncated within $[a,b]$, \texttt{PLP} denotes a smoothed Powerlaw+Gausian Peak, \texttt{BPL} a smoothed broken powerlaw, $\mathcal{N}_{[a,b]}^{\mu, \sigma}$ a Gaussian with mean $\mu$, and variance $\sigma^2$ truncated within $[a,b]$, and $r(z)$ is the ratio of the merger rate at redshift $z$ relative to the local rate~(i.e. at $z=0$).}
\end{table*}

Traditional dynamical assembly in dense stellar clusters is expected to result in isotropic spin orientations with respect to the orbit~\citep{Mapelli:2021gyv, Chattopadhyay:2023pil, Rodriguez_2022}, although accretion from stellar mergers~\citep{Kiroglu:2025bbp}, evolution in gaseous environments~\citep{Tagawa:2020dxe, McKernan:2021nwk, Li:2022cul,Mckernan:2017ssq, Santini:2023ukl, McKernan:2023xio, McKernan:2024kpr, Cook:2024ajp, Fabj:2025vza}, or hierarchical perturbers~\citep{Antonini:2017ash, Liu:2018nrf, Rodriguez:2018jqu} may break this isotropy and lead to a higher propensity for aligned or in-plane spins. 
On the other hand, in isolated BBH formation, binary interactions such as mass transfer and tides are expected to preferentially align BH spin components to the orbital axis~\citep{Kalogera:1999tq, Bavera:2020inc, Gerosa:2018wbw, Steinle:2022rhj}. Hence, in the subpopulation of systems that do not comprise hierarchical mergers, it is possible to further disentangle the contributions of isolated binary evolution and various subchannels of 1G+1G dynamical mergers by modeling these potentially distinct spin properties. Note that these distinct trends in the distribution of component spins will also manifest in that of the effective aligned and effective precessing spin parameters of BBHs, which are often measured better than component spins for each individual detection.

In this \textit{Letter}, we probe the astrophysical origins of the GWTC-4 sample of BBH detections and show that it can emerge from the mixing of three specific subpopulations that are consistent with the expectations of distinct formation channels. Using parametrized mixture models, we constrain the relative abundance of each subpopulation and how they evolve over cosmic time. We provide a physical interpretation of mass-based transition scales reported in GWTC-4 that naturally emerge in our inferred distributions without explicit modeling. We identify new correlations between masses, redshifts, and effective precessing spins, and show that the two robust features in the mass spectrum likely arise from distinct formation channels. Our findings hint towards a self-consistent interpretation of the current observed population as emerging from specific relative abundances of isolated binary evolution, 1G+1G dynamical formation, and hierarchical mergers, with relative contributions that vary across cosmic time. 

\section{Methods}
\label{sec:methods}
 We investigate the hypothesis that subpopulations originating from specific formation channels give rise to particular features in the mass spectrum, whose combined contributions can self-consistently explain the mass-based transitions observed in the distributions of other BBH parameters. To that end, we model the population-level distribution of BBH primary masses~$(m_1)$, mass ratios~$(q)$, effective aligned spins~$(\chi_{eff})$, effective precessing spins~$(\chi_p)$, and redshifts~$(z)$, as a three-component mixture. Each component is constructed from simple parametrizations that are motivated by the findings of data-driven studies and are capable of capturing specific features in the mass-spectrum that have been robustly identified by past investigations~\citep{Banagiri:2025dmy, Sridhar:2025kvi, LIGOScientific:2025pvj}. We delineate these parametrizations in Table~\ref{tab:models}.

Note that we target the hierarchical subpopulation with component 3 but do not enforce its existence~\citep{Farah:2026jlc, Vijaykumar:2026zjy, Plunkett:2026pxt}, while allowing much higher flexibility through the parametrizations of the other two components. We explore variations in modeling assumptions by considering alternative functional forms and by relaxing prior restrictions, and find that our results are robust against such variations. Detailed functional forms of our models, the corresponding hyperpriors, model comparison tests, and additional checks for model systematics are presented in Appendices~ \ref{sec:app:mod-comp}, \ref{sec:app-metric-modcomp}, \ref{sec:app-ppchecks}, and \ref{sec:app-trans}.
\begin{figure*}
\includegraphics[width=0.32\textwidth]{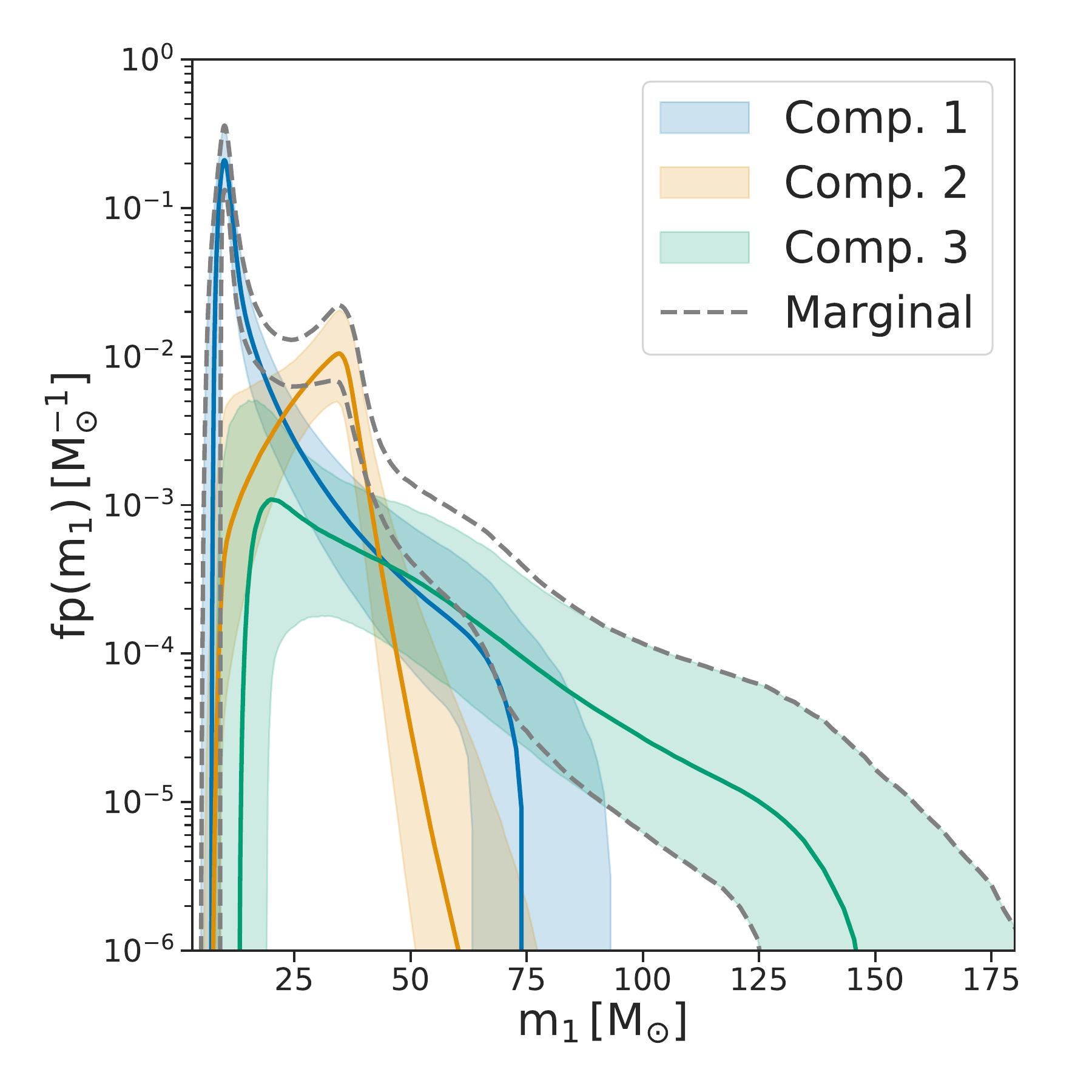}
\includegraphics[width=0.32\textwidth]{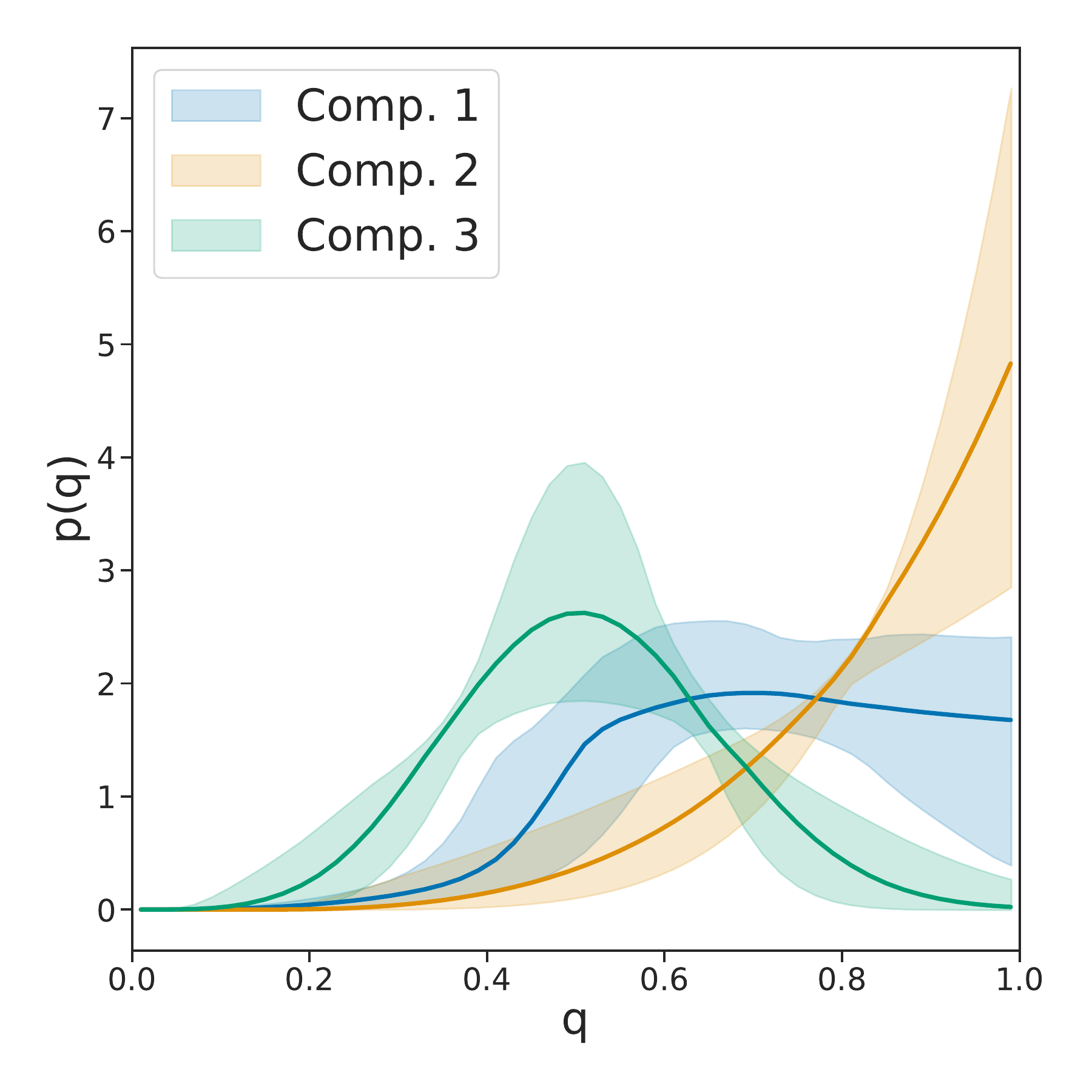}
\includegraphics[width=0.32\textwidth]{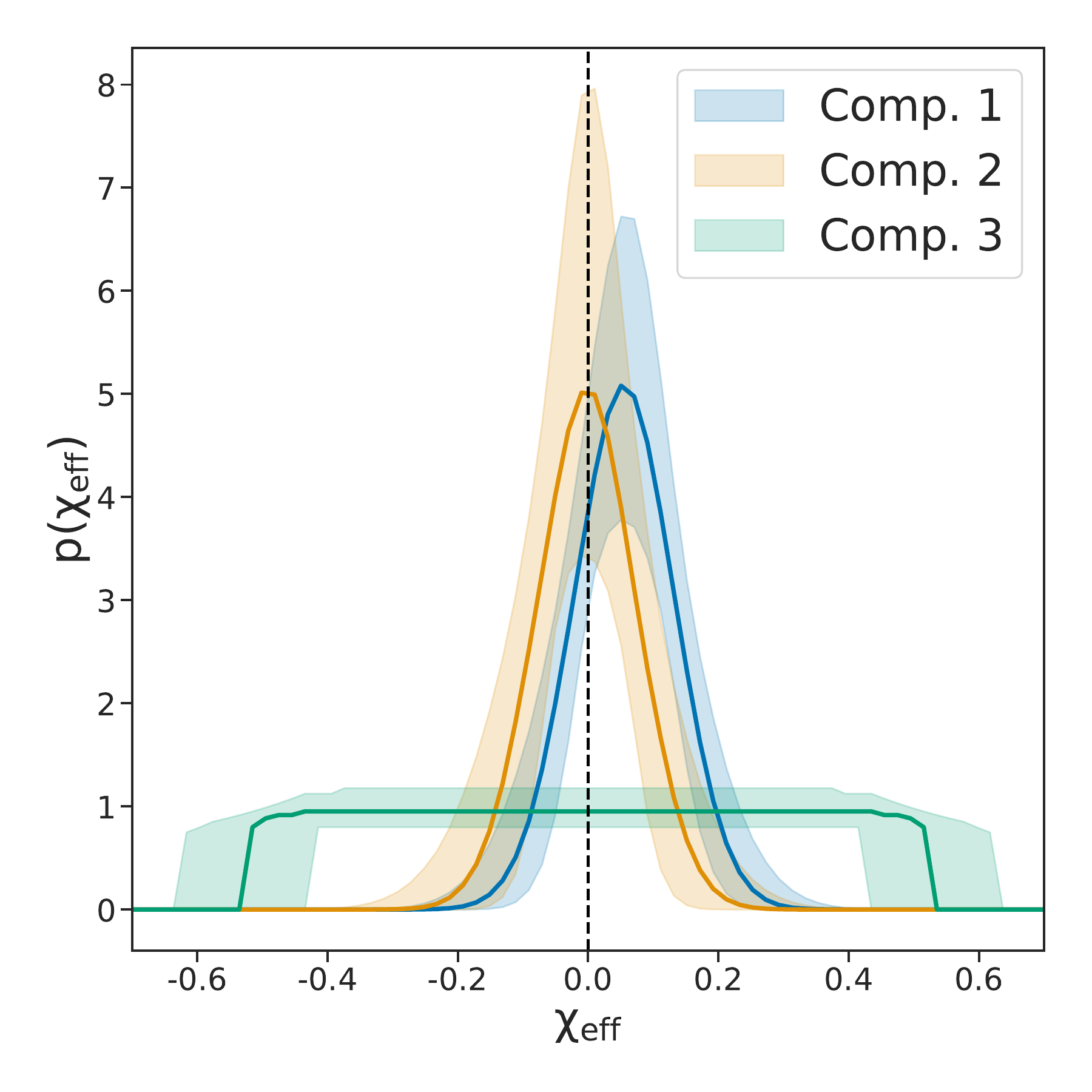}
\includegraphics[width=0.32\textwidth]{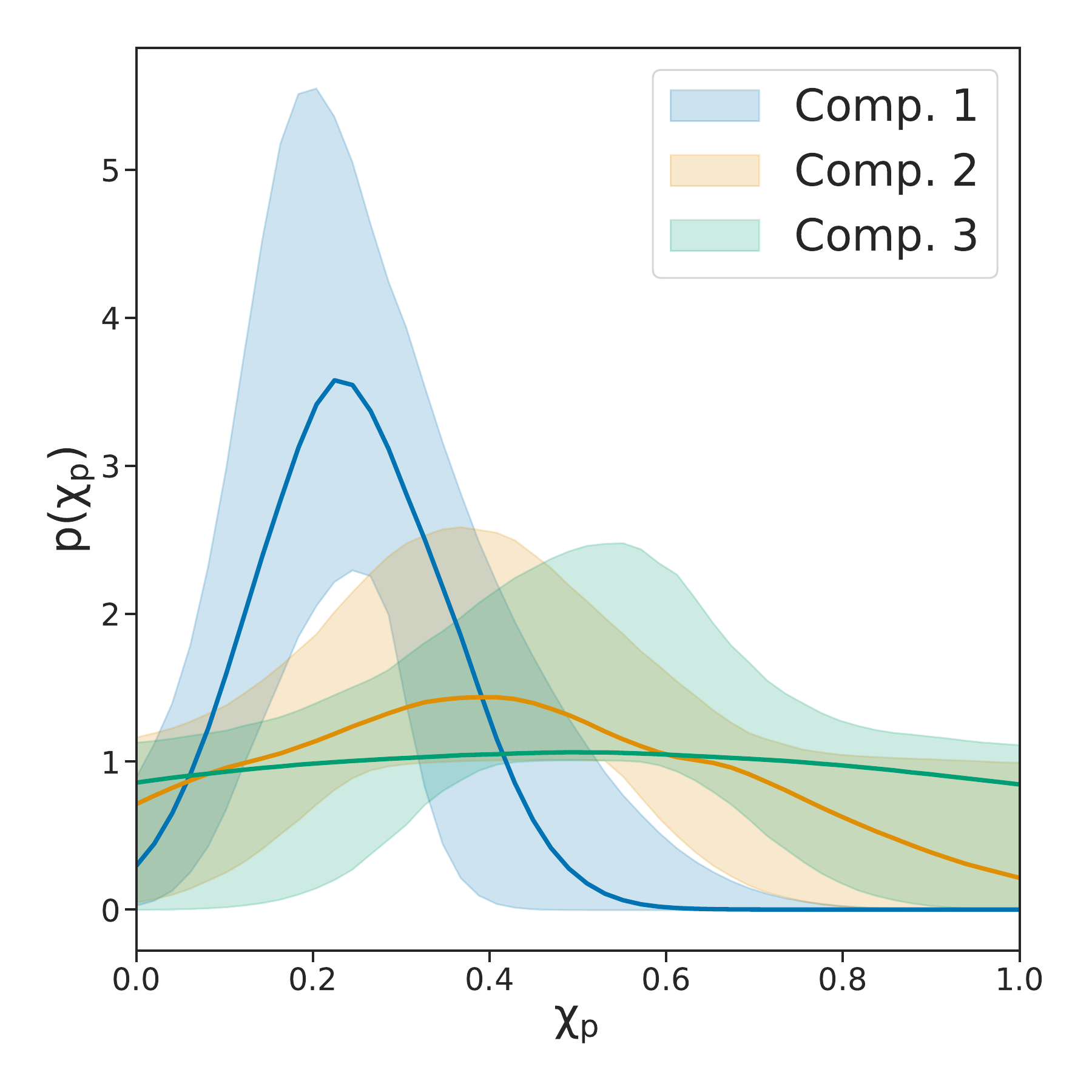}
\includegraphics[width=0.32\textwidth]{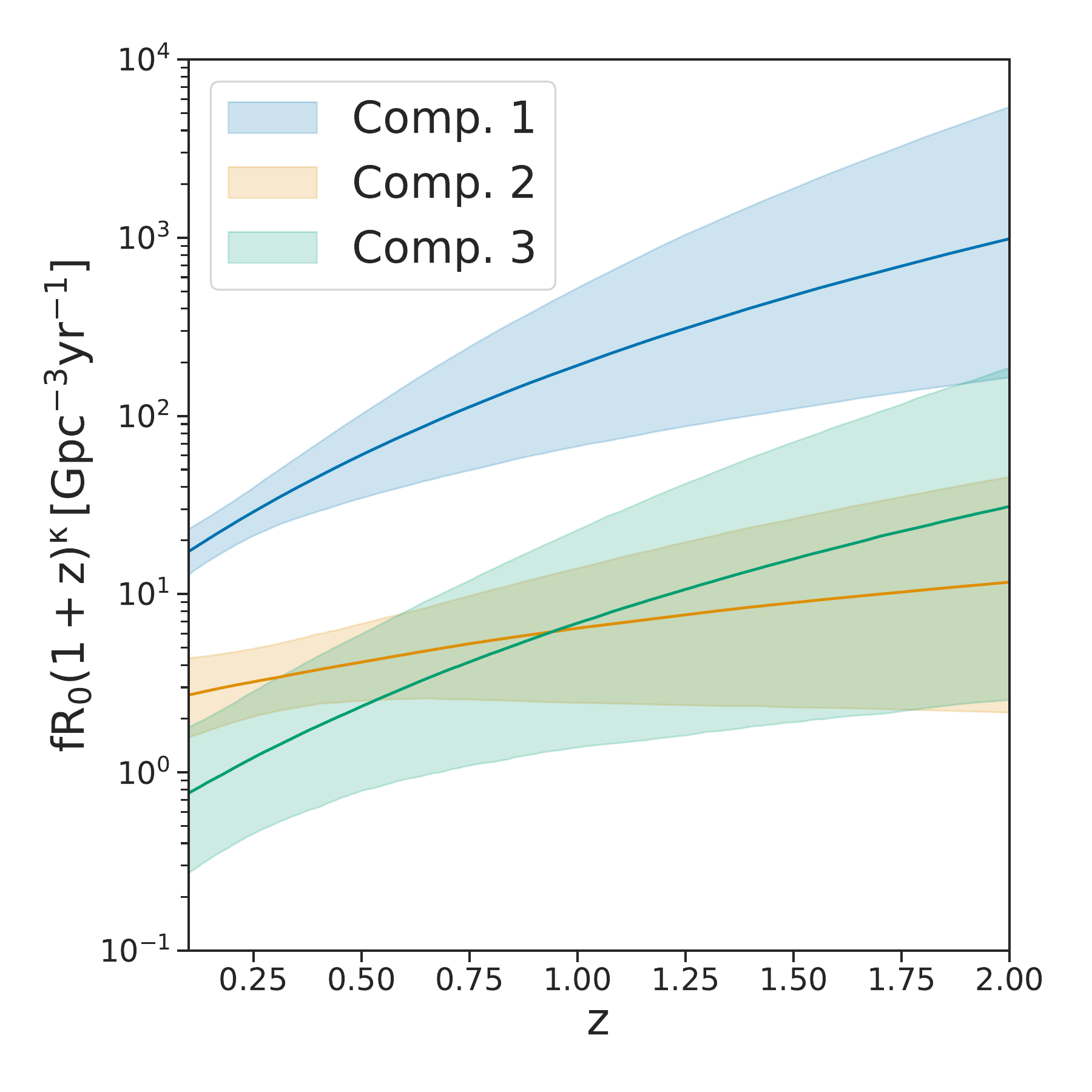}
\includegraphics[width=0.32\textwidth]{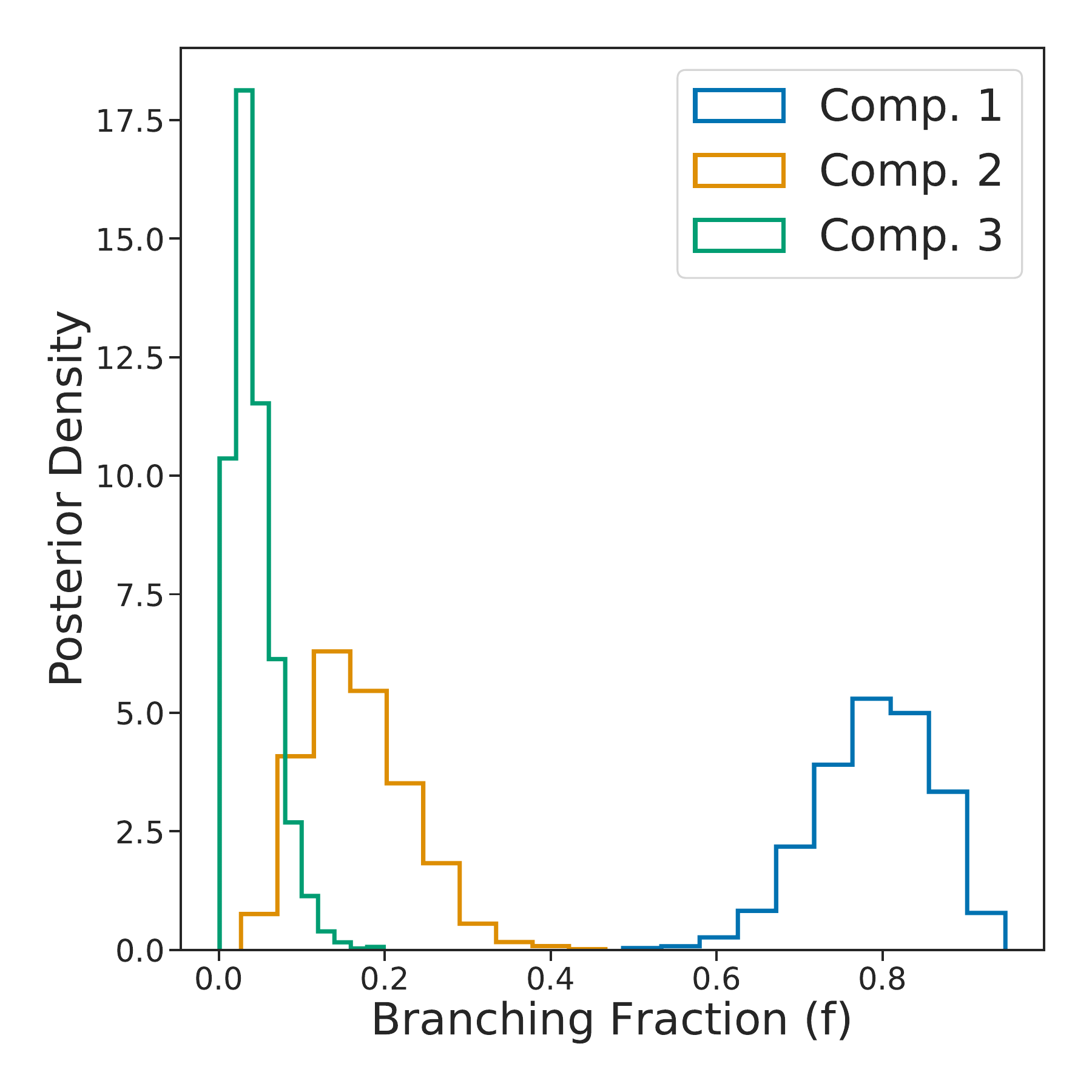}
\caption{\label{fig:subpops}Inferred distributions of primary mass~(weighted by the branching fractions, \textit{top left}), mass-ratio~\textit{(top center)}, effective aligned \textit{(top right)} and preccesing~\textit{(bottom left)} spins for each component. In these panels, the shaded regions denote the $90\%$ credible intervals of the posterior and the solid lines represent the posterior median. The redshift evolution of the merger rate~\textit{(bottom center)} and branching fractions~\textit{(bottom right)} are also shown. See Appendices \ref{sec:app-prpdraws}, and \ref{sec:app-trans} for comparisons with prior-predictive draws and data-driven reconstructions of the corresponding distributions, respectively.} 
\end{figure*}

In a mixture of subpopulations with potentially distinct $q,\chi_{eff},\chi_p,z$ distributions, one can expect the emergence of several mass scales that demarcate transitions in the marginal distributions of these BBH parameters. Such transition scales can distinguish the mass ranges in which different formation channels contribute dominantly to the astrophysical merger rate. We define $m_{t}^{12}$ such that for $m_1\leq m_t^{12}$, the contribution from component 1 dominates that of the others. Similarly $m_t^{23}$ is defined such that for $m_t^{12} \leq m_1 \leq m_t^{23}$, component 2 has the highest relative abundance. We reconstruct these numbers from the inferred distributions in postprocessing\footnote{For each hyperparameter, the minimum $m_1$ on a grid above which $f_2p_2(m_1)>f_1p(m_1)$ is defined as $m_{t}^{12}$ and the maximum $m_1$ below which $f_2p_2(m_1)\geq \max\{f_1p_1(m_1),f_3p_3(m_1)\}$ is defined as $m_t^{23}$}, since these are not hyperparameters that constrain our distribution functions.

\section{Results} 
\label{sec:results}
We constrain our mixture model from GWTC-4 data publicly released by the LVK~\citep{ligo_scientific_collaboration_2025_16740128, ligo_scientific_collaboration_and_virgo_2021_5546663,ligo_scientific_collaboration_and_virgo_2022_6513631,ligo_scientific_collaboration_and_virgo_2025_16053484, LIGOScientific:2019lzm, LIGOScientific:2025snk, KAGRA:2023pio}, using a Bayesian hierarchical analysis~\citep{popgw2,popgw3, pop-vitale, Mandel:2018hfr,Pdet1-Farr, Pdet2-essick, Thrane:2018qnx}, and reconstruct the $m_1,q,\chi_{eff},\chi_p,z$ distributions of each component. From the inferred distributions, we compute the emergent transition mass scales. Our model is preferred by the data over a single-component default distribution that comprises a broken power law plus two peaks in $m_1$~\citep{LIGOScientific:2025pvj}, single powerlaws in $q$ and $z$, and single truncated gaussians for $\chi_{eff}$ and $\chi_p$, by a Bayes factor of $10^6$. For additional details of our population analysis and its implementation, see Appendix~\ref{sec:app-gwpop}.

\begin{table*}[htt]
\centering
\centering
\begin{tabular}{ccccc}
\hline
\hline
$~~~~~~~~~$ Comp. $~~~~~~~~~~$ & $m_1\leq m_t^{12}$  & $~~~~~~m_t^{12}\leq m_1\leq m_t^{23}~~~~~~$ &$~~~~~~~m_t^{23}\leq m_1\leq m_{max,(1 or 2)}~~~~~~$&$m_1>m_{max,(1 or 2)}$\\
\hline
1 & $96.3^{+1.8}_{-5.0}\%$ & $16.3^{+11.9}_{-8.3}\%$ & $45.4^{+26.9}_{-37.7}\%$ & $0.0\%$\\
2 & $2.5^{+4.0}_{-1.2}\%$ & $74.6^{+10.8}_{-14.0}\%$ & $9.1^{+10.1}_{-4.3}\%$ & $0.0\%$\\
3 & $0.34^{+3.67}_{-0.34}\%$ & $5.8^{+11.6}_{-4.8}\mathbf{\%}$ & $51.6^{+29.0}_{-37.1}\mathbf{\%}$ & $100\%$ \\
\hline
\end{tabular}
\caption{\label{tab:relative-frac} Relative abundance of various components in different mass ranges. See Figure~\ref{fig:mass-scales} for the joint posterior distribution of the transition mass scales as well as these relative fractions of each component.}
\end{table*}
In Figure~\ref{fig:subpops}, we show the inferred distributions of BBH parameters for each mixture component (with the shaded regions representing $90\%$) credible intervals of the posterior). There is evidence of three distinct subpopulations, whose unique features in the distributions of intrinsic BBH parameters are presented as follows.
\begin{enumerate}
    \item{ \textit{Subpopulation 1 $(Comp.~1)$}: This component peaks sharply at $10M_{\odot}$ and falls off like a powerlaw up to a maximum mass of $70.0^{+21.7}_{-7.5}M_{\odot}$. The $\chi_{eff}$ distribution narrowly peaks at a small but positive value, with $23.0^{+12.9}_{-10.3}\%$ systems having $\chi_{eff}<0$, demonstrating a preference for slowly spinning aligned systems.(The reported fraction of negative $\chi_{eff}$ systems are better visualized in the cumulutive distributions which are presented in Appendix~\ref{sec:app-additional}, Figure~\ref{fig:cdfs}). The mass ratio distribution has support over a broad range of values: $0.6\text{--}1.0$ which declines steeply for very small~$(q<0.6)$ mass ratio values. Hints of additional features that might be present in the mass-ratio distribution of this component are investigated in Appendix~\ref{sec:app-additional}, Figure~\ref{fig:q-1}}. The $\chi_p$ spin distribution favours significantly smaller values than the other two subpopulations, indicating very few systems with in-plane spin components. These systems comprise $79.0^{+11.5}_{-10.9}\%$ of all astrophysical BBH mergers in the local universe. 
    \item{\textit{Subpopulation 2 $(Comp.~2)$}: This subpopulation displays a very shallow rise in the mass spectrum with no sharp feature near $10M_{\odot}$ but clearly captures the $35M_{\odot}$ peak. At higher masses, the merger rate for this component falls off much more sharply than that of Subpopulation 1, until a similar maximum mass. The mass-ratio distribution peaks strongly at $q=1$, demonstrating a strong preference for equal-mass binaries. The $\chi_{eff}$ values of these systems are narrowly concentrated about zero, with $51.7^{+16.1}_{-17.1}\%$ systems having $\chi_{eff}<0$, indicating small spin magnitudes and equal fractions of aligned and antialigned spin orientations with respect to the orbit.~(The reported fraction of negative $\chi_{eff}$ systems are better visualized in the cumulutive distributions which are presented in Appendix~\ref{sec:app-additional}, Figure~\ref{fig:cdfs})} The $\chi_p$ distribution peaks at significantly higher values than Subpopulation 1, demonstrating a higher fraction of systems with in-plane spin components. This equal mass, slowly spinning subpopulation with isotropically oriented spin components and an overabundance of systems in $30-40M_{\odot}$ contributes $14.5^{+11.6}_{-8.0}\%$ of all local BBHs. 
    \item{ \textit{Subpopulation 3 $(Comp.~3)$}: This component shows a shallow decline in the mass spectrum spanning from $12.7^{+5.1}_{-6.5}M_{\odot}\text{--}140.0^{+43.4}_{-15.0}M_{\odot}$, a mass-ratio distribution strongly preferring values smaller than $q=0.7$, a broad and relatively flat effective spin distribution between -0.5 and 0.5, and marginally higher precession than component 2. The break in the power law of this component~$(58.6^{+15.8}_{-24.3}M_{\odot})$ is inferred to be roughly twice the break mass of component 2~$(36.1^{+2.1}_{-2.5}M_{\odot})$. (The edges of the quoted mass ranges and locations of breaks are overplotted with the inferred mass distributions in Figure~\ref{fig:m1-wide} of Appendix~\ref{sec:app-additional}.) This subpopulation contributes to $2.5^{+5.5}_{-1.8}\%$ of merging BBHs in the local universe.} 
\end{enumerate}
We find that component 2, i.e., the one responsible for the $35M_{\odot}$ peak, has a much shallower redshift evolution~($\kappa_2=1.59^{+1.61}_{-1.84}$) than the other two, whereas components 1 and 3 have redshift evolutions with similar slope~($\kappa_1=3.89^{+2.03}_{-1.61}$, $\kappa_3=3.75^{+2.01}_{-2.34}$). Even though these $90\%$ bounds of $\kappa_i$ overlap, the $1\sigma$ intervals do not, indicating a redshift evolving mass spectrum at more than $68\%$ confidence~(see Appendix~\ref{sec:app-redevol}).

As evident from the weighted mass spectra of the different components, the GWTC-4 detection sample indicates that there are two mass scales at which the $q$, $\chi_{eff}$, $\chi_p$, and $z$ distributions transition into different shapes. We constrain these transition scales from our inferred mass distributions and branching fractions to be $m_t^{12}=23.3^{+4.2}_{-7.0}M_{\odot}$, and 
$m_t^{23}=41.8^{+5.4}_{-4.3}M_{\odot}$~(see Figure~\ref{fig:mass-scales} in Appendix~\ref{sec:app-additional} for the joint posterior distribution of these mass-scales). These are fully consistent with previous investigations that have directly modelled mass-based transitions in the distribution of these parameters~\citep{Banagiri:2025dmy, Ray:2025xti, Tong:2025wpz, Antonini:2024het, Antonini:2025ilj}. We constrain the relative abundance of our three channel-specific subpopulations in each region spanned by the transition masses in Table~\ref{tab:relative-frac}. We compare the marginal distributions of these BBH parameters above and below the transition masses with the previous findings of data-driven analyses~\citep{Sridhar:2025kvi} in Appendix~\ref{sec:app-trans}.  Note that in particular, the region $m_{t}^{23}\leq m_1 \leq m_{max(1or2)}$ has comparable contributions from \textit{Comp. 1} and \textit{Comp. 3}. This leads to a marginal mass-ratio distribution in that region which has substantially higher support in the $q\in(0.6,0.8)$ region~(see Figure~\ref{fig:marginal}, of Appendix~\ref{sec:app-trans}) than one would expect exclusively from astrophysical subpopulations which prefer narrowly peaked mass-ratio distributions about $q=0.5$. This is in agreement with previous investigations by \cite{Ray:2025xti,Sridhar:2025kvi}.

\begin{table*}[htt]
\centering
\centering
\begin{tabular}{cccc}
\hline
Metric  & ~~~~~~Comp. 1~~~~~  & ~~~~Comp. 2~~~~ & ~~~~Comp. 3 ~~~~\\
\hline
$P(\chi_{eff}<0)$ & $23.0^{+12.9}_{-10.3}\%$ & $51.7^{+16.1}_{-17.1}\%$ & $50^{+0}_{-0}\%$\\
$\sigma_{\chi_{eff}}$ & $0.07^{+0.03}_{-0.02}$ & $0.08^{+0.03}_{-0.03}$ & $0.30^{+0.06}_{-0.05}$\\
$\bar{\chi}_p$ & $0.24^{+0.07}_{-0.06}$ & $0.49^{+0.05}_{-0.16}$ & $0.50^{+0.10}_{-0.13}$\\
$P(q>0.7)$ & $50.52^{+20.32}_{-12.20}\%$ & $83.64^{+10.01}_{-15.90}\%$ & $11.00^{+9.98}_{-8.86}\%$\\
\hline
\end{tabular}
\caption{\label{tab:metrics} Metrics for astrophysical interpretation. The hyperposteriors of these metrics are shown in Figure~\ref{fig:metrics} of Appendix~\ref{sec:app-additional}.} For robustness against variation with functional forms and hyperpriors see Appendix~\ref{sec:app-metric-modcomp}~(Figures~\ref{fig:metric-2-comp}, \ref{fig:metric-3-comp}).
\end{table*}
\section{Astrophysical Interpretaion}
\label{sec:conclusion}
Our results reveal new correlations in the BBH populations (such as the distinct redshift evolution and spin precession properties of the $10M_{\odot}$ and $35M_{\odot}$ subpopulations) and provide a self-consistent interpretation of previously reported ones. In particular, mass-based transitions in the distributions of BBH parameters naturally emerge in our inferred distributions without explicit modeling. We have performed rigorous model selection studies and compared our results with those of data-driven analyses to show (in Appendices~\ref{sec:app:mod-comp}, and \ref{sec:app-trans}) that our findings are not driven by prior assumptions. We now discuss the astrophysical interpretation of these features being present in the data, which is largely independent of theoretical uncertainties in binary stellar evolution, BH formation, and properties of host environments. 

To that end, we start by quantifying robust tracers of BBH evolutionary pathways from our inferred distributions~\citep{Fishbach:2022lzq, Vitale:2025lms}, namely, the fraction of $\chi_{eff}<0$ events~($P(\chi_{eff}<0)$), the width of the $\chi_{eff}$ distribution~($\sigma_{\chi_{eff}}$), the median of the $\chi_p$ distributions~($\bar{\chi}_p$), and the fraction of $q>0.7$ events~($P(q>0.7)$), for each subpopulation, which are presented in Table~\ref{tab:metrics}. These metrics can be used to assess whether or not robust astrophysical predictions from various formation pathways are consistent with a particular component.

In particular, $P(\chi_{eff}<0)$ is expected to be $\ll50\%$ for subpopulations that prefer systems with component spins aligned to the orbital axis, whereas ones that prefer isotropic spin orientations are expected to have $P(\chi_{eff}<0)=50\%$~(see, for example, Figure~\ref{fig:chifrac-vs-ctfrac}, in Appendix~\ref{sec:app-eff-comp-spin}). Similarly, for a subpopulation that predominantly consists of slowly spinning black holes~(with dimensionless spin magnitudes~$<0.3$), one would expect small values of $\sigma_{\chi_{eff}}<0.15$, and vice-versa for higher spin-magnitude values~(see, for example, Figure~\ref{fig:sig-vs-a} in Appendix~\ref{sec:app-eff-comp-spin}). In addition, for any given range of spin magnitude values, one can expect $(\bar{\chi}_p)$ to be high if there is a preference for systems with in-plane spin components, and small otherwise (see, for example, Figure~\ref{fig:barchip-vs-finp}, in Appendix~\ref{sec:app-eff-comp-spin}). Finally, for subpopulations which prefer equal mass systems, $P(q>0.7)$ is expected to be large$(>50\%)$, and vice-versa for ones that prefer unequal mass systems since unequal masses correspond to $q\ll1$. Given the inferred values of $(P(\chi_{eff}<0), \sigma_{\chi_{eff}}, \bar{\chi}_p, P(q>0.7) )$ for each component (as presented in Table~\ref{tab:metrics}), we now turn to predictions of BBH formation for these metrics that are largely independent of persisting theoretical uncertainties.  

\begin{table*}[htt]
\centering
\centering
\begin{tabular}{ccccc}
\hline
Channel  & ~~ $P(\chi_{eff}<0)$ ~~ & ~~ $\sigma_{\chi_{eff}}$~~  & ~~ $\bar{\chi}_p$~~ & ~~ $P(q>0.7)$~~ \\
\hline
Isolated binaries & $\ll50\%$ &  small & small & $>50\%$\\
1G+1G in clusters & $=50\%$ &  small & small & $\gg 50\%$\\
Triple systems  & \textbf{?} & small & large & \textbf{?}\\
Hierarchical & $=50\%$ &  large & large & $\ll 50\%$ \\
\hline
\end{tabular}
\caption{\label{tab:predictions}A summary of expected values of various metrics that are likely to be consistent with particular formation channels. Here, ? represents the fact that there remains substantial uncertainty and that robust predictions necessitate further investigation. }
\end{table*}

\begin{table*}[htt]
\centering
\centering
\begin{tabular}{cccc}
\hline
Channel  & ~~ Comp. 1  ~~ & ~~ Comp. 2~~  & ~~ Comp. 3~~ \\
\hline
Isolated binaries & \CheckmarkBold &  \XSolidBold & \XSolidBold\\
1G+1G in clusters & \XSolidBold &  \CheckmarkBold & \XSolidBold\\
Triple systems  & \textbf{?} & \textbf{?} & \XSolidBold\\
Hierarchical & \XSolidBold &  \XSolidBold & \CheckmarkBold \\
\hline
\end{tabular}
\caption{\label{tab:conclusion} Plausiable origins of each subpopulation. Here, \CheckmarkBold~means plausible, \XSolidBold~implies unlikely, and \textbf{?} indicates unlikely but plausible that the corresponding channel contributes dominantly to the respective subpopulation. }
\end{table*}

In isolated binary evolution, binary interactions such as mass transfer and tides are expected to preferentially align spin orientations with the orbit~\citep{Kalogera:1999tq, Bavera:2020inc, Gerosa:2018wbw, Steinle:2022rhj}. Higher fractions of anti-aligned systems are only possible through large natal kicks imparted by the supernovae during BH formation~\citep{Wysocki:2017isg, Callister:2020vyz, Fragione:2021qtg}. However, large kick magnitudes can lead to disruption of the binary and are found to be inconsistent with observations of Galactic BH binary systems~\citep{Mandel:2015eta, Mirabel:2016msh}. Hence, for a subpopulation of isolated binaries, large values of $\bar{\chi}_p$, and $P({\chi}_{eff}<0) \sim 50\%$ are unlikely. This prediction is largely insensitive to uncertainties in binary evolution and stellar collapse except possibly for the potential of off-axis kicks that can toss the BH spin axis~\citep{Tauris:2022ggv}. Note, however, that for certain kick prescriptions, isolated binary evolution can still be consistent with moderate values $P(\chi_{eff}<0)$ such as those exhibited by component 1 but not with the ones obtained for components 2 and 3 (see Appendix~\ref{sec:app-iso}).

On the other hand, as BHs born through stellar collapse, the components of isolated binaries should either exhibit an upper mass gap due to the pair instability process or pollute it through mechanisms (such as super-Eddington accretion during mass-transfer onto the first-born BH or chemical mixing in tight binaries) that also lead to preferentially high spins and aligned spin orientations~\citep{Briel:2022cfl,vanSon:2020zbk, deMink:2016vkw}. Very high mass, isotropic spin~($P(\chi_{eff}<0)=50\%$) systems are, therefore, also difficult to explain through the isolated channel.

For 1G+1G dynamical assembly in dense star clusters, randomization of spin orientations leads to equal preference for aligned and anti-aligned components~\citep{Mapelli:2021gyv, Chattopadhyay:2023pil, Rodriguez_2022}. Proposed methods for efficient spin alignment, such as BBH--star collisions~\citep{Kiroglu:2025bbp}, only comprise at most 10\% of all cluster mergers and can also lead to high spin magnitudes~\citep{Kiroglu:2024xpc}, even though the latter is subject to uncertainties in accretion efficiency. This likely makes dynamical assembly inconsistent with subpopulations that simultaneously have small $\sigma_{\chi_{eff}}$, $P(\chi_{eff}<0)\ll50\%$, and large $\bar{\chi}_p$. This channel also prefers equal-mass binaries due to mass segregation causing heavier objects to sink to the center of the cluster~\citep{Rodriguez:2016vmx, Farr:2017uvj, Antonini2023}, and strong gravitational interactions between three or more bodies preferentially leading to the more massive components forming a bound binary and ejecting less massive components~\citep{Heggie:1997gq}. While mechanisms for polluting the PISN gap through this channel have been proposed, the resulting mergers continue to prefer equal-mass systems~\citep{Kiroglu:2024xpc, Kiroglu:2025bbp}, leading to inconsistency with a high-mass subpopulation that exhibits strong mass asymmetry. 


Similarly, BBH mergers with spins aligned to the orbit are unlikely to occur in triple systems, since Lidov-Kozai resonance in the inner binary induced by a distant tertiary component can preferentially orient binary spins perpendicular to the orbit. The predicted distributions of cosine tilts exhibit a global peak at zero and a heavier tail towards positive values as compared to negative ones~\citep{Antonini:2017ash, Rodriguez:2018jqu, Liu:2018nrf, Stegmann:2025zkb}. This is consistent with high $\bar{\chi}_p$ and a $\chi_{eff}$ distribution that peaks very close to zero and is skewed towards positive values. The width of this $\chi_{eff}$ distribution is expected to be a monotonic function of component spin magnitudes, with small magnitude values leading to suppressed tails and a sharper peak that is narrowly close to $\chi_{eff}=0$~\citep{Antonini:2017ash}. Hence, it is possible that BBHs in triple systems are likely inconsistent with a subpopulation that simultaneously exhibits $P(\chi_{eff}<0) \ll 50\%$ and small values of $\sigma_{\chi_{eff}}$, although further investigation and direct comparison with the predicted distributions is needed to verify this conclusion. Note, however, that triple predictions that are consistent with the small $\sigma_{\chi_{eff}}$ and $P(\chi_{eff}<0) \ll 50\%$ (component 1) will likely be inconsistent with component 2~(small $\sigma_{\chi_{eff}}$ and $P(\chi_{eff}<0)\sim 50\%$) and vice-versa. 

Finally, in the hierarchical channel, components that are remnants of previous mergers, through a robust prediction of numerical relativity simulations, have high spin magnitudes of $\sim 0.67$, with dispersion dependent on spin magnitudes, orientations, and mass ratios \citep{Berti:2008af, Hofmann:2016yih, Rodriguez:2019huv, Borchers:2025sid, Fishbach:2017dwv}. This will lead to inconsistency with subpopulations that have small $\sigma_{\chi_{eff}}$. Furthermore, due to high recoil kicks, the rarity of retained merger remnants in dynamical environments leads to hierarchical mergers being predominantly 2G+1G systems which exhibit strong mass asymmetry, with less than a few percent of systems expected in the $q>0.7$ range~\citep{Fitchett:1983qzq,PortegiesZwart:1999nm,Favata:2004wz,Gonzalez:2006md,Lousto:2009ka,Gerosa:2018qay,Mahapatra:2021hme, Zevin:2022bfa}. This channel can also straightforwardly pollute the PISN gap in $m_1$, extending up to very high masses~\citep{OLeary:2005vqo, Antonini:2016gqe,Tagawa:2020qll, Mapelli:2021syv, Antonini:2022vib, Torniamenti:2024uxl, Vaccaro:2025ogk, Rodriguez:2019huv}.

Given these predictions~(summarized in Table~\ref{tab:predictions}) that are largely robust against uncertainties in binary stellar evolution, BH formation, and properties of host environments, and the trends in our subpopulations summarized in Table~\ref{tab:metrics}, we conclude that only specific channels likely contribute dominantly to each subpopulation, which is presented in Table~\ref{tab:conclusion}.

While these conclusions are reasonably robust, the direct association of subpopulations with single channels remains elusive. In particular, dynamical formation in AGN disks can give rise to aligned, isotropic, and in-plane spin orientations depending on assumptions in the modeling of disk physics~\citep{Tagawa:2020dxe, McKernan:2021nwk, Li:2022cul,Mckernan:2017ssq, Santini:2023ukl, McKernan:2023xio, McKernan:2024kpr, Cook:2024ajp, Fabj:2025vza}. Similarly, this channel is capable of producing both equal and unequal mass systems depending on various underlying assumptions. In other words, this channel can, in principle, give rise to features consistent with all three subpopulations. On the other hand, for Subpopulation 2, it is unclear whether there is an excess of in-plane systems compared to what is expected for an isotropic distribution. While the $\chi_{eff}$, and $\chi_p$ distributions can together elucidate this mystery, measurement uncertainties in GWTC-4 cannot distinguish between triples and dynamical assembly in dense star clusters as the origin of this subpopulation. 

\begin{figure}[htt]
    \centering
    \includegraphics[width=0.98\linewidth]{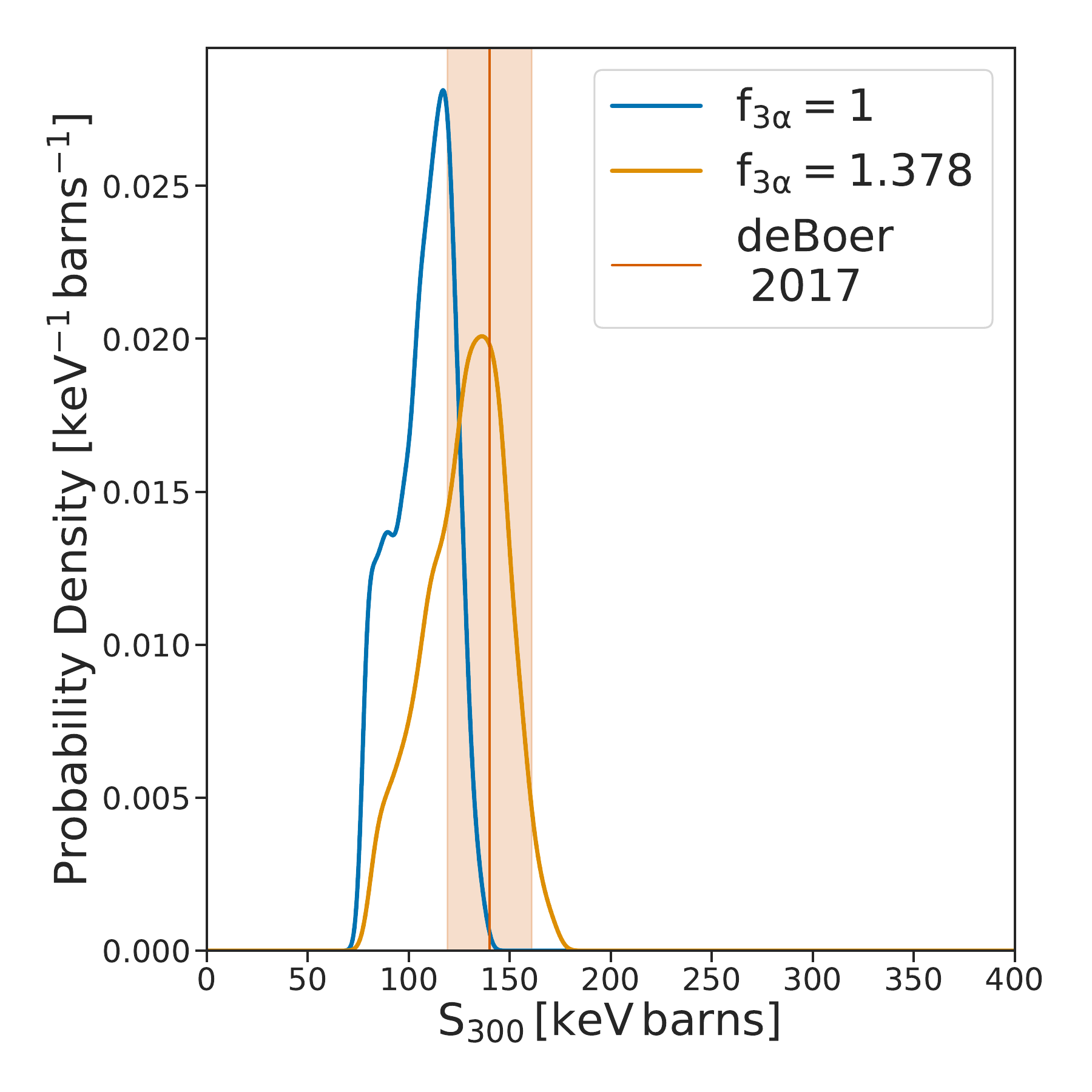}
    \caption{Constraints on the S factor of $^{12}\mathrm{C}(\alpha, \gamma)^{16}O$ at $300~\mathrm{kev}$ obtained using the stellar models of \cite{Farag:2022jcc}, depending on the chosen rate of the $3\alpha$ process, which were taken to be either equal to or $37.8\%$ larger than the values reported in the NCARE compilation~\citep{1999NuPhA.656....3A, Farag:2022jcc}. The shaded region region represents the whole range of values reported by \cite{deBoer:2017ldl} from independent measurements.}
    \label{fig:S2300}
\end{figure}

Despite these degeneracies in interpretation, one possibility consistent with our observed subpopulations merits further discussion. Theoretical simulations show that isolated binary evolution through stable mass transfer can give rise to the $10M_{\odot}$ peak~\citep{vanSon:2022myr}, support for mass ratios in the region $0.6\text{--}1.0$~\citep{vanSon:2020zbk, Olejak:2024qxr, Dorozsmai:2022wff, Broekgaarden:2022nst}, and a preference for slowly spinning components aligned to the orbits~(as exhibited by Subpopulation 1). A sharp peak near $10M_{\odot}$ has not been predicted through AGN disks in existing literature to the best of our knowledge. Similarly, while both triples and 1G+1G mergers in star clusters can be consistent with the spin and mass ratio properties of Subpopulation 2, simulations of dynamical assembly in globular clusters have demonstrated an overdensity consistent with the $35M_{\odot}$ feature~\citep{Bruel:2025sdq}, whereas such predictions are yet to be made for triples. Furthermore, the ratio of branching fractions between Subpopulations 3 and 2 is $0.15^{+0.41}_{-0.11}$, which is consistent with theoretical expectations for the relative abundance of hierarchical and 1G+1G mergers in globular clusters~\citep{Rodriguez:2019huv}. Note, however, that the shallower redshift evolution of Subpopulation 2 relative to 3 would then imply a cluster mass function that likely evolves with redshift~\citep{Farah:2026jlc, Ye:2024ypm, Mai:2025jmk}. 

To summarize, we find that the GWTC-4 BBH sample of detections comprises three distinct subpopulations with local merger rates of $11.3^{+5.3}_{-4.0}\text{Gpc}^{-3}\text{yr}^{-1}$, $2.2^{+1.8}_{-1.1}\text{Gpc}^{-3}\text{yr}^{-1}$, and $0.42^{+0.76}_{-0.32}\text{Gpc}^{-3}\text{yr}^{-1}$, that are likely consistent with specific formation channels. Our narrative interprets the mass-based transitions as emerging from the specific relative abundances of these three channels. Under this hypothesis, we interpret the common maximum mass of Subpopulations 1~(likely mergers from isolated binary evolution) and 2~(likely 1G+1G mergers in globular clusters) as the PISN cutoff and constrain the S factor\footnote{The S-factor is the part of the cross-section for charged-particle nuclear reactions that is dependent on nuclear structure, independent of Coulomb repulsion. The $^{12}\mathrm{C}(\alpha, \gamma)^{16}O$ reaction rate in massive stellar interiors is an uncertain paramter in theoretical models for PISNe and maps directly to the maximum mass of 1G BHs. We use the theoretical fits of \cite{Farag:2022jcc}~(presented in their Table 2) to map our posterior on $m_{max(1or2)}$ to that of the S-factor of $^{12}\mathrm{C}(\alpha, \gamma)^{16}O$ at $300~\mathrm{kev}$, which is presented in Figure~\ref{fig:S2300}.} of $^{12}\mathrm{C}(\alpha, \gamma)^{16}O$ at $300~\mathrm{kev}$~(Figure~\ref{fig:S2300}) to be in strong agreement with independent measurements~\citep{deBoer:2017ldl}. With new data releases imminent by the LVK, and ongoing developments in detailed binary evolution modeling, robust conclusions on this hypothesis and more precise measurements of stellar-evolution uncertainties can be expected in the near future.

\section{Acknowledgements}
We thank Aleksandra Olejak, Fulya K{\i}ro{\u{g}}lu, Jakob Stegman, Fabio Antonini, and Ish Gupta for useful discussions and suggestions. We are further grateful for A.O.'s assistance with parsing population synthesis datasets that are used in the appendix. A.R. was supported by the National Science Foundation (NSF) award PHY-2512923. M.Z. gratefully acknowledges funding from the Brinson Foundation in support of astrophysics research at the Adler Planetarium. V.K. was supported by the Gordon and Betty Moore Foundation (grant awards GBMF8477 and GBMF12341), through a Guggenheim Fellowship, and the D. I. Linzer Distinguished University Professorship fund. We are grateful for the computational resources provided by the LIGO laboratory and supported by National Science Foundation Grants PHY-0757058 and PHY-0823459. This material is based upon work supported by NSF’s LIGO Laboratory, which is a major facility fully funded by the National Science Foundation. We gratefully acknowledge the support of the NSF-Simons AI-Institute for the Sky (SkAI) via grants NSF AST-2421845 and Simons Foundation MPS-AI-00010513.

\appendix

\section{Additional Results: Cumulative Density Functions and Posteriors of Metrics}
\label{sec:app-additional}
\begin{figure}[htt]
    \centering
    \includegraphics[width=0.5\linewidth]{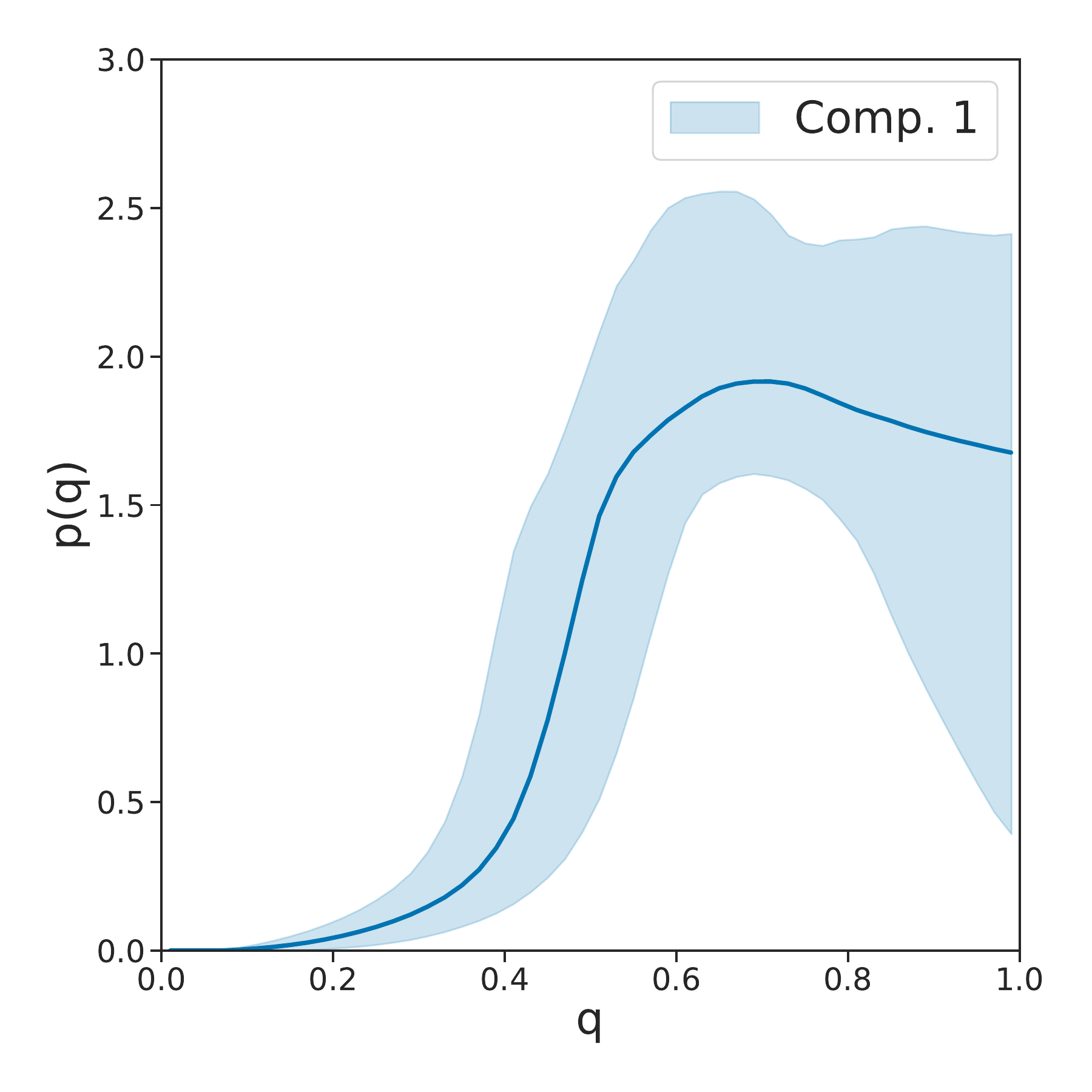}
    \caption{Features in the inferred mass-ratio distribution of \textit{Comp. 1}.} 
    \label{fig:q-1}
\end{figure}
In this section, we present additional results for characterizing the three subpopulations and investigating their astrophysical interpretations, a summary of which was presented in the main text. In particular, we present cumulative density functions corresponding to the population distributions presented in the main text of the paper, the joint posterior distribution of integral metrics whose median values and error bars were used as the basis of our astrophysical interpretation, and the posterior distributions of various mass-scales that demarcate the transitions in the distributions of other BBH parameters, whose central values and uncertainty estimates were also quoted in the main text. Furthermore, we investigate whether or not there are additional features in some of our inferred density functions, which, although more uncertain than the features highlighted in the main text, might hint towards interesting astrophysical implications that might can be explored in more detail with future catalogs.

We first investigate the features of the mass-ratio distribution of \textit{Comp. 1} by zooming into the inferred density function that was presented in the top middle panel of Figure.~\ref{fig:subpops}. As can be seen in Figure~\ref{fig:q-1}, the credible intervals are consistent with a mild overdensity in the $0.6-0.7$ region~(in agreement with the findings of previous studies that have used more flexible population models on the same dataset). Note however that the median is consistent with a broad peak that has support over the entire range of mass-ratio values $q\in(0.6,1.0)$. With larger catalogs to be imminently released by the LVK, we can expect more definitive conclusions on the exact shape of the mass-ratio distribution of \textit{Comp. 1}.
\begin{figure}[htt]
    \centering
    \includegraphics[width=0.32\linewidth]{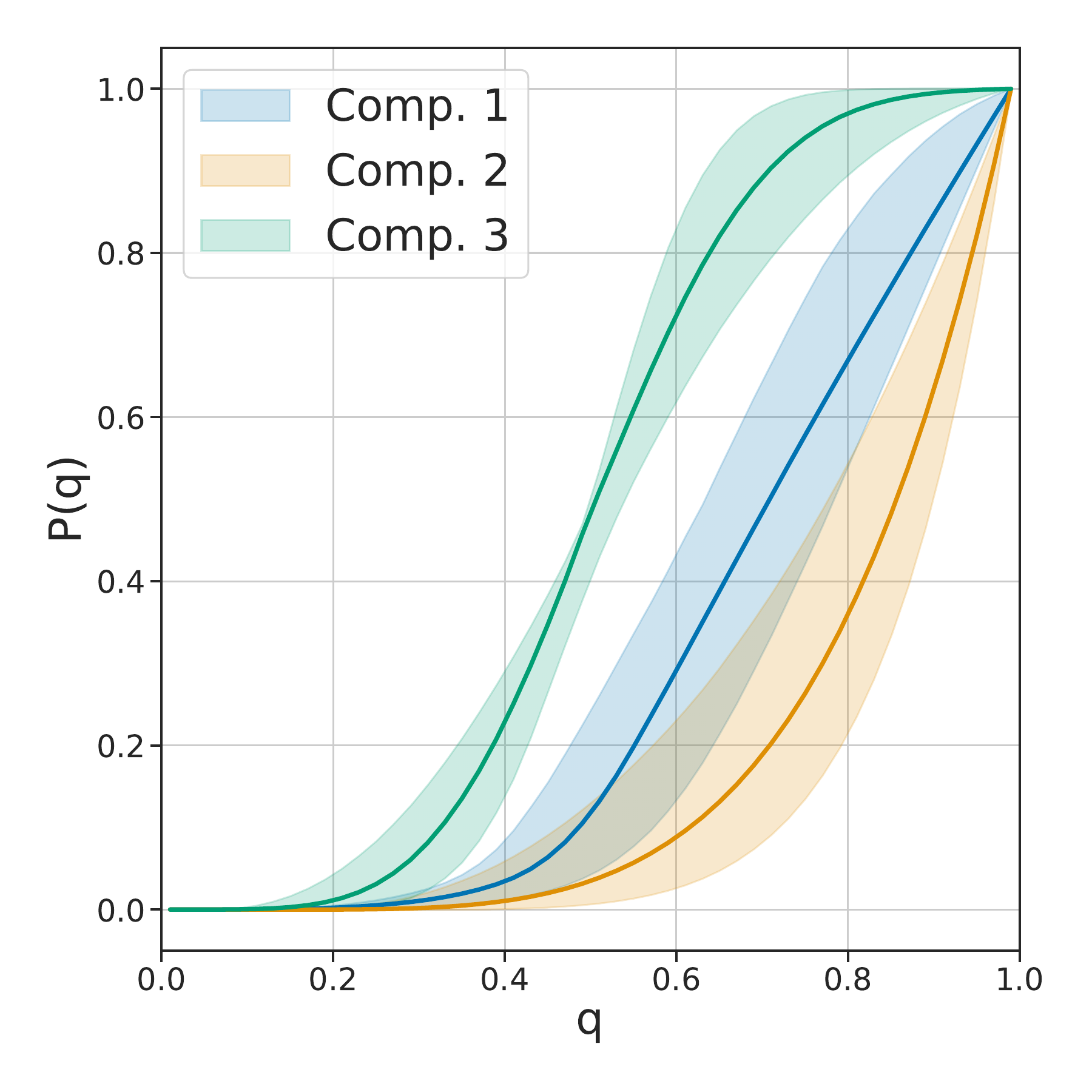}
    \includegraphics[width=0.32\linewidth]{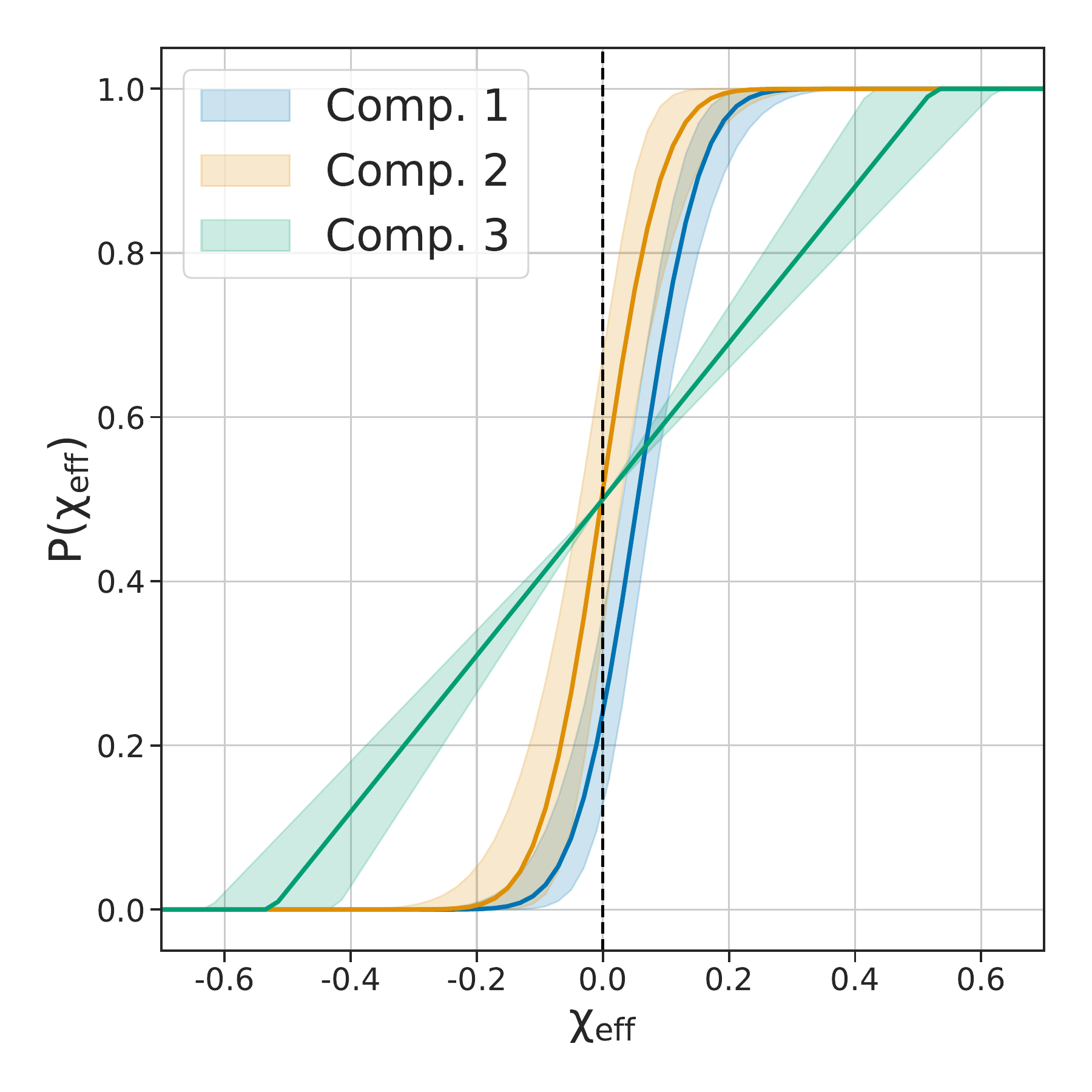}
    \includegraphics[width=0.32\linewidth]{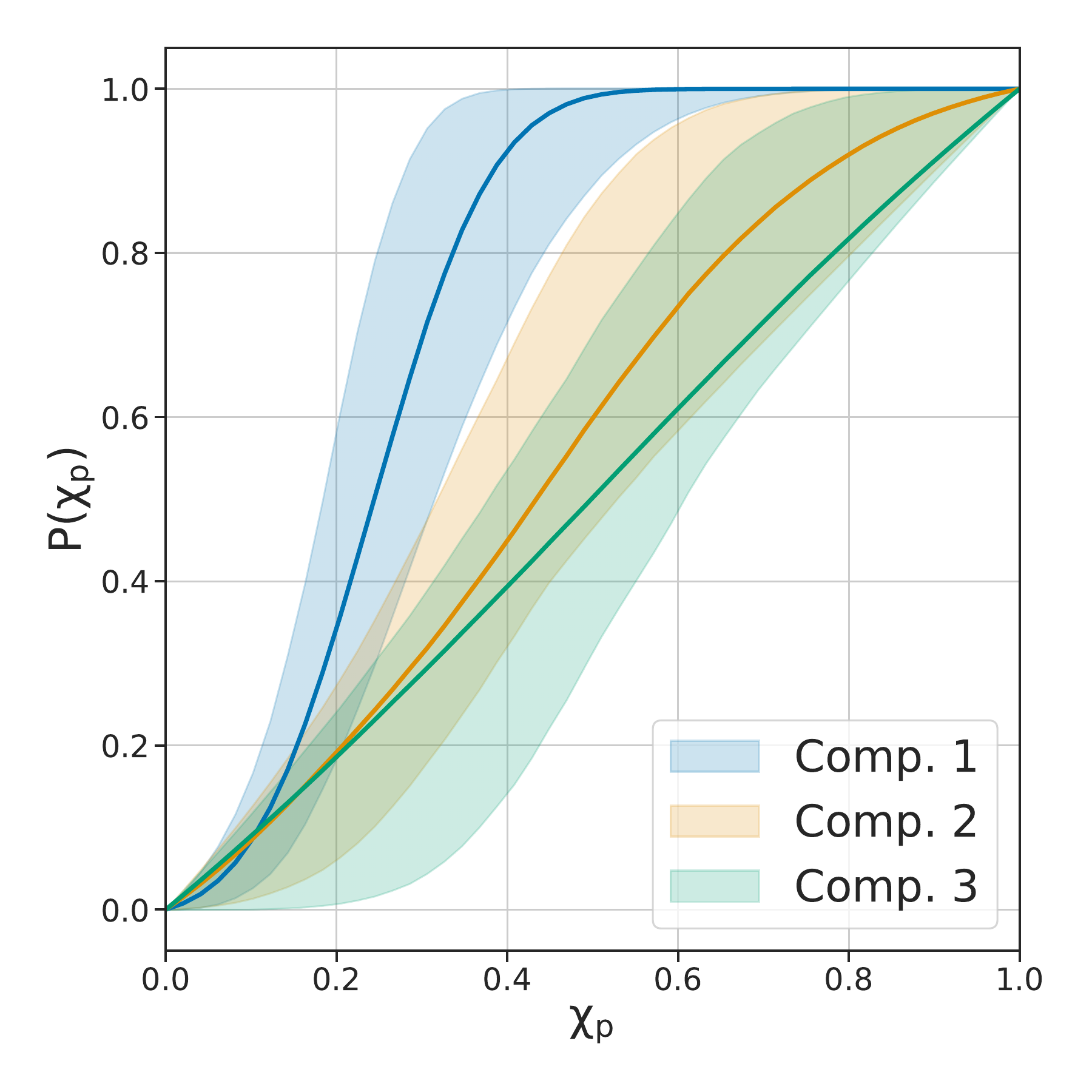}
    \caption{Inferred cumulative density functions of mass ratio~(left), $\chi_{eff}$~(center), and $\chi_p$~(right) corresponding to the population distributions of each component that was presented in Figure~\ref{fig:subpops}.} 
    \label{fig:cdfs}
\end{figure}

\begin{figure}[htt]
    \centering
    \includegraphics[width=0.8\linewidth]{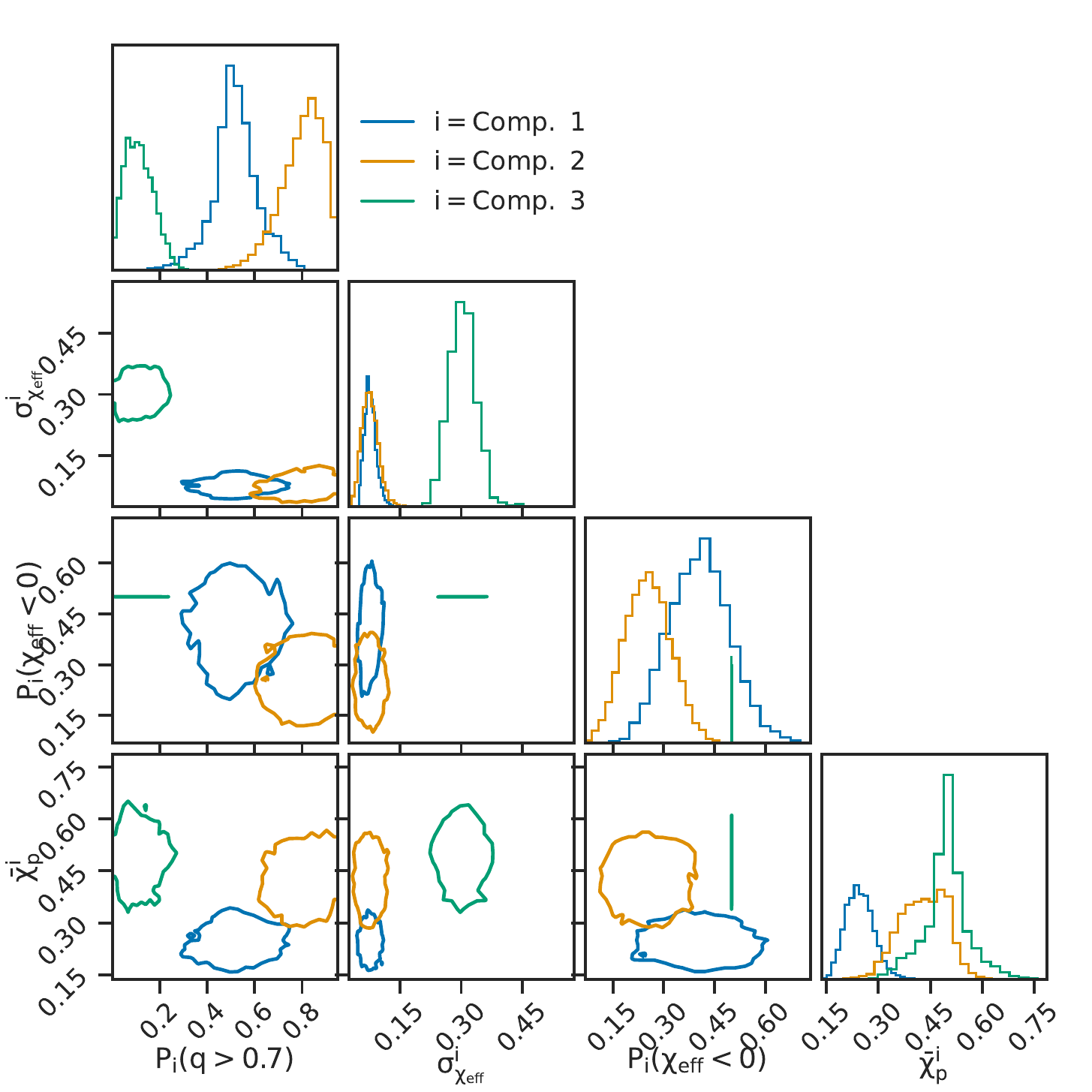}
    \caption{Joint posterior distributions of the integral metrics reported in Table~\ref{tab:metrics} that were used to derive our astrophysical interpretation.} 
    \label{fig:metrics}
\end{figure}

We now present the inferred cumulative density functions of mass ratios and effective spins, whose corresponding population distributions were presented in the main text. While the inferred distributions presented in Figure~\ref{fig:subpops} are illustrative of the distinct trends in each subpopulation, the values of various integral metrics (such as fraction of events in different ranges of parameter space) that are crucial to our astrophysical interpretation are difficult to read off directly from the inferred density functions. Hence, in Figure~\ref{fig:cdfs}, we also present the cumulative density functions corresponding to the distributions presented in the main text, which can be directly compared with Figure~\ref{fig:metrics}, which shows the joint posterior distributions of the metrics reported in Table~\ref{tab:metrics}. 

\begin{figure}[htt]
    \centering
    \includegraphics[width=0.8\linewidth]{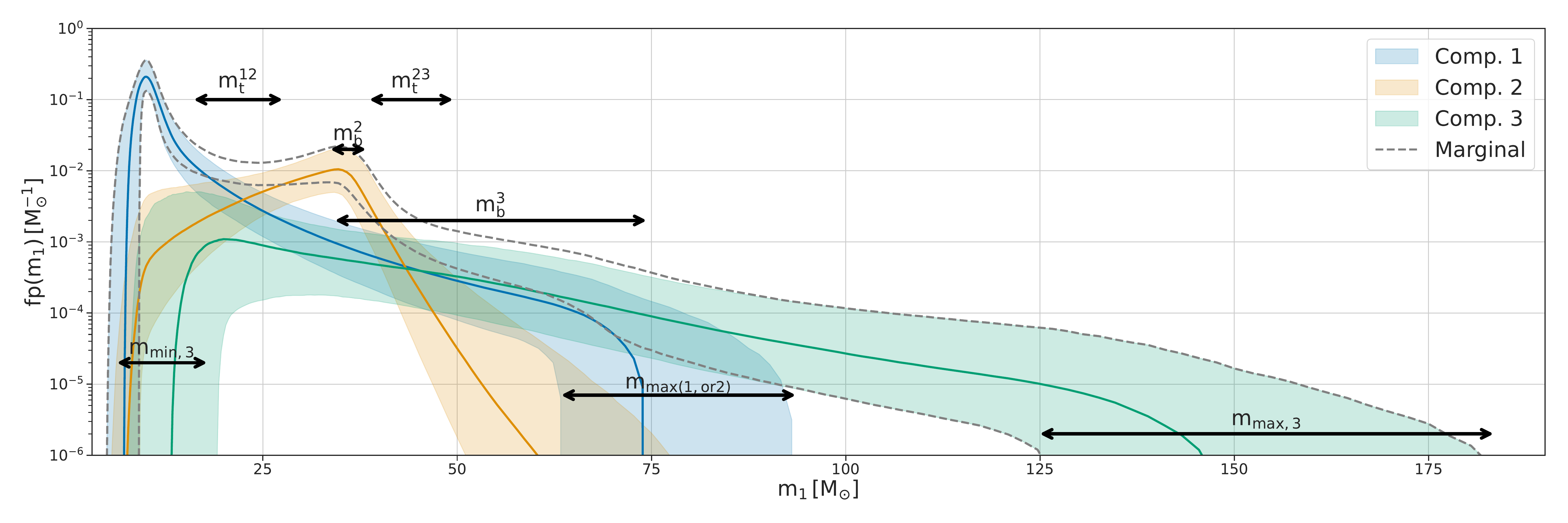}
    \caption{Various mass-scales visualized with the features of the marginal mass-distribution. The arrows corresponding to each quantity span the $90\%$ credible interval.} 
    \label{fig:m1-wide}
\end{figure}

\begin{figure}[htt]
    \centering
    \includegraphics[width=0.8\linewidth]{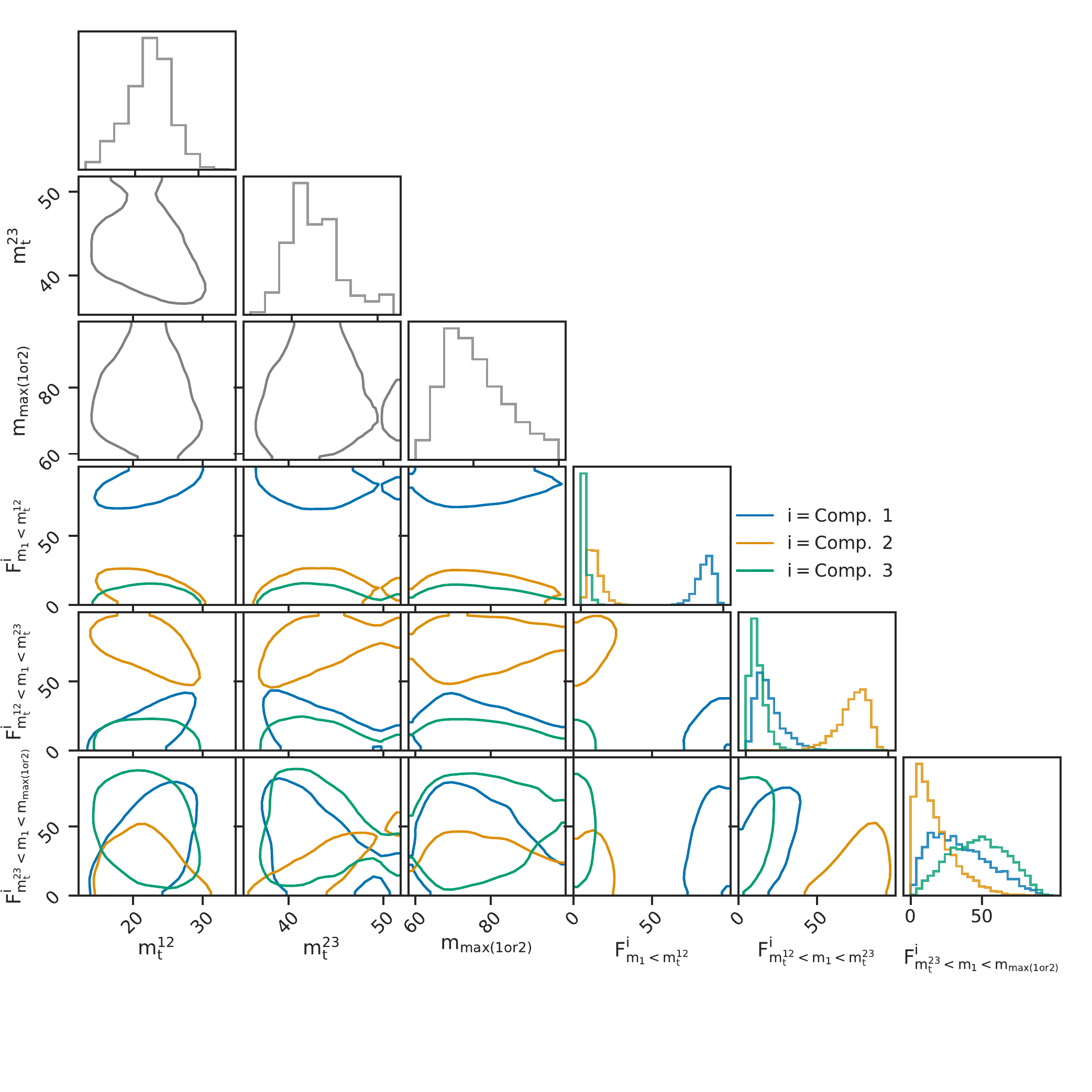}
    \caption{The joint posterior distribution of the three mass scales $(m_{t}^{12}, m_{t}^{23}, m_{max(1or2)})$ that represent transitions in the distribution of other BBH parameters, along with those of the fraction of systems from each component that contribute to the astrophysical merger rate in the different regions demarcated by the transition scales (which were quoted in Table~\ref{tab:relative-frac}).} 
    \label{fig:mass-scales}
\end{figure}

Finally, we present additional details about the various mass-scales recovered in our analyses including hyperparameters such as breaks in the broken powerlaw models of Comp. 2 and 3, the maximum masses of various components as well as derived quantities such as the transition mass-scales whose posteriors were reconstructed in postprocessing from the inferred distributions of each component. In Figure~\ref{fig:m1-wide} we visualize which features of the marginal mass distribution are represented by these various mass-scales. In addition, we show in Figure~\ref{fig:mass-scales}, the joint posterior distribution of the three mass scales $(m_{t}^{12}, m_{t}^{23}, m_{max(1or2)})$ that represent transitions in the distribution of other BBH parameters, along with those of the fraction of systems from each component that contribute to the astrophysical merger rate in the different regions demarcated by the transition scales~(which were quoted in Table~\ref{tab:relative-frac}).

\section{The effective spin distributions of isolated binaries}
\label{sec:app-iso}

\begin{figure}[htt]
    \centering
    \includegraphics[width=0.5\linewidth]{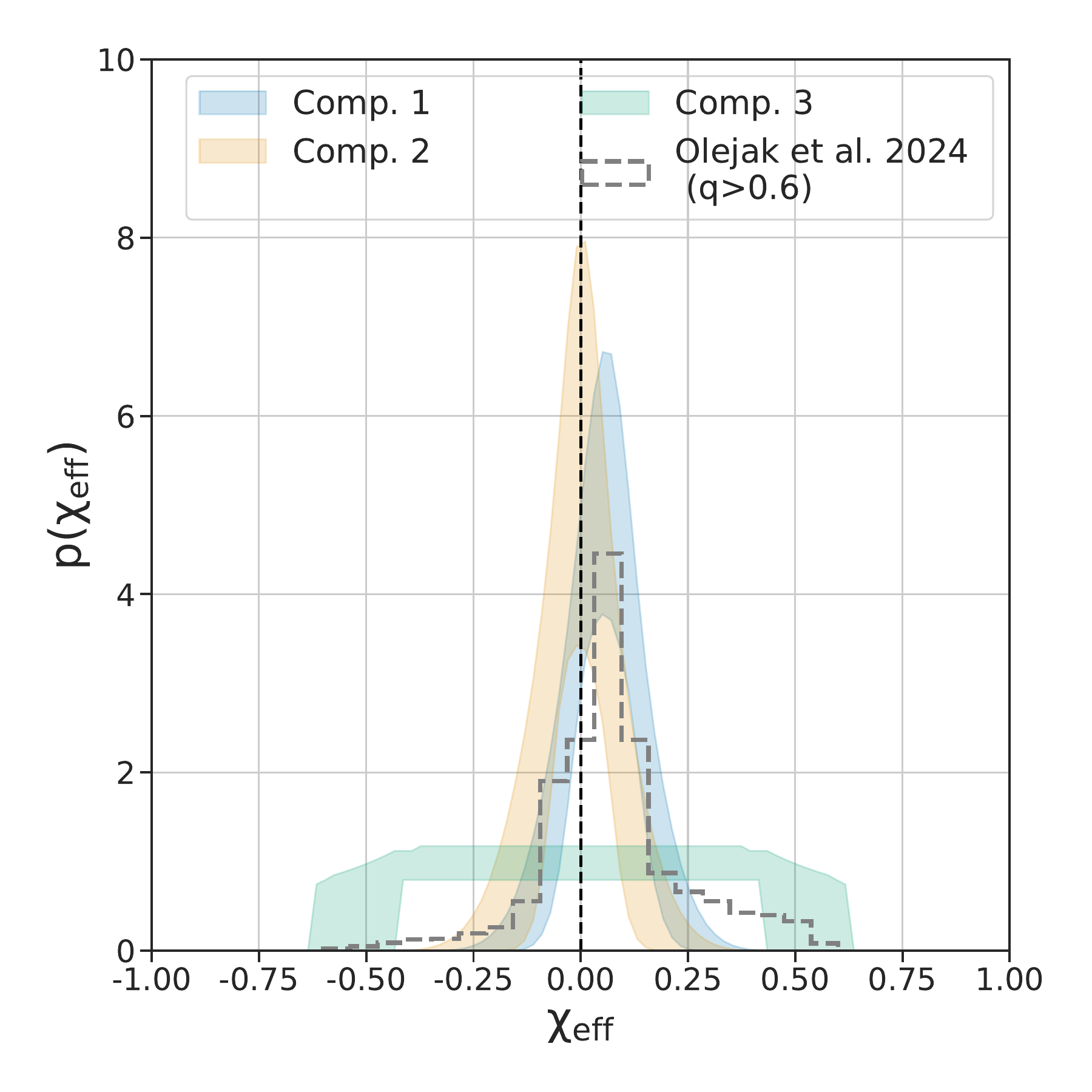}
    \caption{Comparing the theoretical predictions from certain models of isolated binary evolution through the stable mass transfer channel.} 
    \label{fig:chieff-iso}
\end{figure}

In this section, we show that for certain assumptions on binary mass transfer and the natal kicks received by the supernova remnants, isolated binary evolution can, in fact, produce an effective spin population consistent with our inferred distributions for component 1. While isolated binary evolution can strongly prefer systems whose component spins are aligned to the orbital axis, \cite{Olejak:2024qxr} show that for high natal kicks (i.e. with magnitudes drawn from a Maxwellian distribution of width $\sigma = 133~\mathrm{km\,s^{-1}}$ and non-decreased by fallback), the stable mass transfer channel can lead to subpopulation of systems with a small but non-negligible $\chi_{eff}<0$ tail consistent with our findings for component 1.

In Figure~\ref{fig:chieff-iso}, we compare the effective spin distribution of a synthetic population of isolated BBH mergers simulated using the \texttt{StarTrack} code~\citep{Belczynski:2005mr, Belczynski:2017gds, Belczynski:2020bca, Olejak:2021iux, Olejak:2021fti, Olejak:2022zee}, which are taken from the public datasets~\citep{olejak:2024-12530199} released by \cite{Olejak:2024qxr}, with our inferred ones. We find that the population of simulated BBHs with mass-ratios of $0.6$ or higher can be consistent with the trends of effective spins identified for component 1, but not the other two components. 

Note that the width of the tails in the theoretical $\chi_{eff}$ distribution strongly depends on the chosen value of $\sigma$. Furthermore, the impact of $\sigma$ values on the $\chi_{eff}$ distribution can be different depending on the prescriptions used to model the natal kicks of compact remnants~\citep{Olejak:2021iux, Olejak:2024qxr}. However, for any given kick prescription, lower values of $\sigma$ can be expected to give rise to smaller kicks (hence smaller spins and narrower $\chi_{eff}$ distributions) than higher values of $\sigma$. Considering that the within the non-fallback decreased kick models of \cite{Olejak:2024qxr}, $\sigma = 133\mathrm{km\,s^{-1}}$ corresponds to vary high natal kicks, we expect that the small disagreement between the theoretical prediction and the inferred distribution for component 1 near the tails can, in principle, be mitigated by fine-tuning $\sigma = 133~\mathrm{km\,s^{-1}}$ to reasonably lower values. While these predictions are susceptible to additional assumptions about binary evolution, it is clear that component 1 can originate from these particular models of the isolated channel, whereas component 2 and component 3 are less likely to do so.

\section{Population Analysis}
\label{sec:app-gwpop}
The results presented in this work were obtained through Bayesian hierarchical inference using a three-component mixture model for the joint distribution of BBH primary masses~$(m_1)$, mass ratios~$(q)$, effective aligned~$(\chi_{eff})$ and precessing~$(\chi_p)$ spins, and redshifts~$(z)$, which takes the following form: 
\begin{widetext}
\begin{equation}
    \frac{dN}{dm_1dqd\chi_{eff}d\chi_pdz}(\vec{\Lambda}) = R_0\frac{dV}{dz}\frac{T_{obs}}{1+z}\sum_{i=1}^3f_ip_i(q|m_1, \vec{\Lambda}_{q}^i)p_i(m_1|\vec{\Lambda}_{m}^i)p_i(\chi_{eff}, \chi_p|\vec{\Lambda}_{\chi}^i)(1+z)^{\kappa_i},
\end{equation}
\end{widetext}
where $R_0$ is the local rate per comoving volume~$(V)$ and source-frame time~$(\frac{T_{obs}}{1+z})$, $f_i$ is the branching fraction of the formation channel contributing dominantly to the $ith$ component, and $\kappa_i$ its corresponding redshift evolution parameter~\citep{Fishbach:2018edt}. Here, $\vec{\Lambda}_\theta^{i}$ represents additional hyperparameters that control the shape of the distribution function of the BBH parameter $\theta$ for the \textit{ith} component. Our mixture components are constructed from simple functional forms, which are delineated in the next section. See also \cite{Tiwari:2020otp, Tiwari:2021yvr, Tiwari:2025oah}, who use generic mixtrue models for a flexible reconstruction of the joint BBH population distribution.

We constrain our distributions by modeling the occurrence of BBH mergers as an inhomogeneous Poisson process~\citep{Mandel:2018mve, Thrane:2018qnx, Wysocki2019}. We construct the likelihood function of population hyperparameters using individual event parameter estimation data for all GWTC-4 BBH candidates found with a false alarm rate of less than 1 per year. We use the exact same PE samples used by \cite{LIGOScientific:2025pvj}, which are delineated in their section 3.4. We rely on simulated signals injected into detector noise realizations to correct for Malmquist biases that might result from the choice of a stringent detection criterion. Using publicly released LVK data for single event PE runs and detectable injections, we constrain the posterior distribution of population hyperparameters and compute the Bayesian evidence in favour of model assumptions by means of nested sampling techniques~\citep{Skilling:2006gxv}. We ensure the convergence of Monte Carlo sums used to construct our likelihood function using the standard variance-based penalty proposed by \cite{Talbot:2023pex}. We rely on the \texttt{gwpopulation} library~\citep{Talbot2025,2019PhRvD.100d3030T} for implementing our hierarchical inference. The code developed to implement this analysis will be made publicly available after the acceptance of this manuscript, and, in the meantime, will be shared upon reasonable request.

\section{Model Comparison: Functional Forms \& Hyperpriors } 
\label{sec:app:mod-comp}
We now explore model variations to investigate the prior dependence of our results. Our mixture components are constructed from simple parametrizations such as powerlaws~(\texttt{PL}), smoothed powerlaws~(\texttt{SPL}), smoothed broken power laws~(\texttt{BPL}), smoothed broken power laws + Gaussian peak~(\texttt{BPLP}), smoothed power laws mixed with Gaussian peaks~(\texttt{PLP}), truncated Gaussians~($\mathcal{N}$), and uniform distributions~$(U)$ which are expressed below:

\begin{equation}
    \texttt{PL}^{\alpha}_{a,b}(x) = \begin{cases}\frac{(1+\alpha)x^{\alpha}}{b^{1+\alpha}-a^{1+\alpha}}, & a\leq x \leq b\\
    0, & o.w.\end{cases}
\end{equation}
\begin{equation}
    \texttt{SPL}^{\alpha,\delta}_{a,b}(x) \propto S(x,a,b,\delta_m) \texttt{PL}^{\alpha}_{a,b}(x)
\end{equation}
\begin{equation}
  \mathcal{N}_{a,b}^{\mu,\sigma}(x) \propto \begin{cases}
      e^{-\frac{1}{2}\left(\frac{x-\mu}{\sigma}\right)^2}, & a\leq x \leq b\\
    0, & o.w.
  \end{cases}  
\end{equation}
\begin{equation}
    U_{a,b}(x) = \begin{cases}
        \frac{1}{b-a}, & a\leq x \leq b\\
        0, & o.w.
    \end{cases}
\end{equation}
\begin{equation}
    \texttt{PLP}^{\alpha,\delta, \lambda, \mu,\sigma}_{a,b}(x) = (1-\lambda)\texttt{SPL}_{a,b}^{\alpha,\delta}(x) + \lambda \mathcal{N}_{a,b}^{\mu,\sigma}(x)
\end{equation}
\begin{equation}
    \texttt{BPL}^{\alpha_1, \alpha_2, x_b \delta}_{a,b}(x) \propto S(x,a,b,\delta_m) \begin{cases}
        \texttt{PL}_{a,x_b}^{\alpha_1}(x)\frac{\texttt{PL}_{a,x_b}^{\alpha_1}(x_b)}{\texttt{PL}_{x_b,b}^{\alpha_2}(x_b)},& x \leq x_b,\\
        \texttt{PL}_{x_b,b}^{\alpha_2}(x), & o.w.
    \end{cases}
\end{equation}
and,
\begin{align}
    \texttt{BPLP}^{\alpha_1, \alpha_2, x_b, \delta, \lambda, \mu,\sigma}_{a,b}(x) &= (1-\lambda)\texttt{BPL}_{a,b}^{\alpha_1,\alpha_2,x_b,\delta}(x) \nonumber \\
    &+ \lambda\mathcal{N}_{a,b}^{\mu,\sigma}(x)
\end{align}
where,
\begin{equation}
    S(x,a,b,\delta)= \frac{1}{1+e^{\frac{\delta}{\delta - x + a}-\frac{\delta}{x-a}}}
\end{equation}
is a smoothing window for the turn on of power laws. Note that the constants of proportionality are dervided by imposing that each distribution function is normalized. For our most preferred model, the functional forms chosen for each component's distribution functions were listed in the main text (Table~\ref{tab:models}). We also define a Broken Powerlaw plus two Gaussian peaks~\texttt{BPL2P} mass function, which is the default distribution of \cite{LIGOScientific:2025pvj}, used as a benchmark for our model comparison studies, as follows:

\begin{align}
    \texttt{BPL2P}^{\alpha_1, \alpha_2, x_b, \delta, \lambda_1, \mu_1,\sigma_1, \lambda_2, \mu_2, \sigma_2}_{a,b}(x) &= \lambda_1 \mathcal{N}_{a,b}^{\mu_1,\sigma_1}(x) \nonumber \\
    &+ (1-\lambda_1)\nonumber \\
    &~\texttt{BPLP}^{\alpha_1, \alpha_2, x_b, \delta, \lambda_2, \mu_2,\sigma_2}_{a,b}(x).
\end{align}

We explore several variations which are listed (along with their Bayes factors and changes in maximum log likelihood, both with respect to the default model that does not allow for subpopulations) in Table~\ref{tab:mdels}. For each model, we have explored further variations in the choices of hyperpriors. The Bayes factors represented in Table~\ref{tab:mdels} correspond to the most preferred hyperprior choice for each set of functional forms.

Note that previous studies such as \cite{Banagiri:2025dmy} have shown that among models which account for three subpopulations with distinct mass-ratio and effective spin distributions are favoured by the data over ones that allow for a smaller number of features in the underlying population. Data driven studies by \cite{Sridhar:2025kvi} have shown that similar trends of three distinct subpopulations can be recovered without making strong modeling assumptions. Hence, instead of repeating a similar model comparison study as \cite{Banagiri:2025dmy} with our models, we instead investigate the sensitivity of our conclusions to various comparable choices of functional forms for the shapes of the distributions corresponding to each subpopulation.

With models II and III, we explore how the inferred shapes of $p_i(m_1)$ depend on the choice of functional forms. In particular, with II, we verify whether the $35M_{\odot}$ feature can be well described with a change in the powerlaw index or if it requires an additional Gaussian to account for the overabundance. In III, we explore whether the same functional form of $p(m_1)$ for all three components is preferred by the data over our specific choices in I. With model IV, we explore whether the mass-ratio distribution of component 2 is equally well described by a truncated Gaussian similar to components 1 and 3, or if it necessitates the Powerlaw as chosen for I. Finally, with models V and VI, we verify whether the $p(\chi_{eff})$ of component III is better described by a Gaussian or a more flexible Uniform distribution as compared to our choice in I. In all these variations, we find that the data either prefer Model I or are unable to distinguish between the deviations from Model I. 
\begin{table*}[ht]
\centering
\begin{tabular}{|c|c|c|c|c|c|c|c|}
\hline
Model & Comp. &  $p(m_1)$ & $p(q|m_1)$ & $p(\chi_{eff})$ & $p(\chi_p)$ & $\log_{10}BF$ & $\Delta\log \mathcal{L}_{max}$ \\
\hline
~& 1 & $\texttt{PLP}^{\alpha^1_m,\delta_m, \lambda^1_m, \mu^1_m,\sigma_m^1}_{m_{1,min}^1,m_{1,max}^1}$ & $\mathcal{N}_{\frac{m_{2,min}^1}{m_1},1}^{\mu^1_q,\sigma^1_q}$ & $\mathcal{N}_{-1,1}^{\mu^1_{\chi_{eff}},\sigma^1_{\chi_{eff}}}$ & $\mathcal{N}_{0,1}^{\mu^1_{\chi_{p}},\sigma^1_{\chi_{p}}}$
 & ~ 
 & ~ \\ \cline{2-6}

I & 2 & $\texttt{BPL}^{\alpha^2_{m,1}, \alpha^2_{m,2}, m_b^2, \delta_m}_{m_{1,min}^2,m_{1,max}^2}$ & $\texttt{PL}_{\frac{m_{2,min}^2}{m_1},1}^{\beta^2_q,\frac{\delta_m}{m_1}}$ & $\mathcal{N}_{-1,1}^{\mu^2_{\chi_{eff}},\sigma^2_{\chi_{eff}}}$ & $\mathcal{N}_{0,1}^{\mu^2_{\chi_{p}},\sigma^2_{\chi_{p}}}$& 6 & 24
 \\ \cline{2-6}

~& 3 & $\texttt{BLP}^{\alpha^3_{m,1}, \alpha^3_{m,2}, m_b^3, \delta_m}_{m_{1,min}^3,m_{1,max}^3}$ & $\mathcal{N}_{\frac{m_{2,min}^3}{m_1},1}^{\mu^3_q,\sigma^3_q}$ & $U_{-w_{\chi_{eff}}^3,w_{\chi_{eff}^3}}$ & $\mathcal{N}_{0,1}^{\mu^3_{\chi_{p}},\sigma^3_{\chi_{p}}}$ 
 &  &  \\ \hline
 ~& 1 & Same as I & Same as I & Same as I & Same as I
 & ~ 
 & ~ \\ \cline{2-6}

II & 2 & $\texttt{BPLP}^{\alpha^2_{m,1}, \alpha^2_{m,2}, m_b^2, \delta_m, \lambda_{m}^2, \mu_m^2, \sigma_m^2}_{m_{1,min}^2,m_{1,max}^2}$ & Same as I & Same as I & Same as I& 6 & 23  \\ \cline{2-6}

 ~& 3 & Same as I & Same as I & Same as I & Same as I
 & ~ 
 & ~ \\ \hline
  ~& 1 & Same as I & Same as I & Same as I & Same as I
 & ~ 
 & ~ \\ \cline{2-6}

III & 2 & $\texttt{PLP}^{\alpha^2_m,\delta_m, \lambda^2_m, \mu^2_m,\sigma_m^2}_{m_{1,min}^2,m_{1,max}^2}$ & Same as I & Same as I & Same as I& 5.6 
 & 22.7 \\ \cline{2-6}

 ~& 3 & $\texttt{PLP}^{\alpha^3_m,\delta_m, \lambda^3_m, \mu^3_m,\sigma_m^3}_{m_{1,min}^3,m_{1,max}^3}$ & Same as I & Same as I & Same as I
 & ~ 
 & ~ \\ \hline
 ~& 1 & Same as I & Same as I & Same as I & Same as I
 & ~ 
 & ~ \\ \cline{2-6}

IV & 2 &Same as I & $\mathcal{N}_{\frac{m_{2,min}^2}{m_1},1}^{\mu^2_q,\sigma^2_q}$ & Same as I & Same as I& 6 & 23 \\ \cline{2-6}

 ~& 3 & Same as I & Same as I & Same as I & Same as I
 & ~ 
 & ~ \\ \hline


  ~& 1 & Same as I & Same as I & Same as I & Same as I
 & ~ 
 & ~ \\ \cline{2-6}

V & 2 &Same as I & Same as I & Same as I & Same as I& ~ 
 & ~ \\ \cline{2-6}

 ~& 3 & Same as I & Same as I & $U_{a_{\chi_{eff}}^3, b_{\chi_{eff}}^3}$ & Same as I
 & 6 
 & 23 \\ \hline
   ~& 1 & Same as I & Same as I & Same as I & Same as I
 & ~ 
 & ~ \\ \cline{2-6}

VI & 2 &Same as I & Same as I & Same as I & Same as I& 6 
 & 21 \\ \cline{2-6}

 ~& 3 & Same as I & Same as I & $\mathcal{N}_{-1,1}^{\mu^3_{\chi_{eff}},\sigma^3_{\chi_{eff}}}$ & Same as I
 & ~ 
 & ~ \\ \hline
 Default & -- & BPL2P~\citep{LIGOScientific:2025pvj} &$\texttt{PL}_{\frac{m_{2,min}}{m_1},1}^{\beta_q,\frac{\delta_m}{m_1}}$ & $\mathcal{N}_{-1,1}^{\mu_{\chi_{eff}},\sigma_{\chi_{eff}}}$ & $\mathcal{N}_{0,1}^{\mu_{\chi_{p}},\sigma_{\chi_{p}}}$& 0 
 & 0  \\ \hline
\end{tabular}
\caption{\label{tab:mdels} Functional forms for various models. The results presented in the main text correspond to our most preferred model (I).}
\end{table*}

We next turn to variations in hyperpriors for our most preferred model. We start by listing all the hyperpriors that have the highest Bayes factors in Table~\ref{tab:prior-preferred}, along with the prior predictive and posterior predictive credible intervals for each free hyperparameter. As mentioned in the main text, we target the subpopulation of hierarchical mergers with component 3 without enforcing its existence. However, additional restrictions were imposed for components 1 and 2 in Table~\ref{tab:prior-preferred}. In Table~\ref{tab:prior-flex}  we list flexible alternatives for these strongly restrictive hyperpriors along with the corresponding Bayes factors and changes in maximum log likelihood with respect to the default model. It can be seen that relaxing prior restrictions either does not change the inferred values of the corresponding hyperparameter or leads to an increase in prior volume without any change in the maximum likelihood. In other words, the hyperparameters that were fixed in the most preferred prior choice, when allowed to vary, are mostly uninformative over the prior range, leading to comparable shapes in the inferred distributions and slightly smaller or comparable Bayes factors. 

Additional hyperpriors for models II-VI are presented in Tables~\ref{tab:prior-modII},~\ref{tab:prior-modIII,},~\ref{tab:prior-modIV},~\ref{tab:prior-modV}, and~\ref{tab:prior-modVI} respectively.

\begin{table*}[ht]
\centering
\begin{tabular}{cccc}
\hline\hline
Hyperparameter   & Hyperprior & Prior $90\%$& Posterior $90\%$ \\
\hline
$\alpha^1_{m_1}$ &   $U(-8,4)$ & $(-3.49, 7.18)$ & $(2.33, 4.28)$ \\
$m_{1,min}^{1}$ &  $U(3,10)$ & $(3.00, 9.30)$ & $(4.50, 9.12)$ \\
$m_{2,min}^1$ &  $U(3,m_{1,min}^1)$ & $(3.00, 7.13)$ & $(3.88, 5.76)$\\
$m_{1,max}^{1}$ &  $U(58,100)$ & $(61.83, 99.85)$ & $(62.48, 91.69)$ \\
$\lambda_{m_1}^1$ & $U(0,1)$& $(0.10, 0.99)$ & $(0.21, 0.91)$ \\
$\mu_{m_1}^{1}$ &  $U(5,20)$ & $(5.78, 19.19)$ & $(5.63, 10.69)$ \\
$\sigma_{m_1}^{1}$ &  $U(0,10)$ & $(0.31, 9.24)$ & $(0.42, 2.94)$ \\
$\delta_m$ & $U(0,10) $ & $(0.05, 9.01)$ & $(0.01, 8.03)$ \\
$\mu_{q}^1$ &  $U(0.3,0.8)$ & $(0.31, 0.76)$ & $(0.31, 0.74)$ \\
$\sigma_{q}^1$ &  $\mathcal{N}_{0,\infty}^{0,0.1}$ & $(0.00, 0.16)$ & $(0.00, 0.15)$ \\
$\mu_{\chi_{eff}}^1$ &  $U(-0.1,0.2)$ & $(-0.09, 0.18)$ & $(0.03, 0.08)$ \\
$\mu_{\chi_{p}}^1$ &  $U(0.001,0.4)$ & $(0.04, 0.40)$ & $(0.18, 0.31)$ \\
$\sigma_{\chi_{eff}}^1$ &  $U(0.05, 1.0)$ & $(0.15, 1.00)$ & $(0.06, 0.10)$ \\
$\sigma_{\chi_{p}}^1$ &  $U(0.05, 1.0)$ & $(0.02, 0.29)$ & $(0.06, 0.16)$ \\
$\kappa^1$ & $U(-6,8)$ & $(-4.68, 7.95)$ & $(2.28, 5.91)$\\
$f_1$ & $U(0,1) $ & $(0.00, 0.90)$ & $(0.68, 0.91)$ \\
\hline
$m_{1,min}^{2}$ &  $\delta(m_{1,min}^1)$ & - & - \\
$m_{2,min}^1$ &  $\delta(m_{2,min}^1)$ & - & - \\
$m_{1,max}^{1}$ &  $\delta(m_{1,max}^1)$ & - & - \\
$\alpha^2_{1,m_1}$ &   $U(\alpha_{m_1}^1,8)$ & $(-5.00, 2.68)$ & $(-4.76, -0.63)$ \\
$\alpha^2_{2,m_1}$ &   $U(-30,4)$ & $(-3.99, 26.33)$ & $(9.33, 27.27)$ \\
$m_{1,b}^{2}$ &  $U(30,m_{1,max}^{2})$ & $(31.96, 49.88)$ & $(33.61, 38.20)$ \\
$\beta_{q}^2$ &  $U(-2,7)$ & $(-1.15, 6.96)$ & $(1.65, 6.07)$ \\
$\mu_{\chi_{eff}}^2$ &  $U(-1,1)$ & $(-0.89, 0.90)$ & $(-0.04, 0.03)$ \\
$\mu_{\chi_{p}}^2$ &  $U(0.05,2.0)$ & $(0.12, 0.97)$ & $(0.12, 0.59)$ \\
$\sigma_{\chi_{eff}}^2$ &  $U(0.01, 0.2)$ & $(0.03, 0.20)$ & $(0.04, 0.11)$ \\
$\sigma_{\chi_{p}}^2$ &  $U(0.05, 2.0)$ & $(0.07, 1.82)$ & $(0.09, 1.25)$ \\
$\kappa_2$ & $U(-6,8)$ & $(-5.06, 7.47)$ & $(-0.24, 3.20)$\\
$f_2$ & $U(0,1-f_1) $ & $(0.00, 0.58)$ & $(0.07, 0.26)$ \\
\hline
$\alpha^3_{1,m_1}$ &   $U(-4,4)$ & $(-3.99, 3.20)$ & $(-1.39, 3.63)$ \\
$\alpha^3_{2,m_1}$ &   $U(-4,4)$ & $(-3.45, 7.35)$ & $(1.89, 6.67)$ \\
$m_{1,min}^{3}$ &  $U(m_{1,min}^{1}+m_{2,min}^{1},2m_{1,min}^{1})$  & $(6.01, 17.20)$ & $(6.19, 17.85)$ \\
$m_{1,max3}^{3}$ &  $\delta(2m_{1,max}^1)$& - & - \\
$\mu_{q}^3$ &  $\delta(0.5)$ & - & - \\
$\sigma_{q}^3$ &  $\mathcal{N}_{0,\infty}^{0,0.1}$ & $(0.00, 0.17)$ & $(0.10, 0.24)$ \\
$m_{1,b}^{3}$ &  $U(30,80)$ & $(31.42, 76.53)$ & $(34.25, 74.38)$ \\
$\mu_{\chi_{p}}^3$ &  $U(0.01, 0.2)$ & $(0.05, 0.91)$ & $(0.10, 0.84)$ \\
$\sigma_{\chi_{p}}^3$ &  $U(0.05, 2.0)$ & $(0.05, 1.79)$ & $(0.08, 1.70)$ \\
$w_{\chi_{eff}}^3$ & $U(0.01, 1.0)$ & $(0.03, 0.91)$ & $(0.43, 0.62)$ \\
$\kappa^3$ & $U(-6,8)$ & $(-5.97, 6.60)$ & $(1.41, 5.77)$\\
$f_3$ & $U(0,1-f_1-f_2) $ & $(0.00, 0.60)$ & $(0.01, 0.08)$ \\
\hline
\end{tabular}
\caption{\label{tab:prior-preferred} Hyperpriors for our most preferred model. Here $\delta(\Lambda_0)$ implies a delta function prior which fixes the corresponding hyperparameter to $\Lambda_0$. See Table~\ref{tab:prior-flex} where we explore flexible alternatives for these restrictive hyperpriors.}
\end{table*}




\begin{table*}[ht]
\centering
\begin{tabular}{ccccccc}
\hline\hline
Model & Hyperparameter   & New Hyperprior & Prior $90\%$& Posterior $90\%$ & $\log_{10}BF$ & $\log \mathcal{L}_{max}$\\
\hline
 
 Ia & $\sigma_{q}^1$ & $U(0.01,1)$ & $(0.10, 0.91)$ & $(0.14, 0.93)$ & 6 & 22 \\
 Ib &$m_{1,max}^2$ & $U(30, 100)$ & $(30.14, 94.73)$ & $(38.32, 94.03)$ & 6 & 22 \\
 Ic &$m_{1,min}^2$ & $U(3,10)$& $(3.71, 9.99)$ & $(3.55, 9.61)$ & 6 & 23 \\
 Id &$m_{1,max}^2$ & $U(58, 100)$ & $(59.90, 97.29)$ & $(59.11, 95.61)$ & 6 & 23\\
 Ie &$m_{1,max}^3$ & $U(100, 200)$ & $(100.71, 189.45)$ & $(130.81, 198.66)$ & 6 & 23\\
 If &$\mu_{q}^3$ &   $U(0.01,1.0)$ & $(0.03, 0.91)$ & $(0.11, 0.94)$ & 6 & 21 \\
\hline
\end{tabular}
\caption{\label{tab:prior-flex} Flexible alternatives for certain hyperpriors used with Model I.}
\end{table*}

\begin{table}[ht]
\centering
\begin{tabular}{cccc}
\hline\hline
 Hyperparameter   & Hyperprior & Prior $90\%$& Posterior $90\%$ \\
\hline
 
 $\mu_m^2$ & $U(30, 50)$ & $(31.90, 49.94)$ & $(30.53, 34.72)$  \\
 $\sigma_m^2$ & $U(0, 10)$ & $(0.89, 9.75)$ & $(2.02, 6.28)$  \\
 $\lambda_m^2$ & $U(0, 1)$ & $(0.08, 0.98)$ & $(0.00, 0.68)$  \\
\hline
\end{tabular}
\caption{\label{tab:prior-modII} Additional hyperpriors for model II.}
\end{table}

\begin{table}[ht]
\centering
\begin{tabular}{cccc}
\hline\hline
 Hyperparameter   & Hyperprior & Prior $90\%$& Posterior $90\%$ \\
\hline
$\alpha_m^2$ & $U(\alpha_m^1, 8)$  & $(-4.00, 2.97)$ & $(2.02, 8.00)$ \\
$\alpha_m^3$ & $U(-4, 8)$  & $(-2.65, 8.00)$ & $(-0.13, 3.96)$  \\
$\mu_m^2$ & $U(25, 40)$  & $(26.01, 39.36)$ & $(29.67, 34.79)$  \\
$\mu_m^3$ & $U(30, 80)$ & $(32.25, 77.24)$ & $(31.68, 60.68)$  \\
$\sigma_m^2$ & $U(0, 10)$  & $(0.51, 9.49)$ & $(2.48, 7.20)$  \\
$\sigma_m^3$ & $U(5, 20)$  & $(6.14, 19.52)$ & $(7.99, 20.00)$  \\
$\lambda_m^2$ & $U(0, 1)$ & $(0.10, 0.99)$ & $(0.29, 1.00)$   \\
$\lambda_m^3$ & $U(0, 1)$  & $(0.06, 0.96)$ & $(0.00, 0.74)$  \\
\hline
\end{tabular}
\caption{\label{tab:prior-modIII} Additional hyperpriors for model III.}
\end{table}

\begin{table}[ht]
\centering
\begin{tabular}{cccc}
\hline\hline
 Hyperparameter   & Hyperprior & Prior $90\%$& Posterior $90\%$ \\
\hline
 $\mu_q^2$ & $U(0, 2)$  & $(0.21, 1.99)$ & $(0.77, 1.03)$  \\
 $\sigma_q^2$ & $\mathcal{N}_{0,\infty}^{0,0.1}$ & $(0.00, 0.17)$ & $(0.07, 0.21)$ \\
\hline
\end{tabular}
\caption{\label{tab:prior-modIV} Additional hyperpriors for model IV.}
\end{table}

\begin{table}[ht]
\centering
\begin{tabular}{cccc}
\hline\hline
 Hyperparameter   & Hyperprior & Prior $90\%$& Posterior $90\%$ \\
\hline
 $a_{\chi_{eff}}^3$ & $U(-1, b_{\chi_{eff}}^3)$  & $(-1.00, 0.22)$ & $(-0.92, -0.16)$  \\
 $b_{\chi_{eff}}^3$ & $U(-1,1)$ & $(-0.76, 0.84)$ & $(0.42, 0.60)$  \\
\hline
\end{tabular}
\caption{\label{tab:prior-modV} Additional hyperpriors for model V. }
\end{table}

\begin{table}[ht]
\centering
\begin{tabular}{cccc}
\hline\hline
 Hyperparameter   & Hyperprior & Prior $90\%$& Posterior $90\%$ \\
\hline
 $\mu_{\chi_{eff}}^3$ & $U(-1, 1)$   & $(-0.83, 0.93)$ & $(-0.36, 0.34)$  \\
 $\sigma_{\chi_{eff}}^3$ & $U(0,1)$  & $(0.05, 0.94)$ & $(0.09, 0.75)$  \\
\hline
\end{tabular}
\caption{\label{tab:prior-modVI} Additional hyperpriors for model VI. }
\end{table}

\begin{figure*}[htt]
    \centering
    \includegraphics[width=0.32\linewidth]{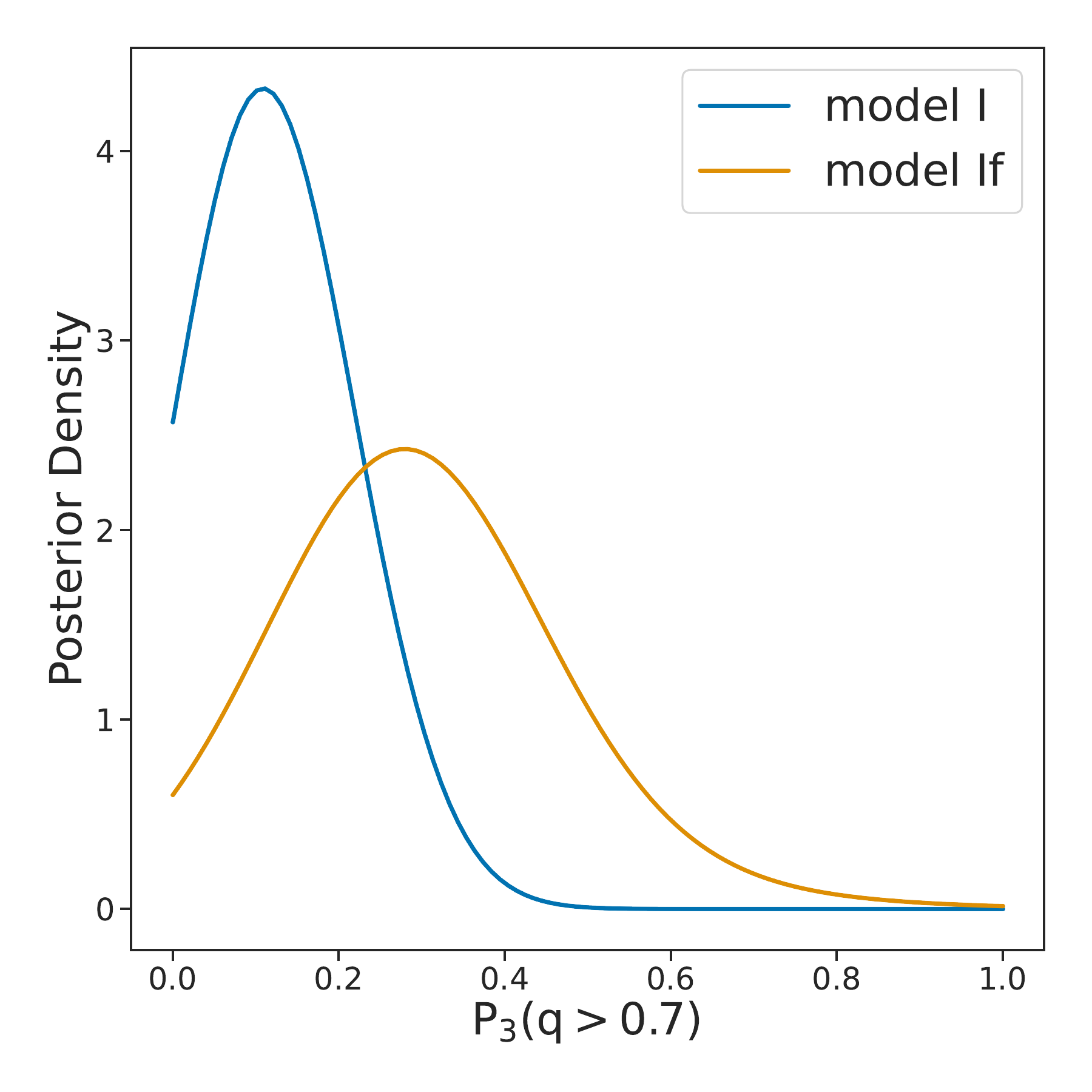}
    \includegraphics[width=0.32\linewidth]{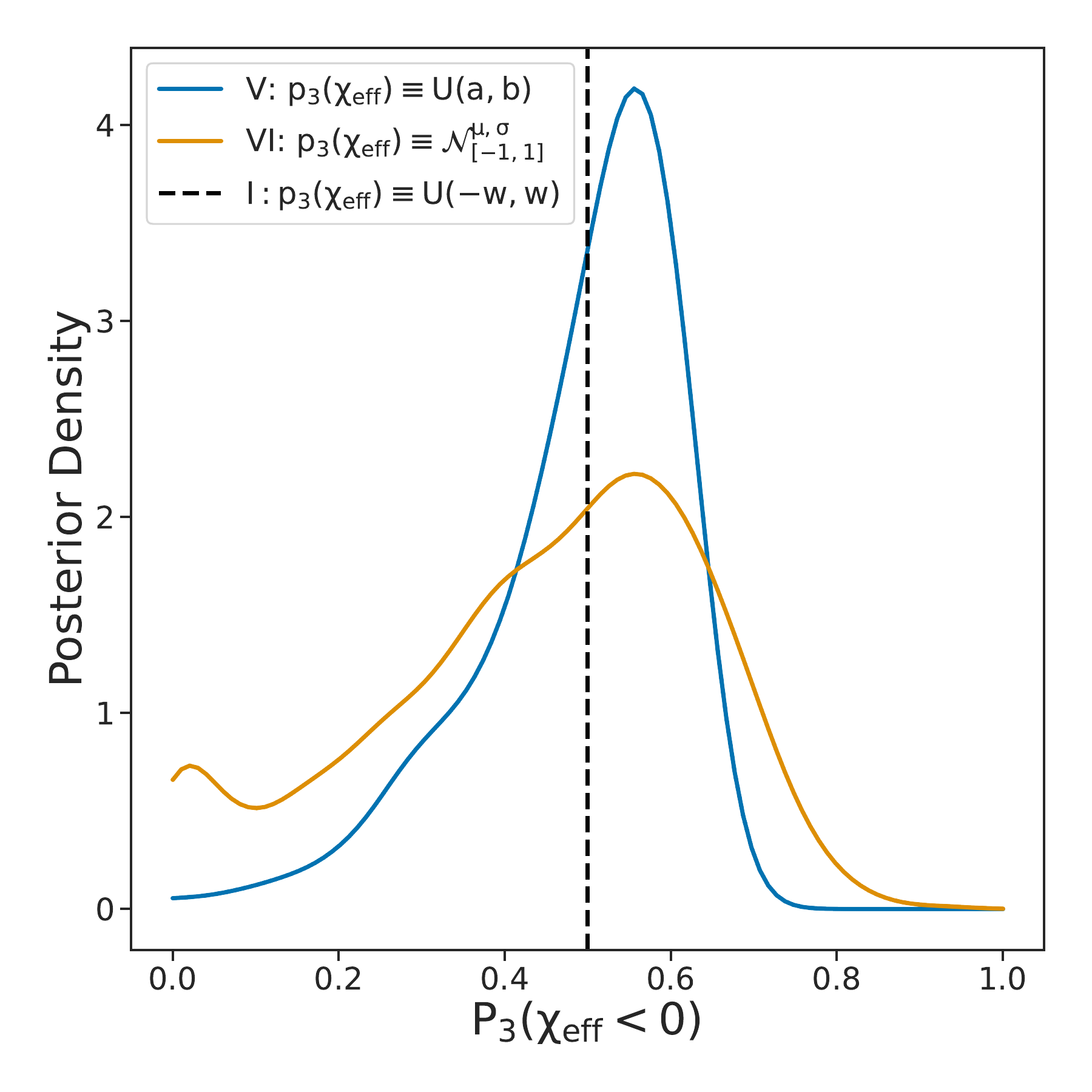}
    \includegraphics[width=0.32\linewidth]{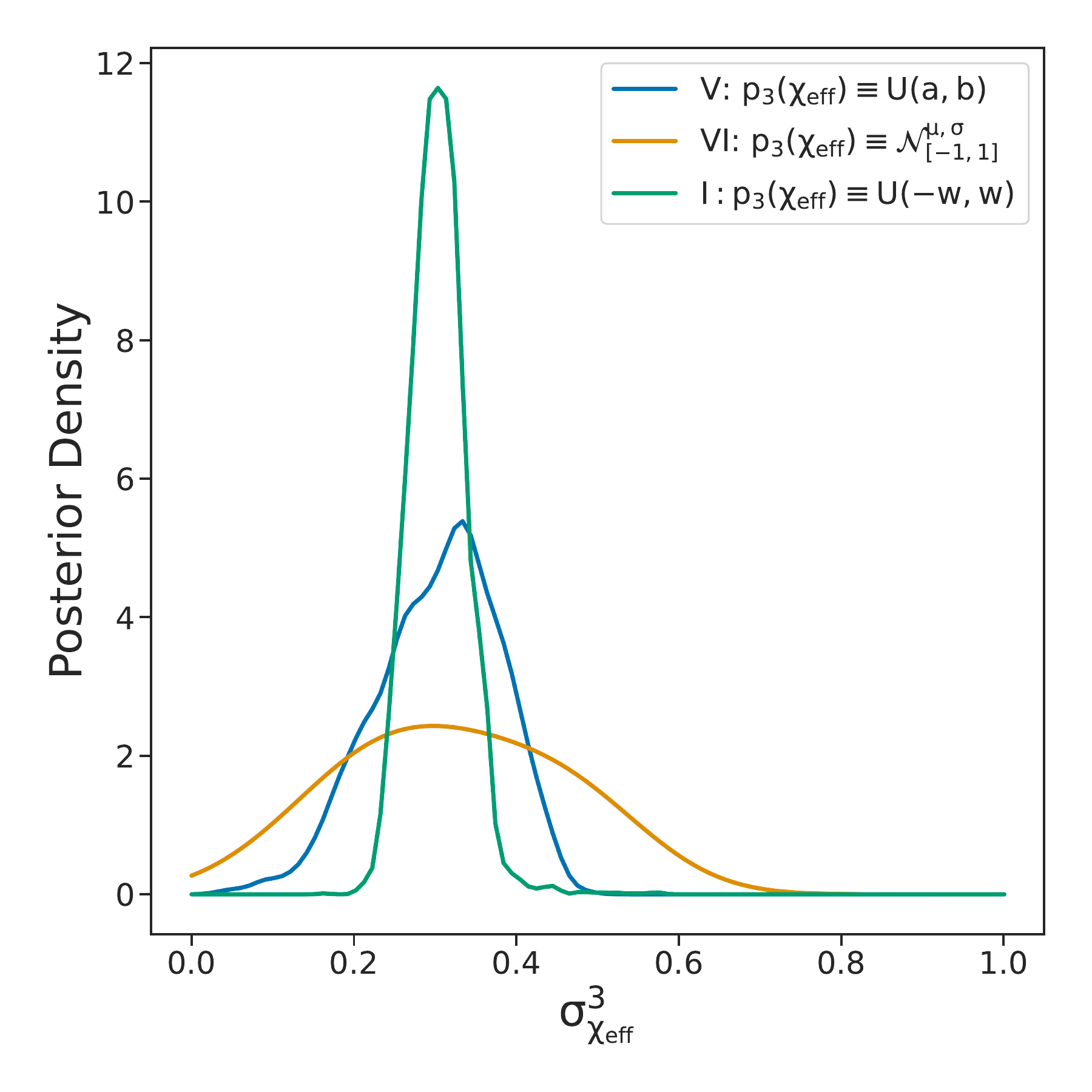}
    \caption{Variations of the metrics $P_3(q>0.7),\sigma_{\chi_{eff}}^3,$ and $P_3(\chi_{eff}<0)$ for Comp. 3 with functional forms and hyperpriors. It can be clearly seen that flexible modeling assumptions do not lead to any qualitative changes in the inferred values and only lead to a broadening of measurement uncertainties~(Note that for $P_3(q>0.7)$, the mode of model I is within the $50\%$ highest density interval of model If, which indicates no statistically significant discrepancy between the claims based on model I and those based on the more flexible prior choices of model If).}
    \label{fig:metric-3-comp}
\end{figure*}

\begin{figure}[htt]
    \centering
    \includegraphics[width=0.98\linewidth]{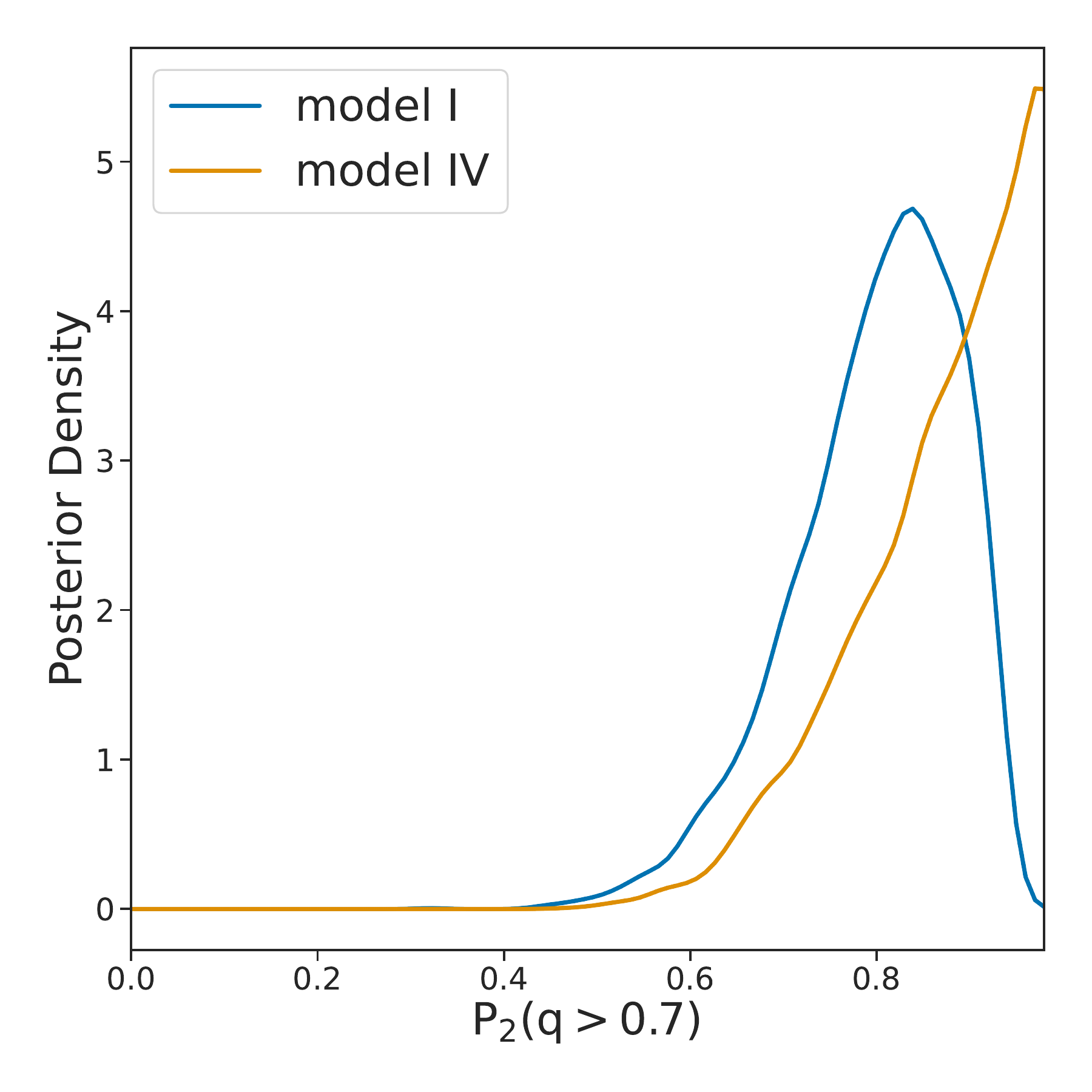}
    \caption{Variations of the metrics $P(q>0.7)$ for Comp. 2 with functional forms (model I vs model IV).} 
    \label{fig:metric-2-comp}
\end{figure}

\section{Variation of astrophysical metrics with functional forms \& priors} 
\label{sec:app-metric-modcomp}
We now demonstrate that the metrics used to establish our astrophysical interpretation, which are summarized in Table~\ref{tab:metrics} of the main text, are robust against variations in modeling assumptions. First, we compare the inferred values of $\sigma^3_{\chi_{eff}}$, $P_3({\chi_{eff}<0})$ and $P_3(q>0.7)$ obtained using model I with those that employ much more flexible functional forms and/or hyperpriors for the corresponding distributions, namely models If, V, and VI. We show in Figure~\ref{fig:metric-3-comp} that more flexible modeling assumptions lead to values consistent with the ones presented in the main text, with the only difference being an expected increase in measurement uncertainty. In other words, our conclusions regarding Comp. 3 remain unchanged upon lifting certain restrictions previously employed for targeting the hierarchical subpopulation. 

In addition to these metrics for Comp. 3, we also explore variations in $P(q>0.7)$ for Comp. 2. Since Comp. 2 is the only one with a \texttt{PL} mass-ratio distribution, as compared to the truncated Gaussians used with the other two components, we investigate whether or not $P_2(q>0.7)$ is affected by this choice of $p_2(q)$. We hence compare the inferred values of $P_2(q>0.7)$ obtained using Model I and Model IV. We show in Figure~\ref{fig:metric-2-comp} that despite being strongly disfavoured by the data, model IV is consistent with model I's measurement of the fraction of systems with $q>0.7$ that was presented in the main text.

We note here that the other astrophysically informative metrics for Comp. 1 and Comp. 2 are unlikely to be affected by strong modeling assumptions since the corresponding distributions are reasonably flexible. Furthermore, we find our results to be consistent with data-driven reconstructions of the joint distribution of BBH parameters (as we show later on), indicating that the inferred conclusions are likely not driven by prior assumptions.

\section{Prior Predictive Draws} 
\label{sec:app-prpdraws}
\begin{figure*}
\includegraphics[width=0.92\textwidth]{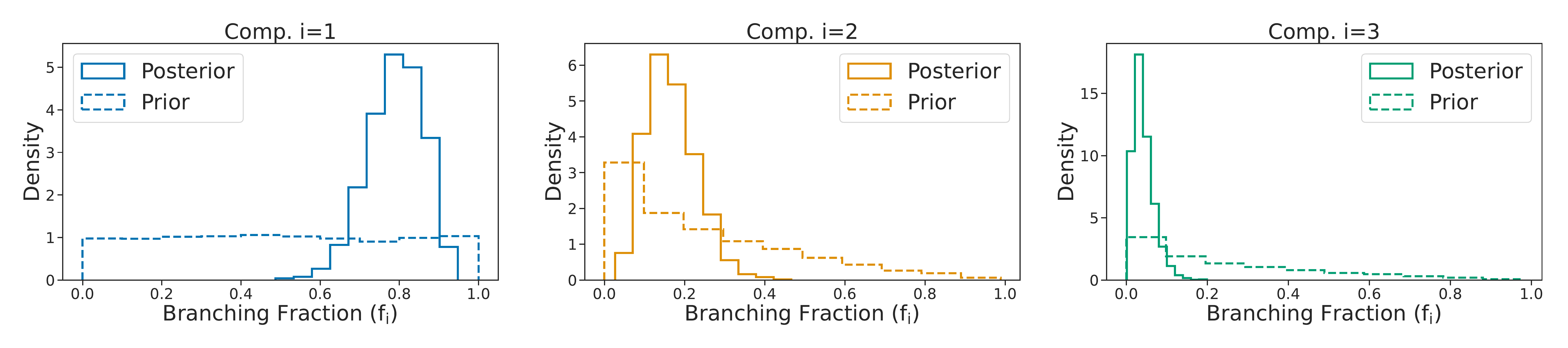}
\includegraphics[width=0.92\textwidth]{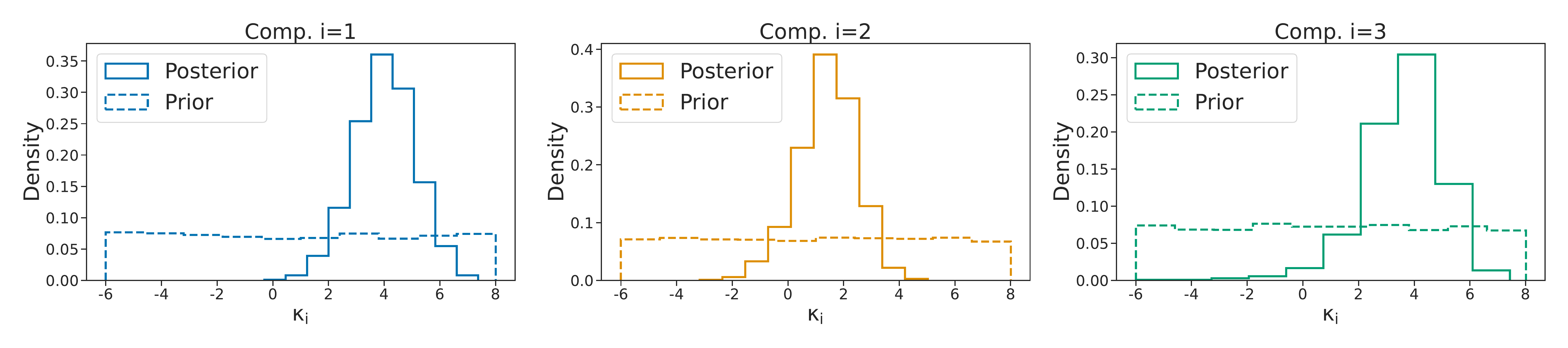}
\includegraphics[width=0.92\textwidth]{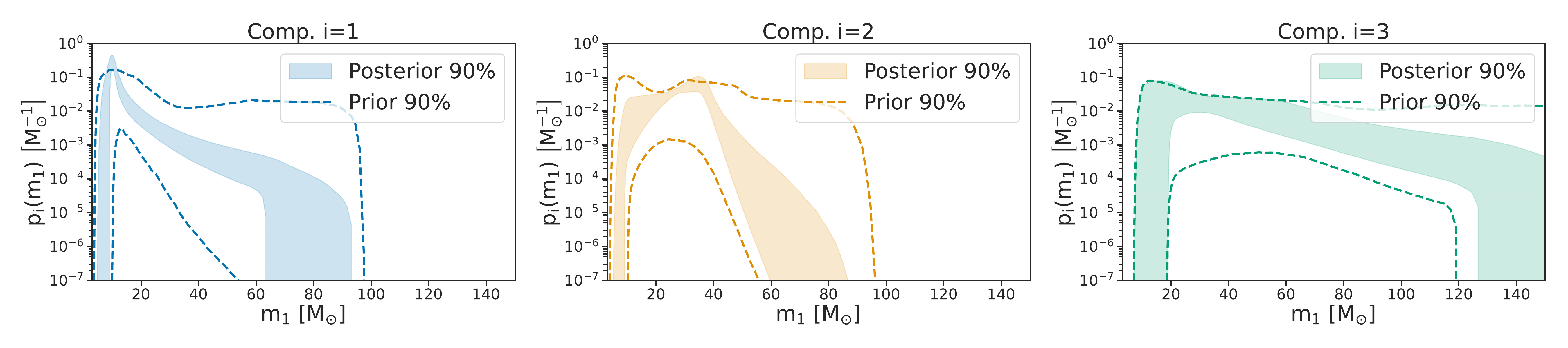}
\includegraphics[width=0.92\textwidth]{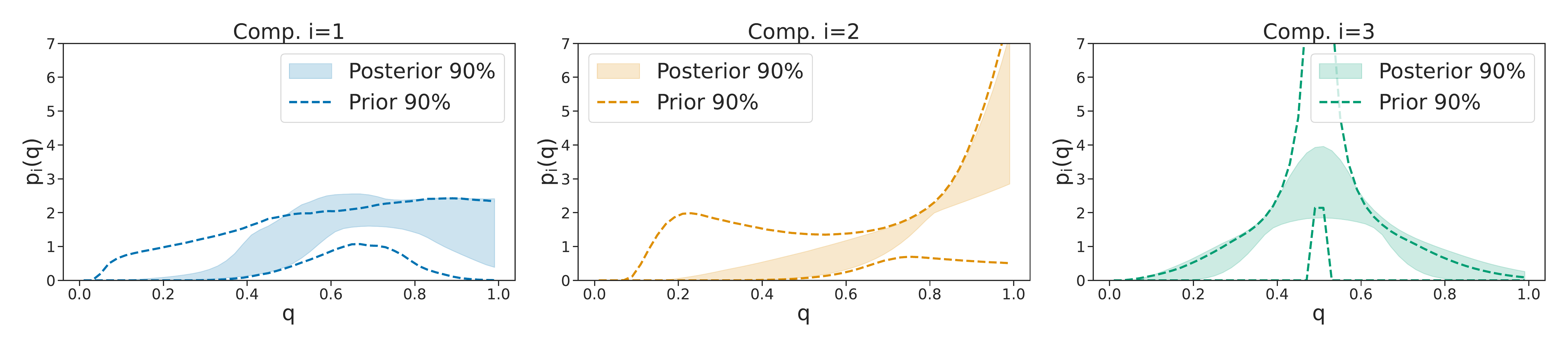}
\includegraphics[width=0.92\textwidth]{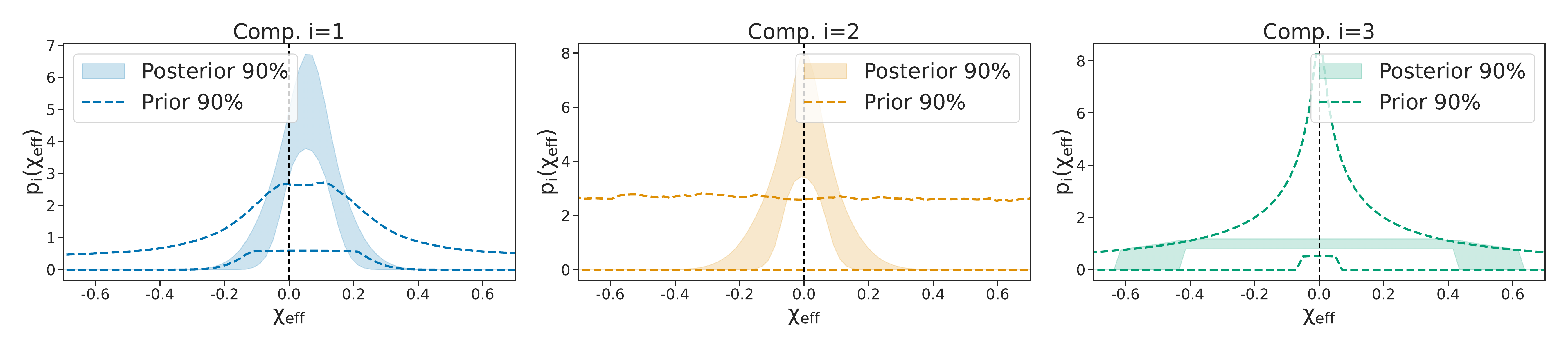}
\includegraphics[width=0.92\textwidth]{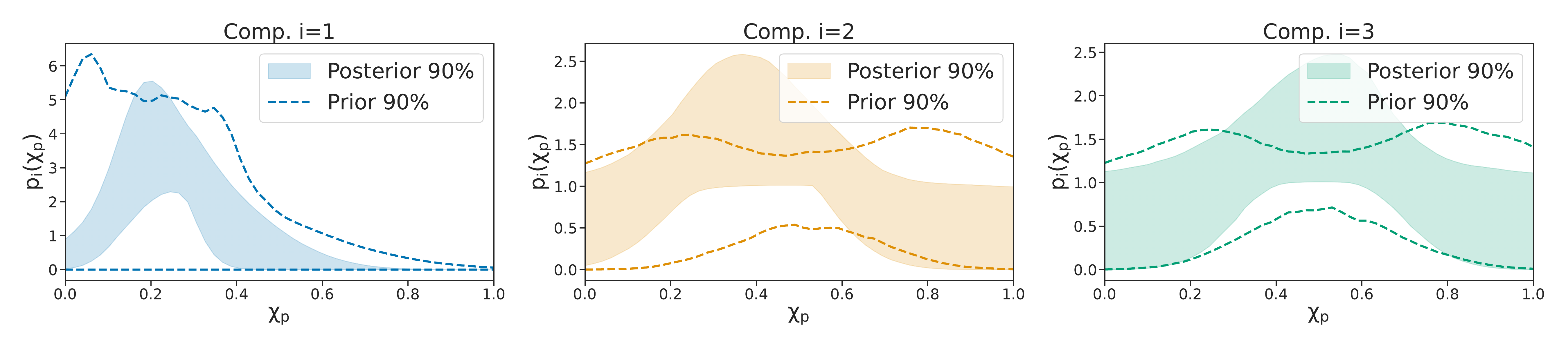}
\caption{\label{fig:comp-prior}Posterior  vs Prior. In the top two panels, the solid lines represent the posterior density where as the dashed lines represent the prior. For the other panels, the shaded region represents the $90\%$ credible interval of the posterior where as dashed lines enclose the $90\%$ credible interval of the prior.} 
\end{figure*}

We now turn to comparisons of our inferred distributions for the most preferred model, to those corresponding to the prior predictive draws of hyperparameters. In Figure~\ref{fig:comp-prior}, it can be seen that the posterior is more informative than the prior for each distribution function and hyperparameter. Coupled with our model comparison results presented before, this indicates that our conclusions are driven by the data and not by modeling assumptions. We later on compare, where available, the marginal distributions of some BBH parameters conditioned on others, with corresponding data-driven reconstructions, to show that our results are consistent with flexible analyses that impose very few modeling assumptions.

\begin{figure*}
\includegraphics[width=0.98\textwidth]{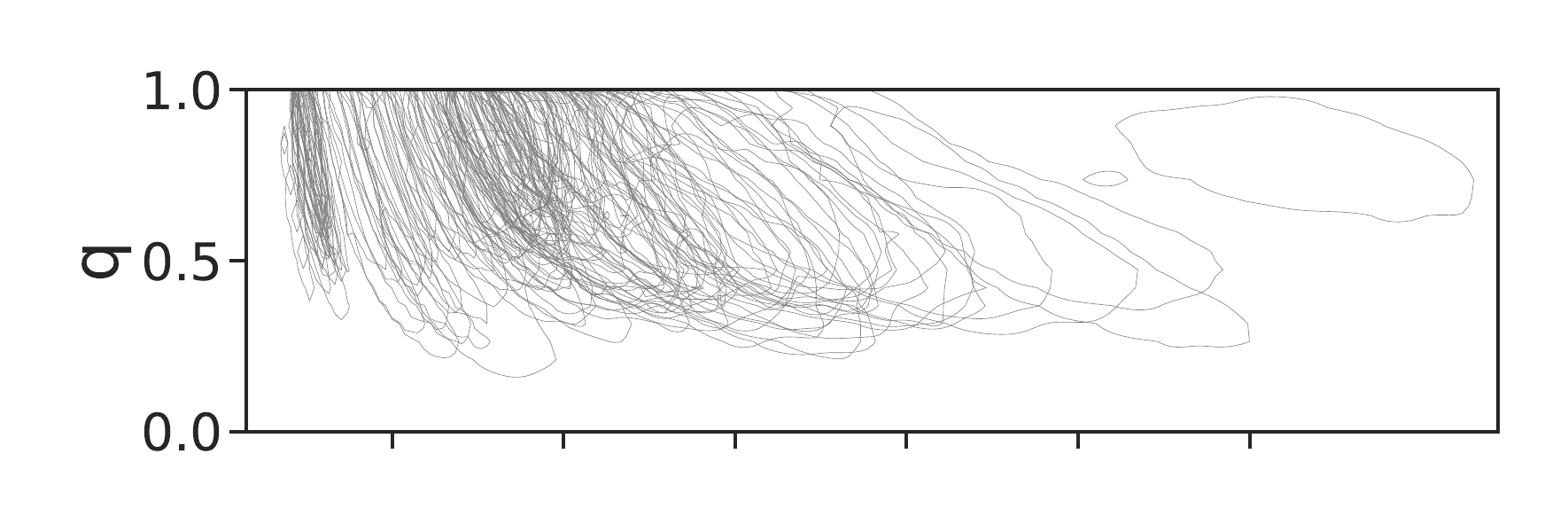}
\includegraphics[width=0.98\textwidth]{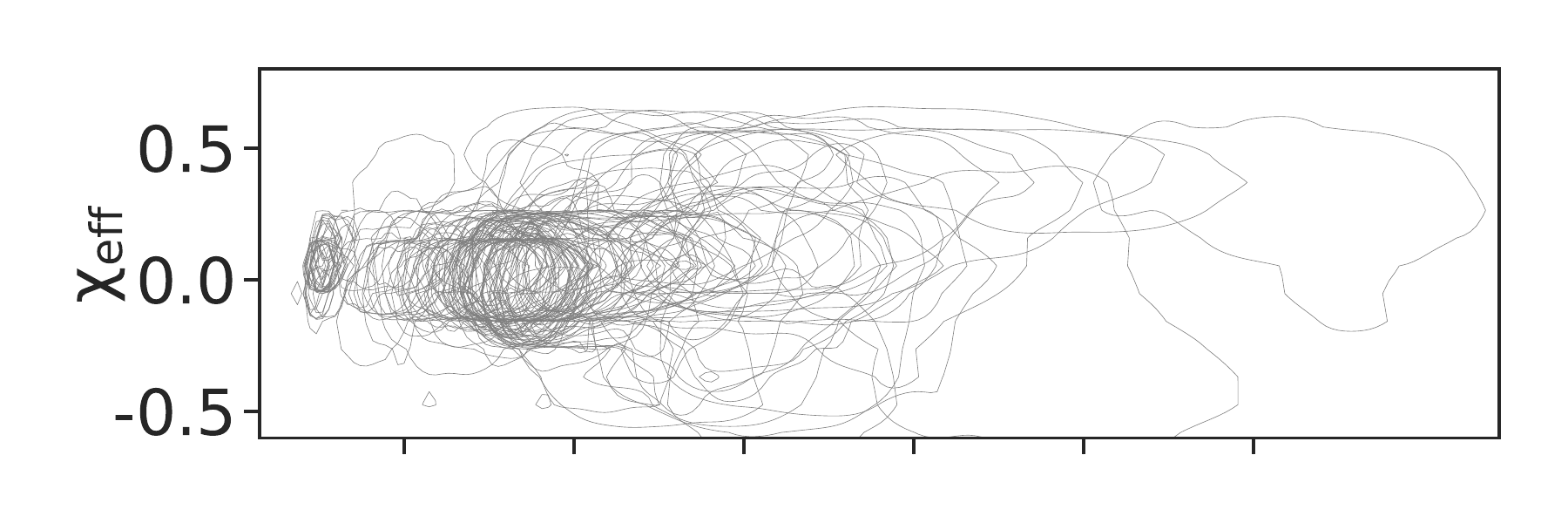}
\includegraphics[width=0.98\textwidth]{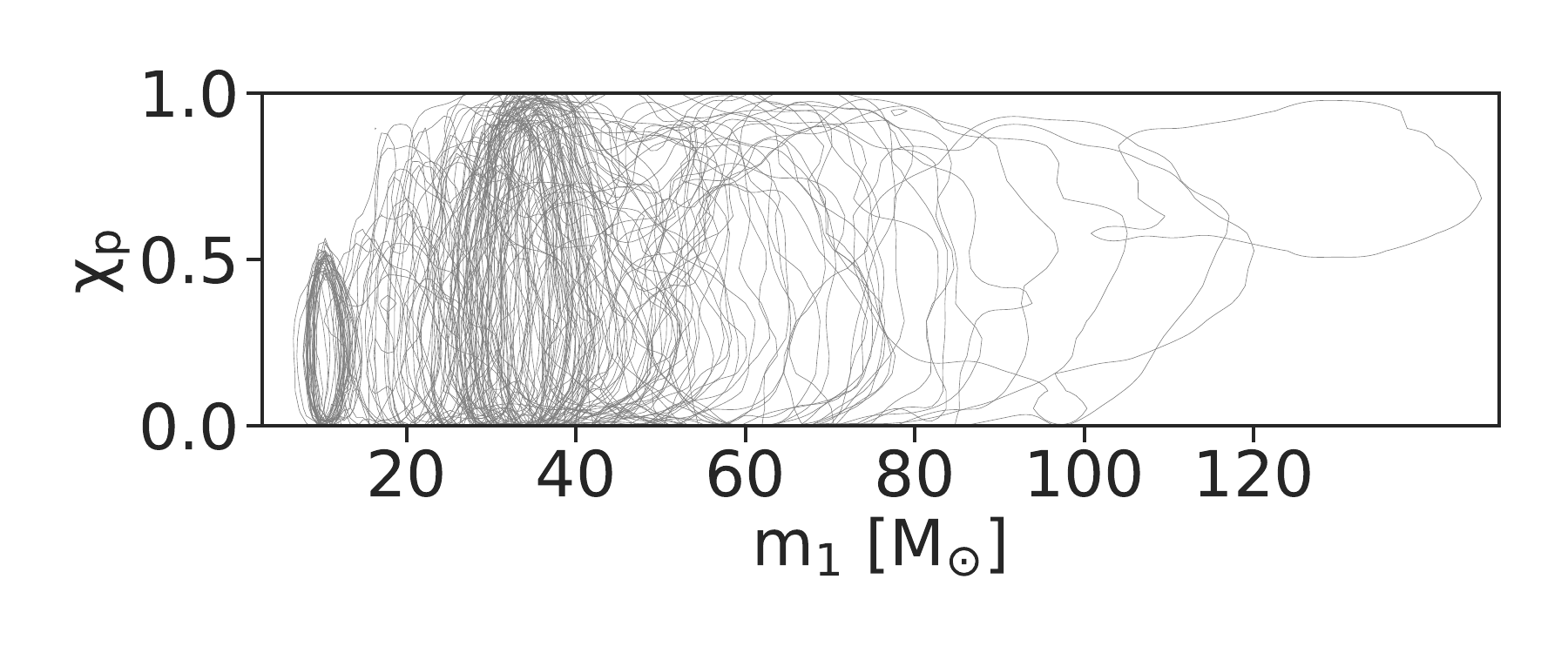}
\caption{\label{fig:reweighted} Individual event measurements of the BBH parameters reweighted to the population-informed prior. Clusters of observed events that represent each of the three subpopulations and their corresponding trends can be clearly identified.} 
\end{figure*}

\begin{figure*}
\includegraphics[width=0.4\textwidth]{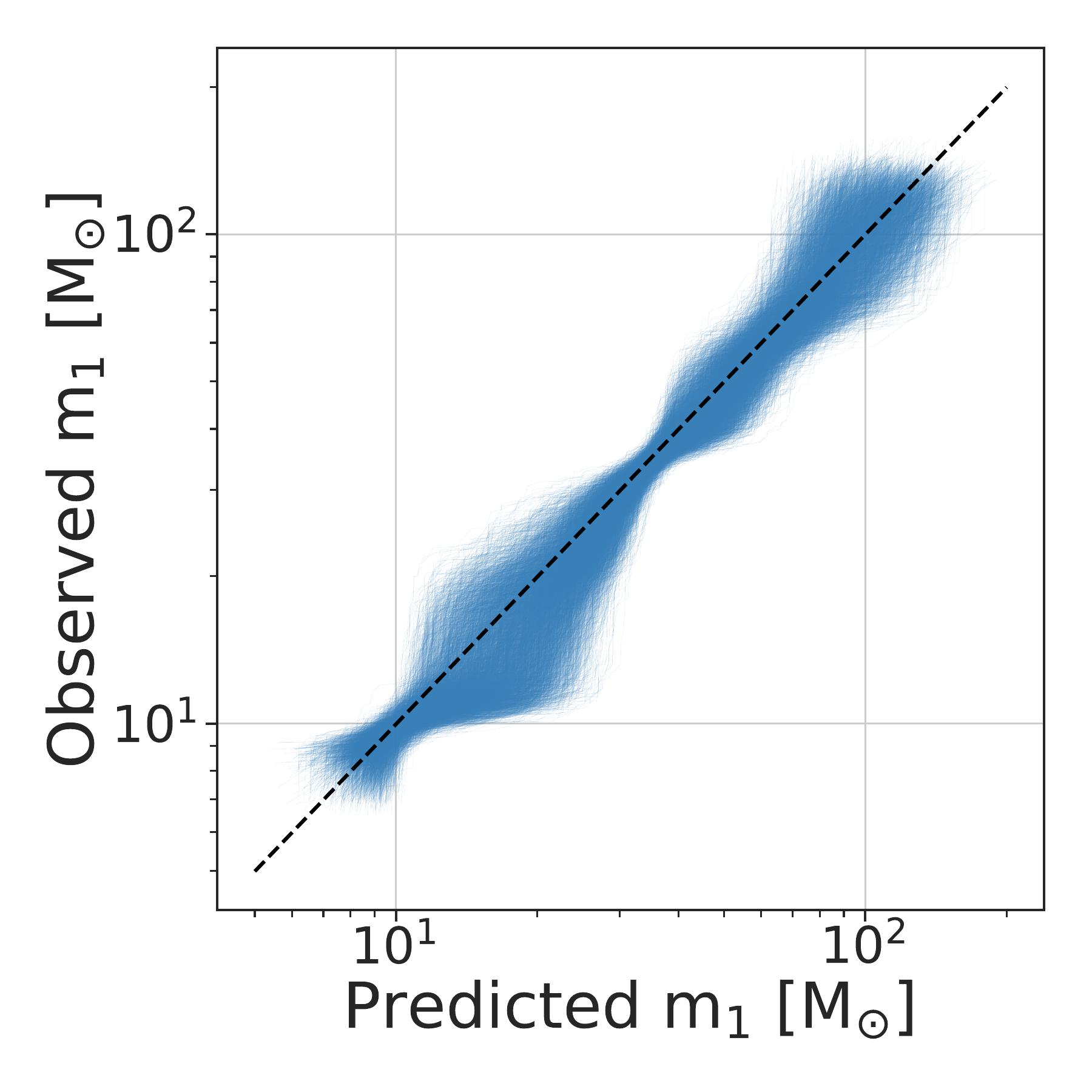}
\includegraphics[width=0.4\textwidth]{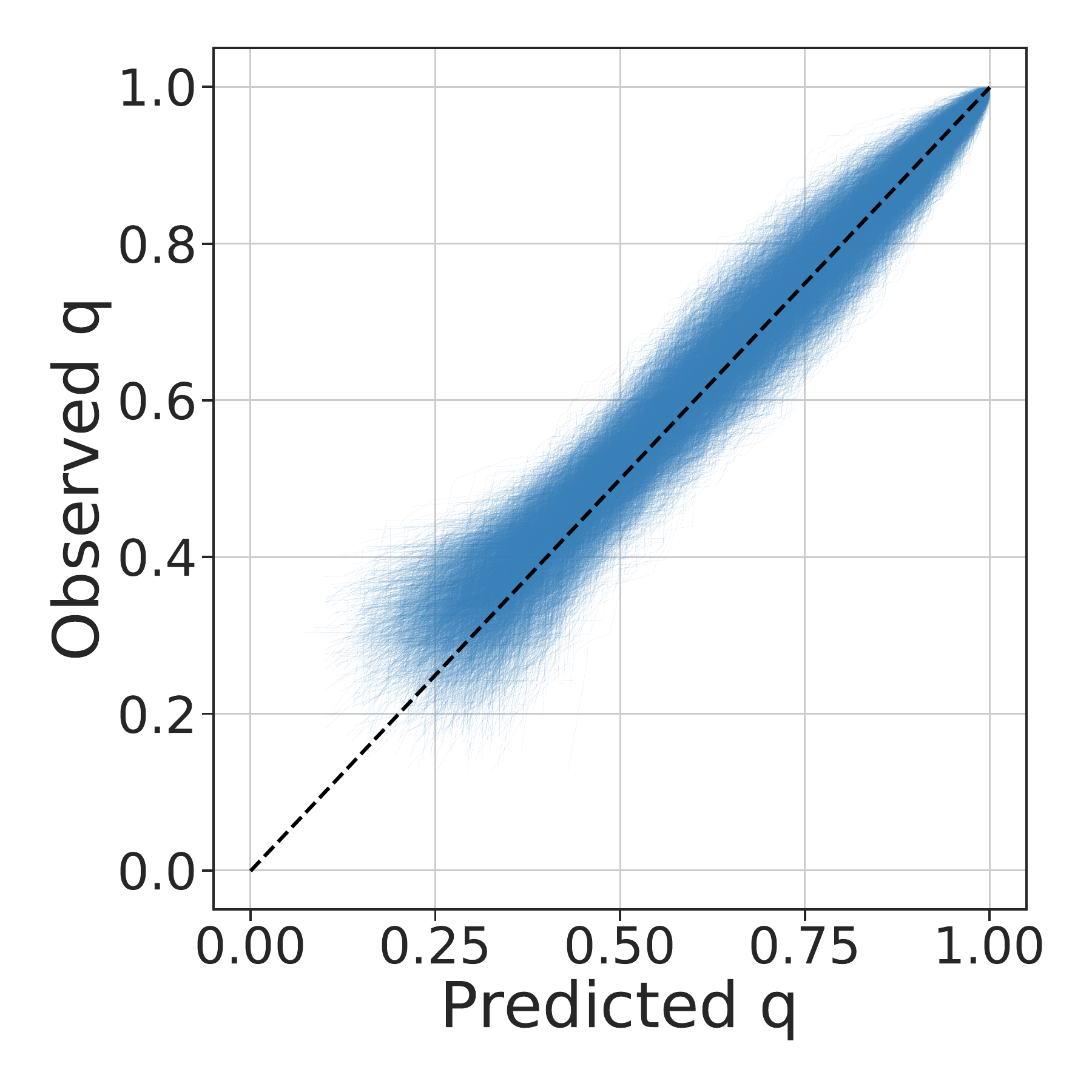}
\includegraphics[width=0.4\textwidth]{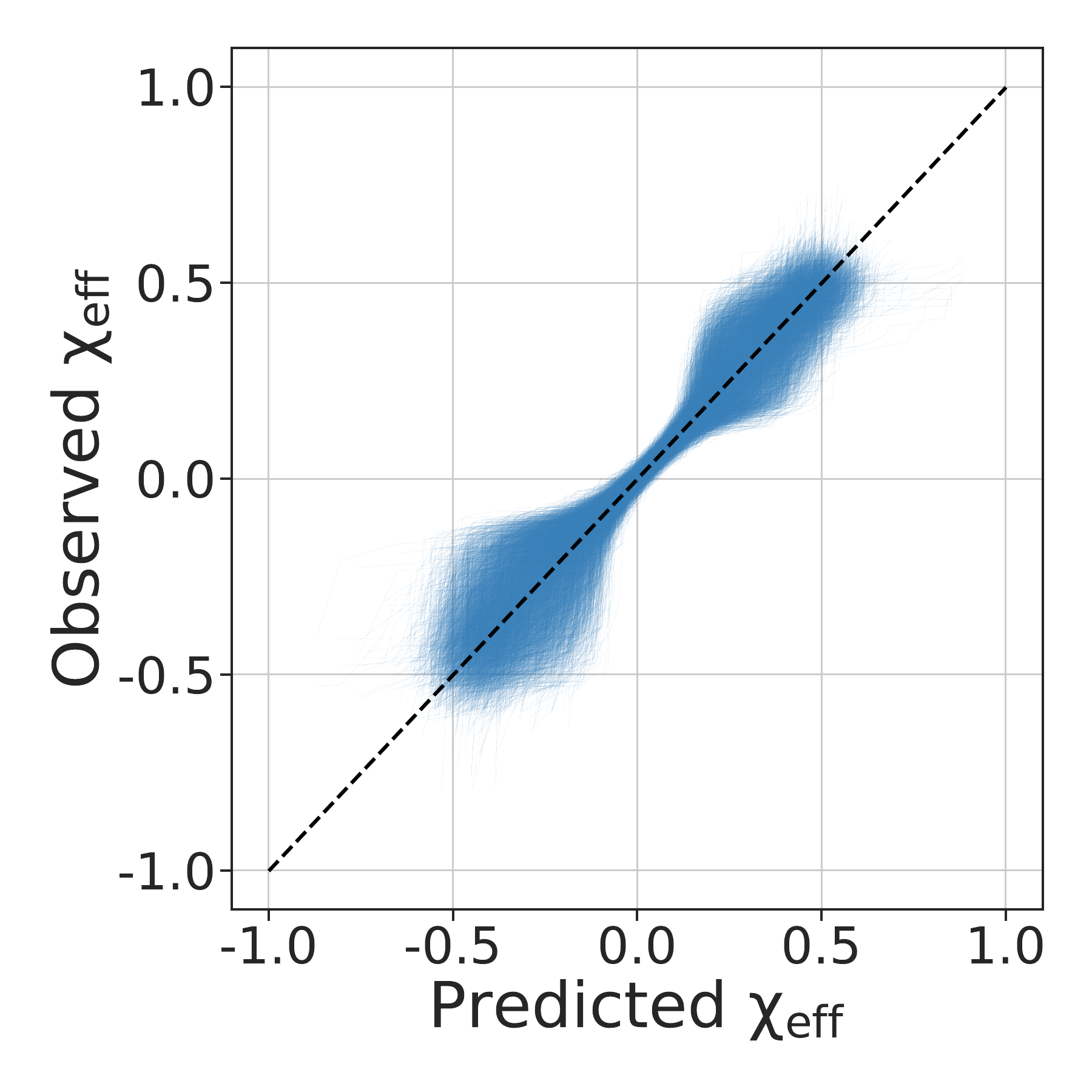}
\includegraphics[width=0.4\textwidth]{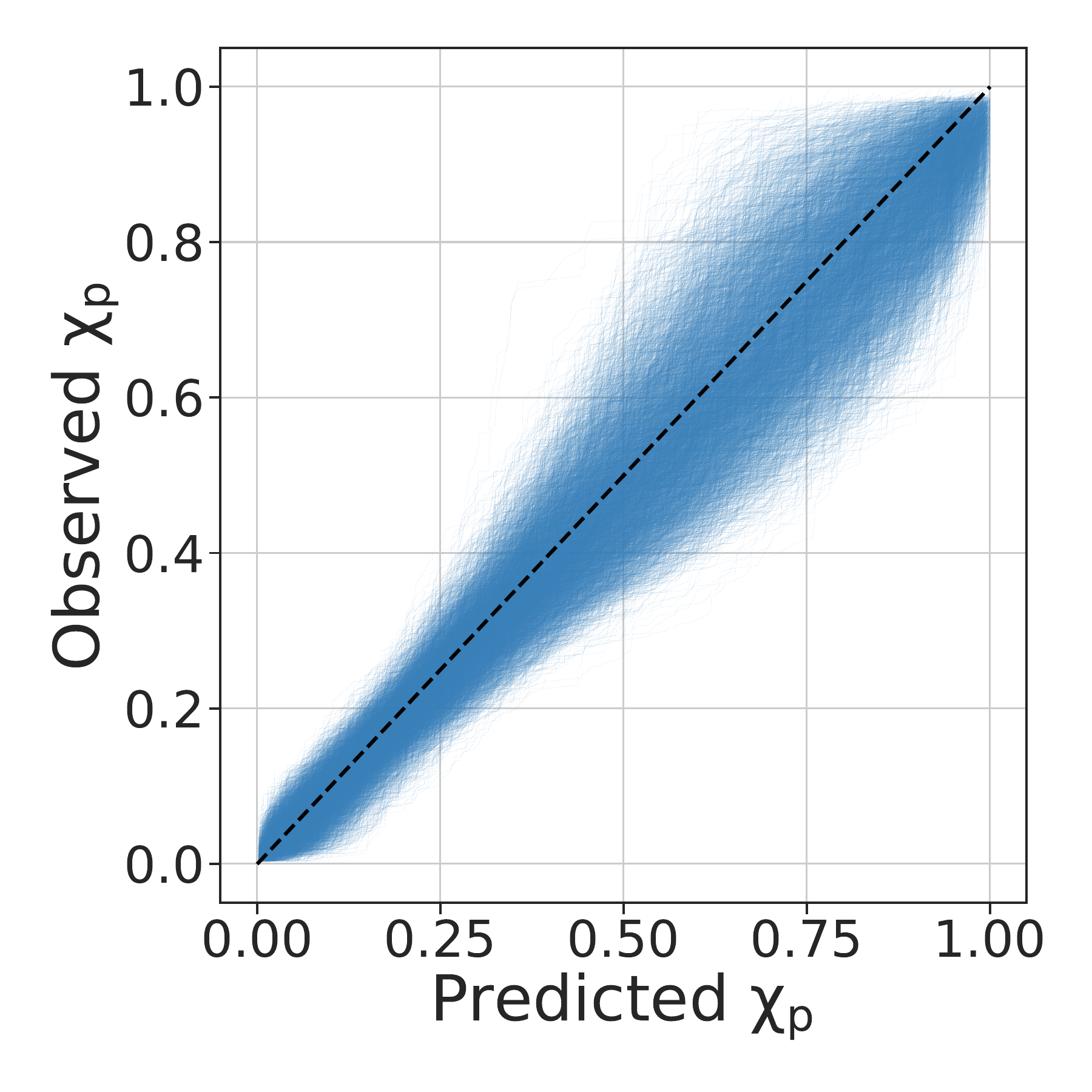}
\includegraphics[width=0.4\textwidth]{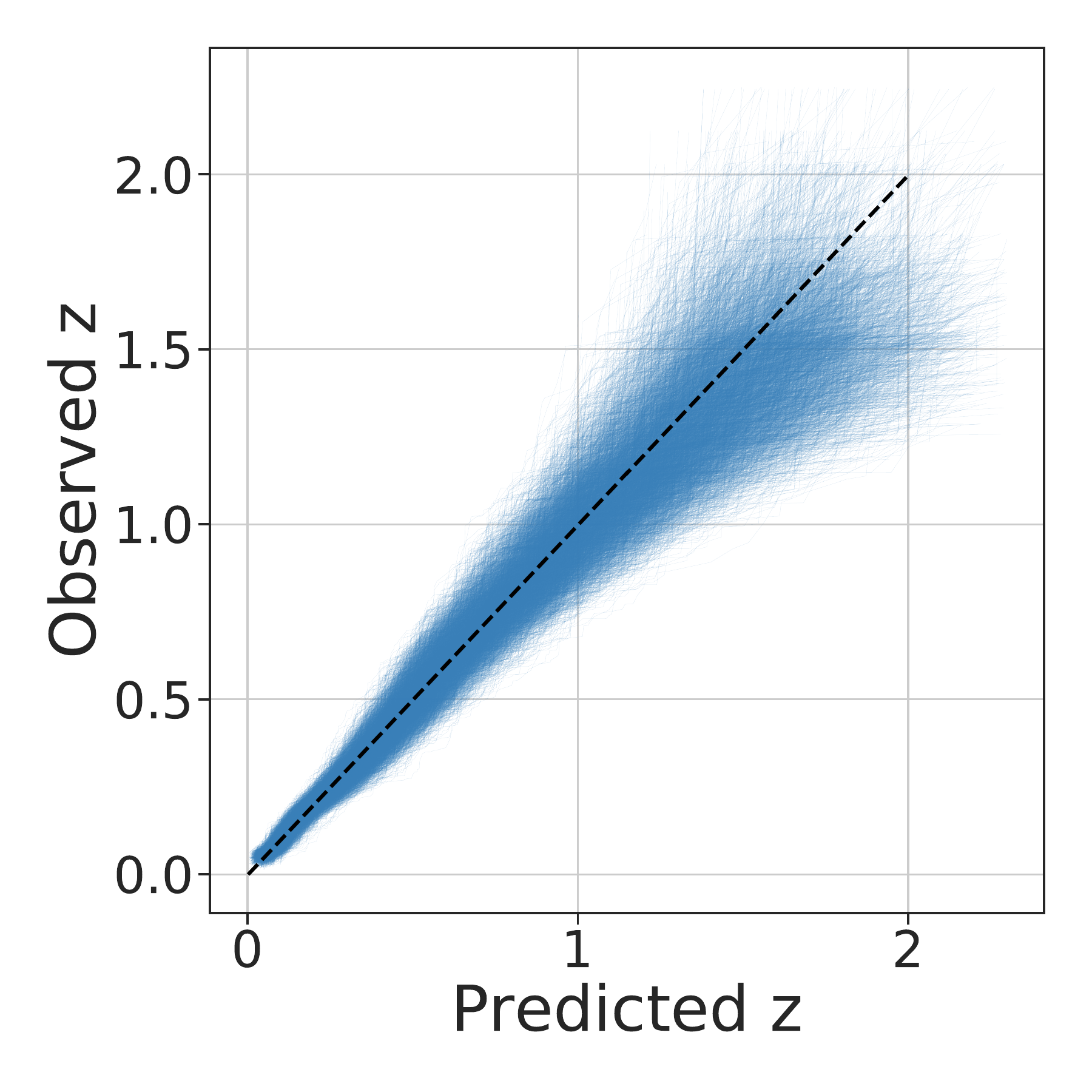}
\caption{\label{fig:ppd} Posterior predictive checks for the inferred $m_1,q,\chi_{eff},\chi_p,$ and $z$ distributions.} 
\end{figure*}

\section{Posterior Predictive Checks} 
\label{sec:app-ppchecks}
Next, we look for additional signatures of model-misspecification in our analysis by performing posterior predictive checks~\citep{Fishbach:2019ckx, Callister:2023tgi, Miller:2024sui}. We start by reweighting the posterior of each individual BBH's parameters to the population-informed prior, as displayed in Figure~\ref{fig:reweighted}. It can be seen that clusters identical to the subpopulations identified in the main text appear in each combination of observed BBH parameters. 
We then compare draws from this reweighted sample of observations to the posterior predictive population of detections. The latter is obtained by reweighting the detectable simulations, which were used to estimate selection effects, to the population-informed prior. As shown in Figure~\ref{fig:ppd}, the traces corresponding to each draw are concentrated about the diagonal, indicating no signatures of systemic model misspecification.

\begin{figure*}
\includegraphics[width=0.32\textwidth]{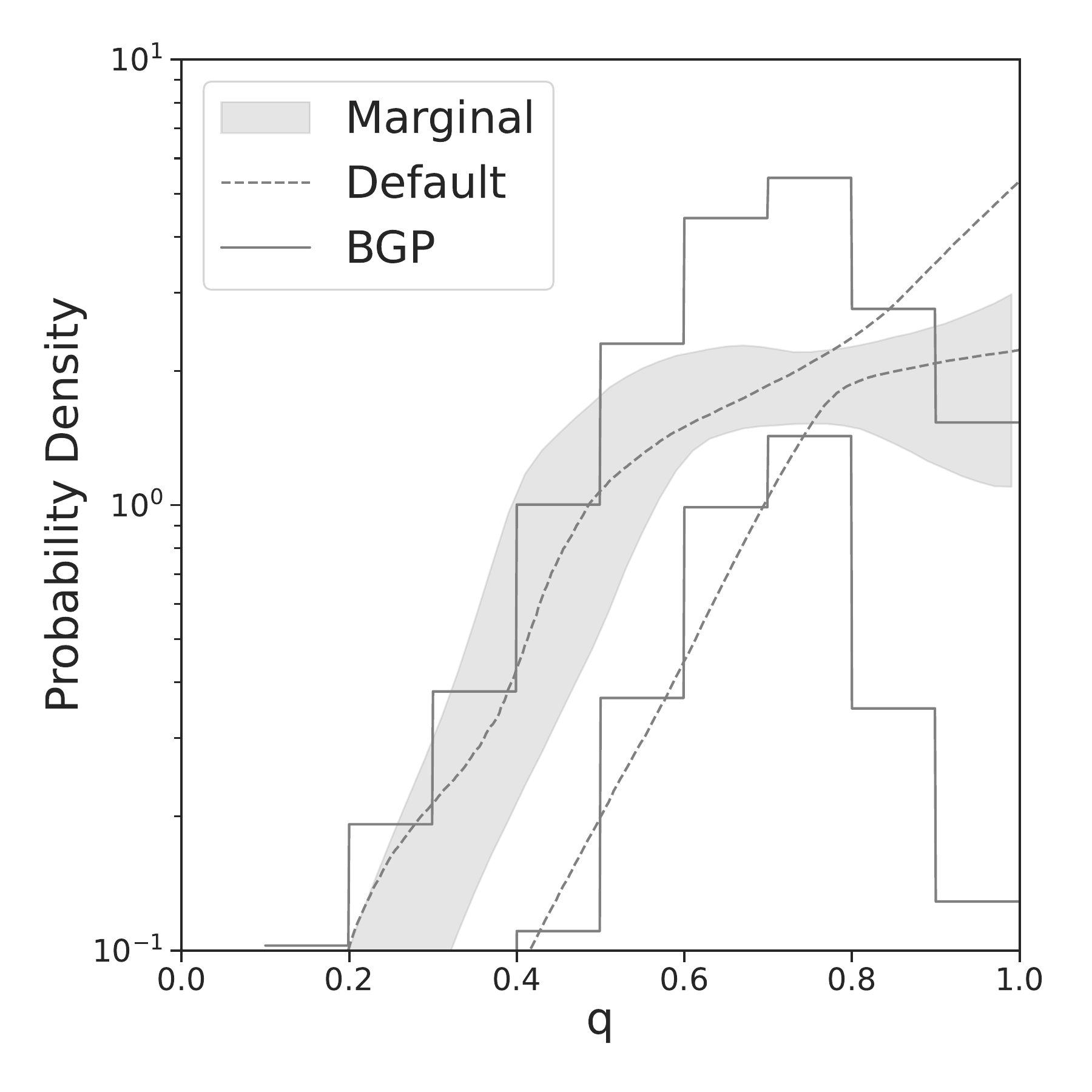}
\includegraphics[width=0.32\textwidth]{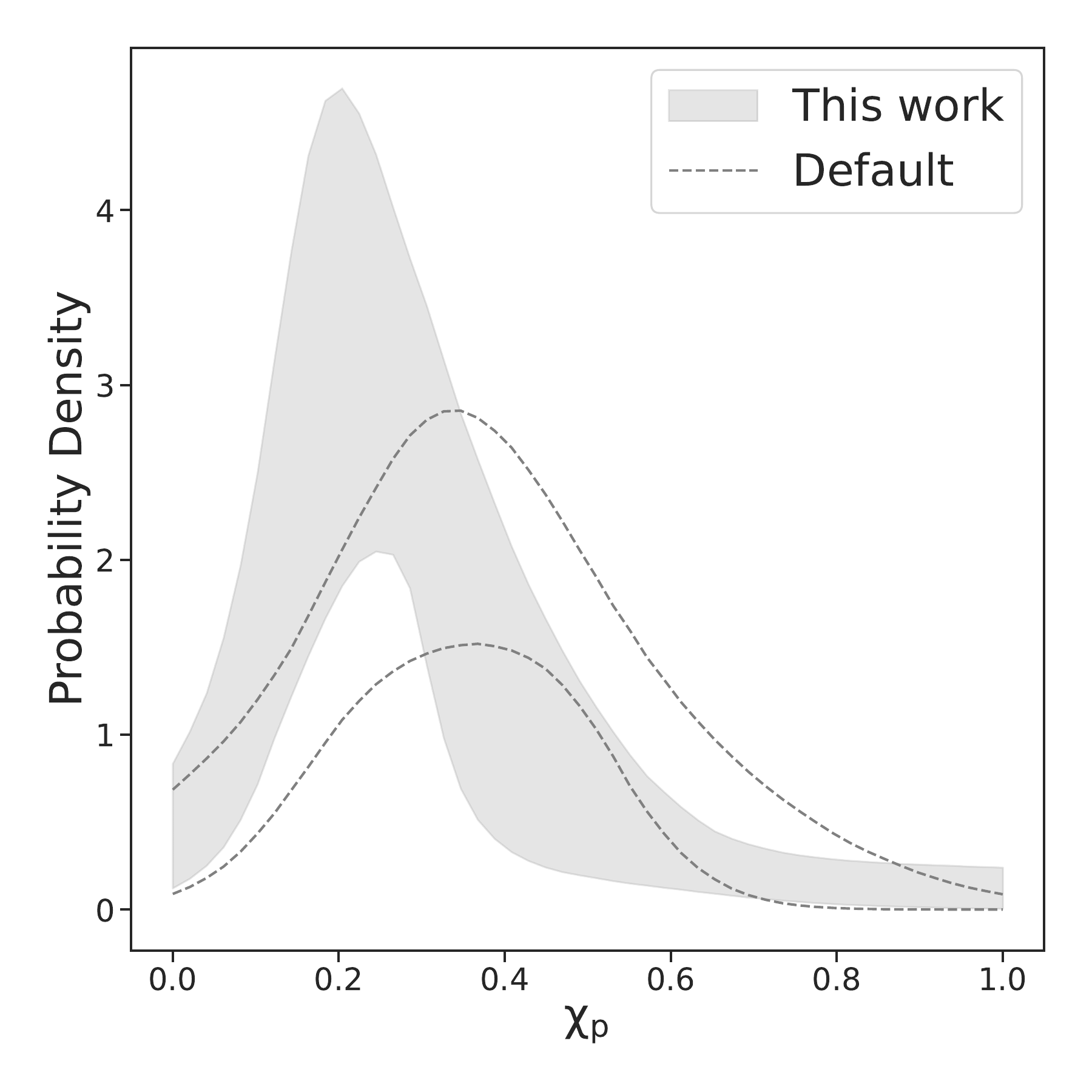}
\includegraphics[width=0.32\textwidth]{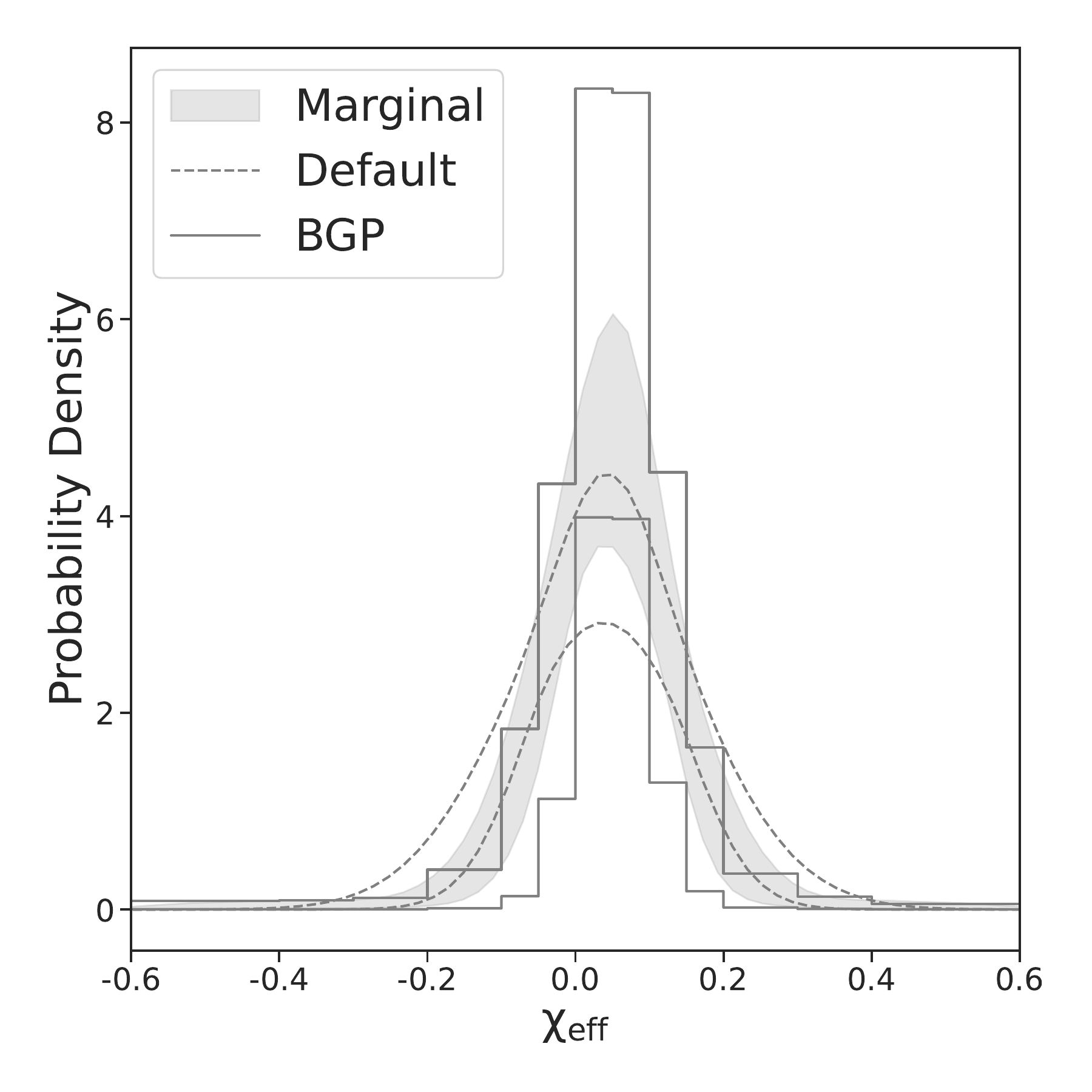}
\includegraphics[width=0.98\textwidth]{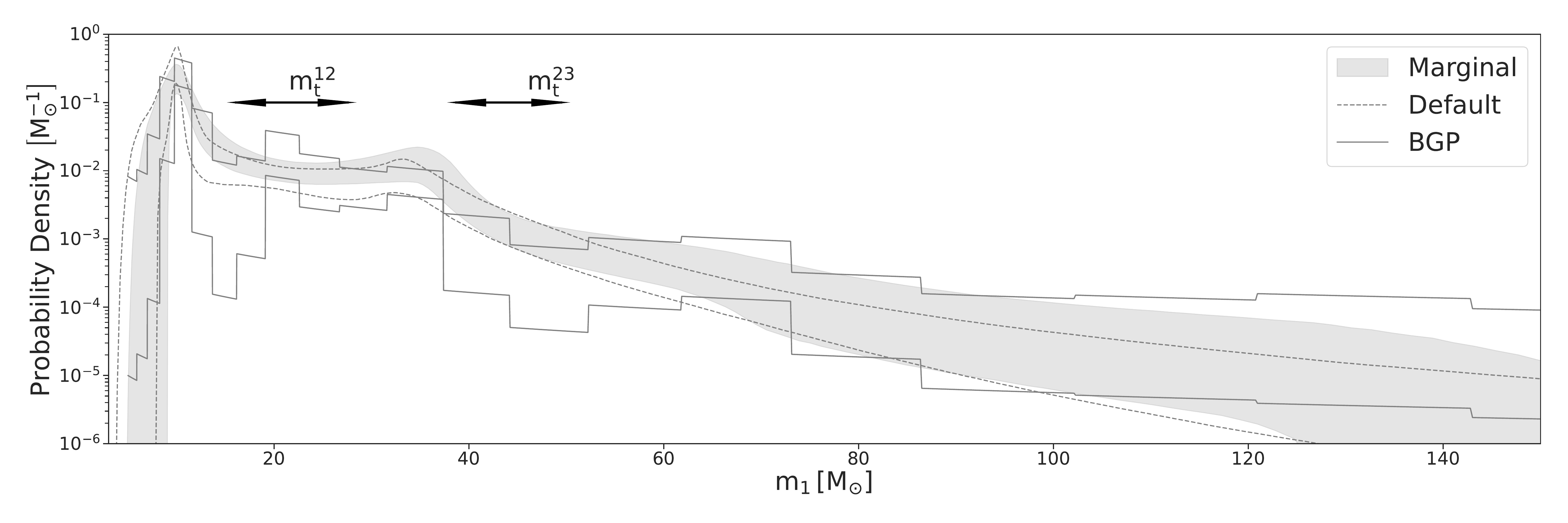}
\includegraphics[width=0.98\textwidth]{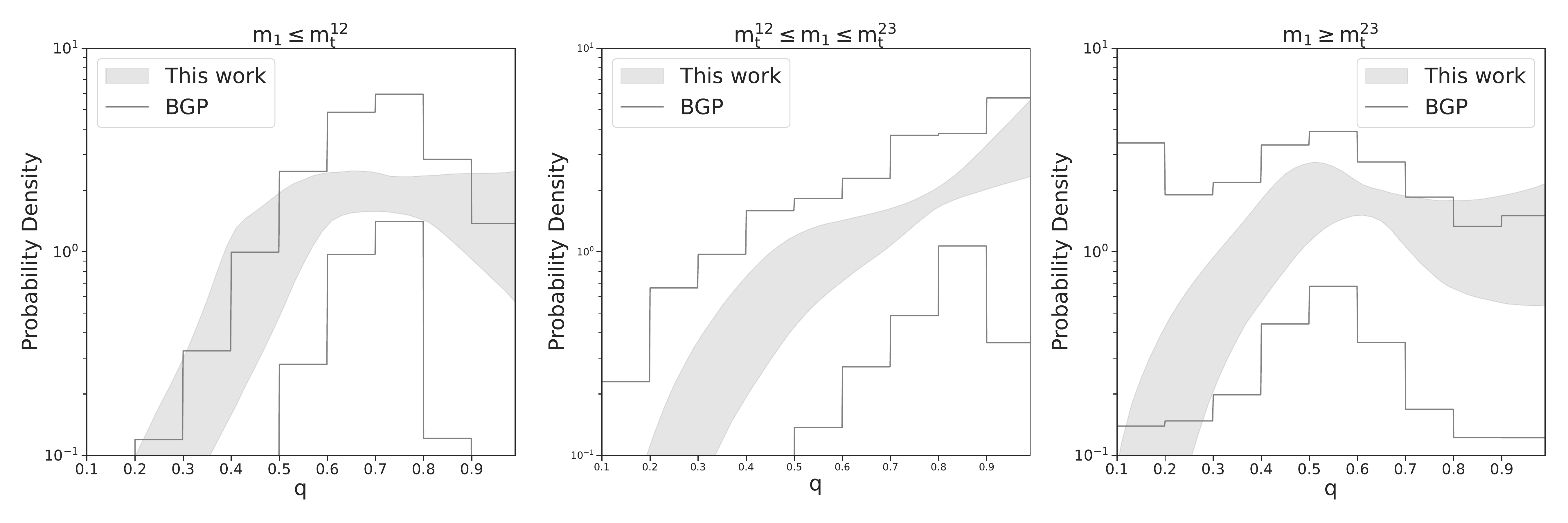}
\includegraphics[width=0.98\textwidth]{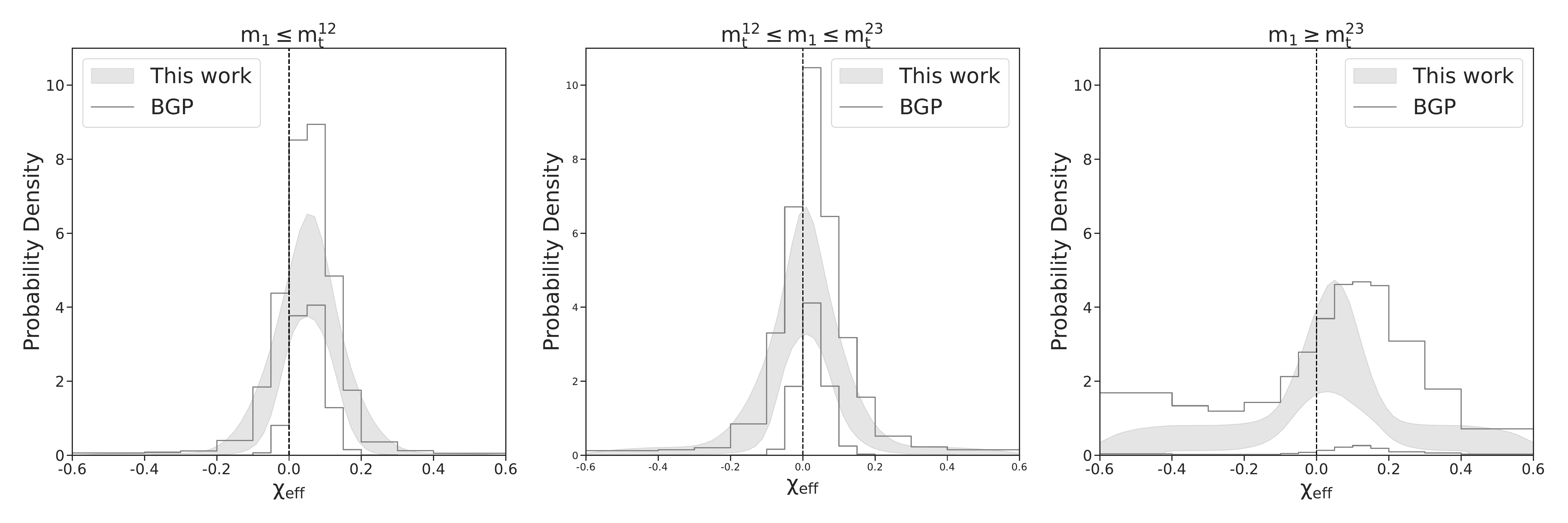}

\caption{\label{fig:marginal} Marginal Distributions of $q, \chi_{eff},\chi_p$ (\textit{first/top row}) and $m_1$~(second row), compared with the default and BGP models. Reconstructed mass-based transitions in the $q$~(\textit{third row}) and $\chi_{eff}$~(\textit{fourth row}) distributions are also compared with the corresponding BGP results.} 
\end{figure*}

\section {Marginal Distributions and Transitions} 
\label{sec:app-trans}
Finally, we show that our reconstructed distributions are consistent with alternative analyses that either do not account for subpopulations (such as the default case) or search for mass-based transitions in the distribution of other BBH parameters using flexible models for the joint distribution. We compare the marginal distributions of $m_1,q,\chi_{eff},\chi_p$ with that of the default analysis as well as a data-driven inference framework based on Binned Gaussian processes~(BGPs)~\citep{Sridhar:2025kvi, Ray:2024hos}. We further compare the inferred distributions of $q$ and $\chi_{eff}$ in the mass ranges separated by our reconstructed transition scales with the corresponding distributions yielded by the flexible BGP analysis. Since current implementations of the BGP are inherently three dimensional and unavailable for $\chi_p$, we could not compare our inferred transitions in the $\chi_p$ distribution with data-driven analogues. Note that our model, while being far more flexible than the default one is less so than the BGP framework, which can, in principle, reconstruct any feature in the joint distribution upto the resolution limit imposed by the choice of binning.

Both of these results are presented in Figure~\ref{fig:marginal}. We find the marginal distributions from our analysis and the ones recovered by the default and BGP models are consistent within uncertainties and in agreement over major features and trends in the population. Furthermore, we find the transitions in $q$ and $\chi_{eff}$ revealed by our model are consistent with the corresponding trends recovered by the flexible BGP framework. This indicates that our parametrizations are not leading to prior-driven conclusions regarding the astrophysical origins of BBH subpopulations. 


\begin{figure}[htt]
    \centering
    \includegraphics[width=0.98\linewidth]{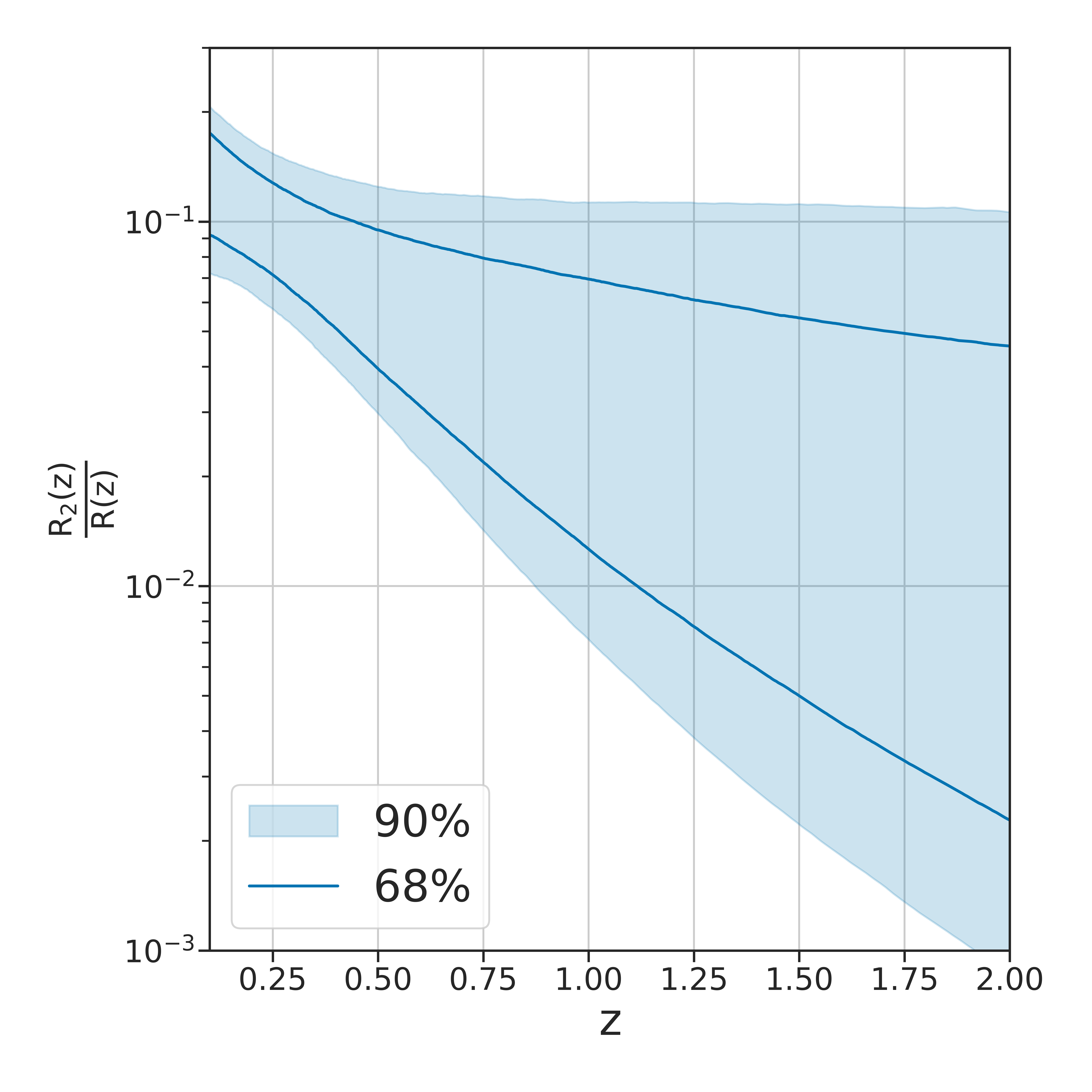}
    \caption{The relative merger rate of Comp. 2 as a function of redshift. It can clearly be seen that the the relative abundance of Comp. 2 (which is responsible for the $35M_{\odot}$) peak is decreasing with redshift at more than $1\sigma$ confidence.} 
    \label{fig:Rz-ratio}
\end{figure}
\begin{table}[ht]
\centering
\begin{tabular}{ccc}
\hline\hline
 Hyperparameter  & Posterior $90\%$ & Posterior $1\sigma$\\
\hline
 $\kappa_1$ & (2.3, 5.9) & (3.1, 5.2) \\
 $\kappa_2$ & (-0.2, 3.2) & (0.4, 2.4)\\
 $\kappa_3$ & (1.4, 5.8) & (2.6, 5.1)\\
\hline
\end{tabular}
\caption{\label{tab:kappa} Credible intervals~($90\%$ and $1\sigma$) on the redshift evolution parameter of each component. }
\end{table}
\section{Redshift evolution of the mass-spectrum}
\label{sec:app-redevol}
We now demonstrate how our results indicate a redshift evolving mass-spectrum for BBHs. As mentioned in the main text, the three components of our mixture model have distinct redshift evolution parameters. Since different components contribute dominantly to different features in the mass spectrum, our results imply that the resulting mass distribution will have distinct shapes at different redshift slices. 

We first list in Table~\ref{tab:kappa}, the $90\%$ and $1\sigma$ credible intervals of $\kappa_i$ for each component. It is clear that $\kappa_2$ is smaller than $\kappa_1$ and $\kappa_3$ by more than $1\sigma$. In other words, the relative abundance of BBHs in the $35M_{\odot}$ feature (which are predominantly Comp. 2), decreases with redshift at the $1\sigma$ level, as seen in Figure~\ref{fig:Rz-ratio}.

Here we note that the $(1+z)^{\kappa}$ model for redshift evolution of the merger rate is inherently limited in the sense that it can often extrapolate $R(z)$ based on measurements of $\kappa$ that are only informed by low redshift observations. In particular, for every mass-range, there is a maximum redshift~(from here on the \textit{detectibility horizon}) above which detector sensitivity drops to zero for any finite (non-zero) ranking statistic threshold~\citep{Heinzel:2024hva}. Hence, for low-mass events, the redshift evolution of the merger rate should be completely uninformed beyond the prior for high redshift values, whereas, for higher mass events, $R(z)$ can be further informed by high redshift observations. This is difficult to capture using a $(1+z)^{\kappa}$ which will always extrapolate $R(z)$ confidently, even for low mass subpopulations, and could likely yield incorrect results.

Hence, for correctly capturing the redshift evolution of different mass ranges, it is necessary to use flexible models for $R(z)$ such as the functional forms used in the phenomenological fits of \cite{Madau:2014bja}, or non-parametric approaches such as those employed by \cite{Ray:2023upk, Heinzel:2024hva}. Furthermore, since marginal distributions are dominated by features of the subpopulation corresponding to the $10M_{\odot}$ peak, whose detectability horizon redshift is very small, imposing the same $R(z)$ for the entire mass-spectrum, in fact, can be problematic. In such an analysis, restrictive~($(1+z)^{\kappa}$) models could bias $R(z)$ at high redshifts since it will always extrapolate based on the low redshift observations in the $10M_{\odot}$ peak, whereas flexible models will yield uninformative $R(z)$ for high redshifts despite there being additional information in the higher mass subpopulations. In other words, modeling $R(z)$ for all BBHs independent of mass will, in general, either yield biased results or discard information present in the data. A detailed investigation of these systematics associated with inferring redshift evolution of the merger rate is ongoing.

\begin{figure*}
\includegraphics[width=0.98\textwidth]{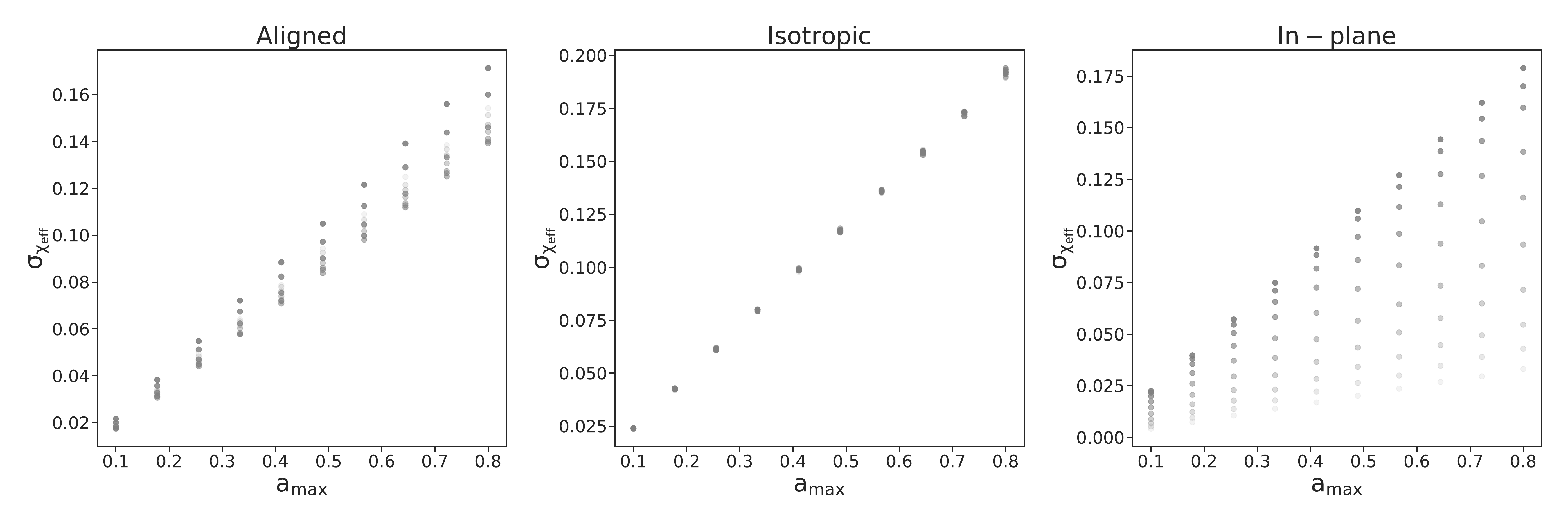}
\includegraphics[width=0.98\textwidth]{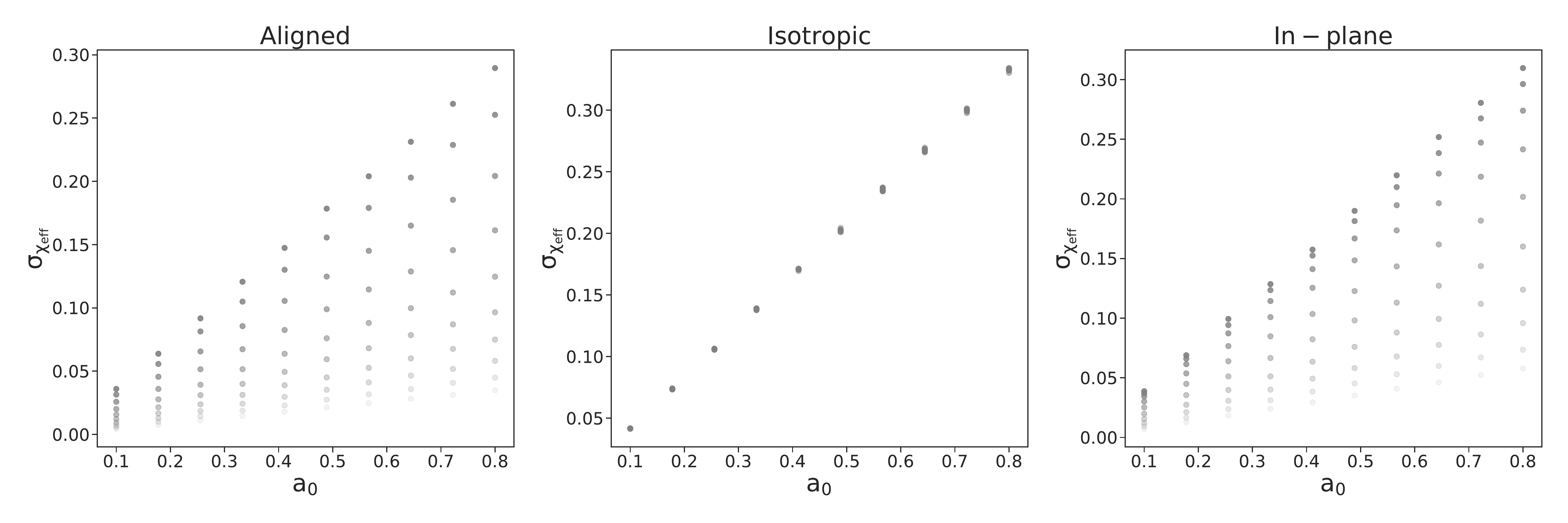}
\caption{\label{fig:sig-vs-a} The variation of $\sigma_{\chi_{eff}}$ with spin magnitude properties for the uniform~(top pannel) and fixed magnitude distributions. The opacity of points increases with the $\sigma_{\cos \theta}$ grid point.} 
\end{figure*}
\section{Effective and Component Spin Populations} 
\label{sec:app-eff-comp-spin}
We now discuss the connection between properties of the effective spin distributions and those of component spin magnitudes and tilts, which is crucial to our astrophysical interpretation in the main text, particularly for the astrophysical origins of the $35M_{\odot}$ subpopulation. Specifically, we show how the width of the effective spin distribution increases monotonically with the preference for higher spin magnitudes, irrespective of spin orientation. We further show that for fixed magnitude values, the median of the $\chi_p$ distribution increases monotonically with the fraction of systems with in-plane spin orientations. Finally, we also show that the fraction of systems with $\chi_{eff}<0$ is a direct measure of the ratio of aligned vs anti-aligned spin orientations with respect to the orbit.

To demonstrate these effects, we consider toy models of spin-magnitude~$(a_{1,2})$ and cosine tilt~$(\cos\theta_{1,2})$ distributions for various different cases. For spin magnitudes, we consider both fixed~$(a_{1,2}\sim \delta(a_0))$ and uniform~$(a_{1,2}\sim U_{0,a_{max})}$ distributions. For spin tilt angles with respect to the orbit, we explore three different cases, which are defined as follows.
\begin{enumerate}
    \item Preferentially \texttt{Aligned}: $\cos\theta_{1,2}\sim \mathcal{N}_{-1,1}^{1, \sigma_{\theta}}$,
    \item \texttt{Isotropic}: $\cos\theta_{1,2}\sim U(-1,1)$,
    \item and preferentially \texttt{in-plane}: $\cos\theta_{1,2}\sim \mathcal{N}_{-1,1}^{0, \sigma_{\theta}}$
\end{enumerate}
Finally, to obtain effective spin metrics, we use all possible combinations of these magnitude and orientation distributions with three different mass-ratio distributions:
\begin{enumerate}
    \item{\texttt{Case A}: $q\sim U(0.6,1.0)$},
    \item{\texttt{Case B}: $q\sim U(0.8,1.0)$}, and
    \item{\texttt{Case C}: $q\sim \mathcal{N}_{0,1}^{0.5,\sigma_q}$},
\end{enumerate}
which are analogous to the trends we find in our inferred subpopulations. We find that all of the trends in the effective spin distributions are, overall, independent of the choice of $q$ distributions.

\begin{figure*}
\includegraphics[width=0.48\textwidth]{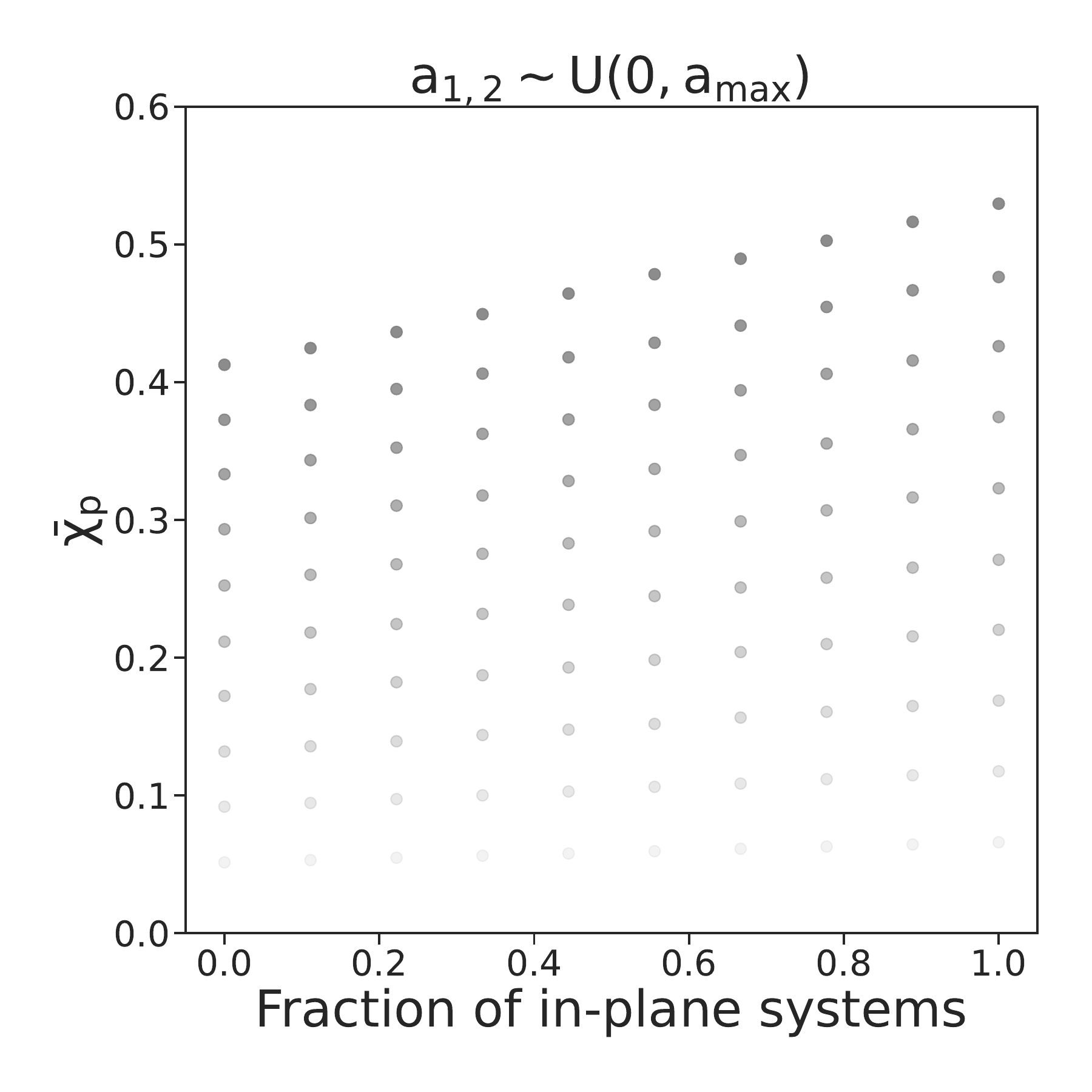}
\includegraphics[width=0.48\textwidth]{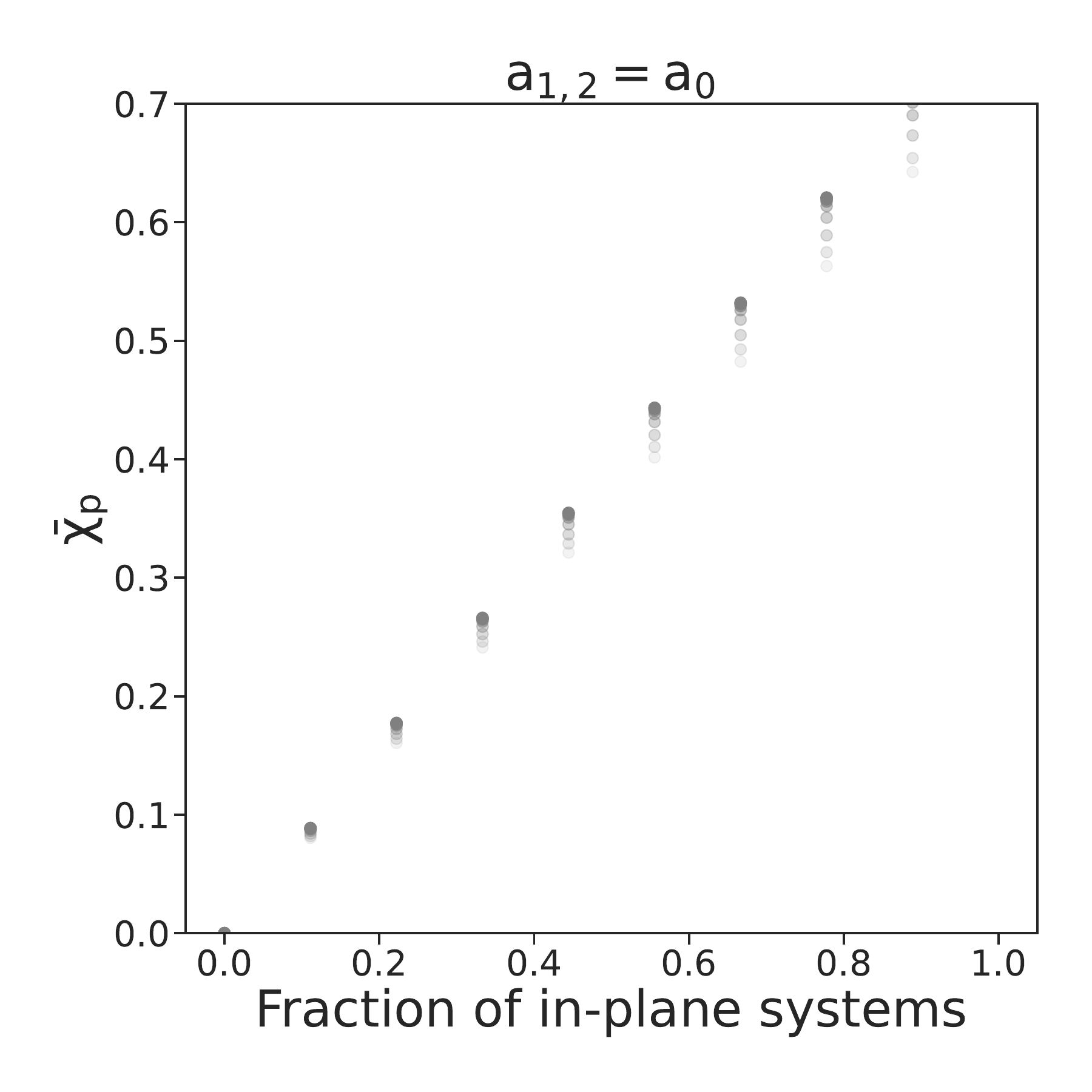}
\caption{\label{fig:barchip-vs-finp} The variation of $\bar{\chi}_p$ with the fraction of in-plane systems in a mixture of isotropic and in-plane populations with $\sigma_{\cos\theta}=0.1$. The opacity of points increases with $a_{max}$ or $a_0$. We use $q\sim U(0.8,0.1)$ for this plot even though the trends are identical for cases A and C.} 
\end{figure*}

\begin{figure*}
\includegraphics[width=0.48\textwidth]{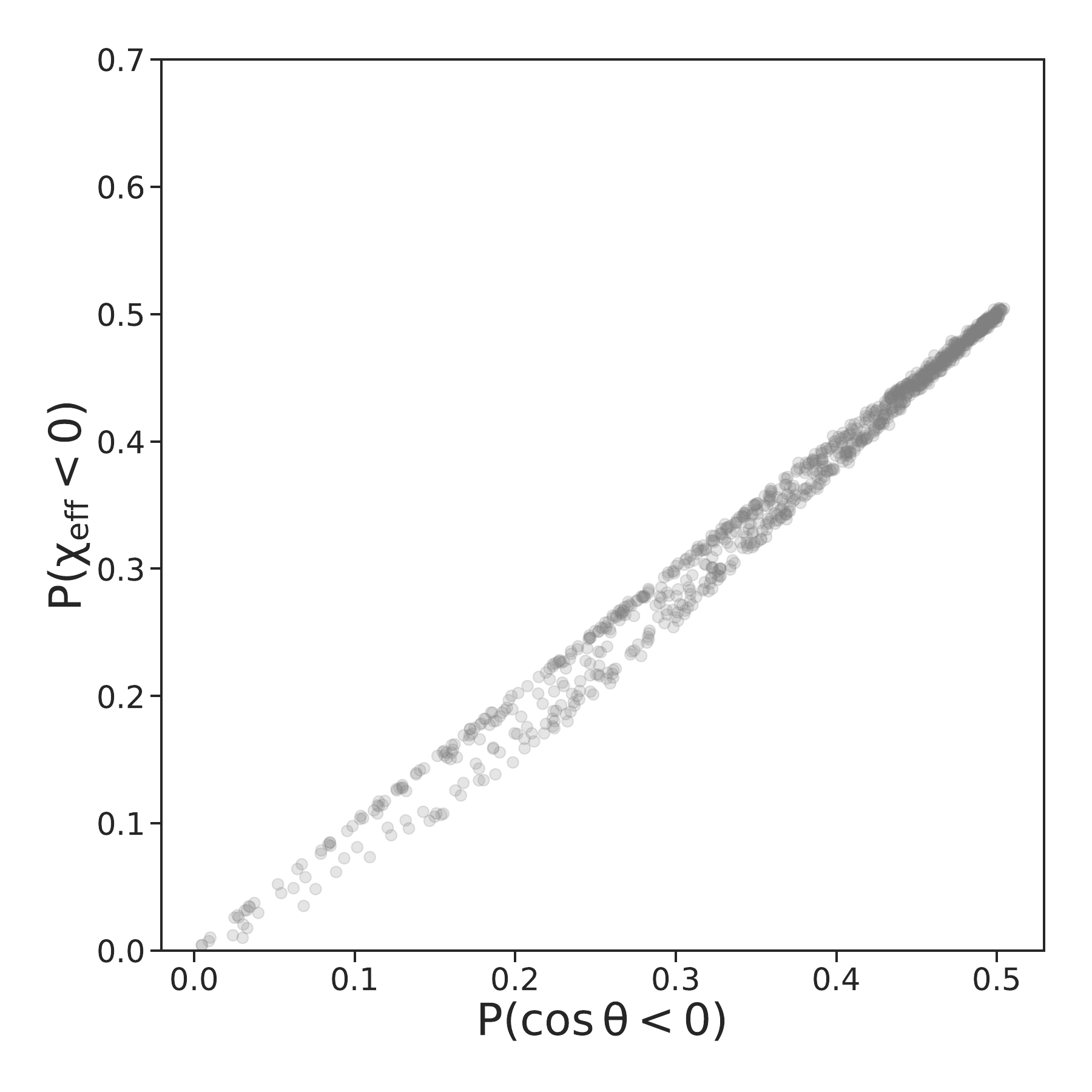}
\caption{\label{fig:chifrac-vs-ctfrac} The variation of $P(\chi_{eff}<0)$ with $P(\cos\theta <0)$ in a random mixture of \texttt{Aligned}, \texttt{Isotropic}, and \texttt{In-plane} populations. For each realization we randomly draw a value of $a_{max}\sim U(0,0.8)$, $\sigma_{\cos\theta}\sim U(0.1,1)$ and mixing fractions $(f_1,f_2,f_3)\sim \texttt{Dir}(0.33)$. Given these values we construct a mixed population of spins with $f_1,f_2,f_3$ being the fraction of \texttt{Aligned}, \texttt{Isotropic}, and \texttt{In-plane} systems respectively, while drawing $q\sim U(0.6,1)$. We then compute $P(\chi_{eff}<0)$ and $P(\cos\theta <0)$ for each realization. We plot the resulting trend over 1000 realizations (which is found to be identical for case B and c $q$ distributions.   } 
\end{figure*}

We first vary $a_{max}/a_{0}$, and, $\sigma_{\theta}$ on a grid to show, in Figure~\ref{fig:sig-vs-a}, that $\sigma_{\chi_{eff}}$ is a monotonic function of either $a_{max}$ or $a_0$ depending on the choice of the spin magnitude distribution, regardless of the distribution of spin-tilt angles. We then consider mixtures of isotropic and in-plane tilts with $\sigma_{\cos \theta} =0.1$ and vary the mixing fraction on a grid. We show in Figure~\ref{fig:barchip-vs-finp} that for each value of $a_0$ or $a_{max}$, $\bar{\chi_p}$ is a monotonic function of the fraction of systems drawn from the in-plane population. Finally, we show in Figure~\ref{fig:chifrac-vs-ctfrac} that for random mixtures of aligned, isotropic, and in-plane populations, $P(\chi_{eff}<0)$ is a monotonic function of $P(\cos\theta <0)$.


\begin{figure*}
\includegraphics[width=0.48\textwidth]{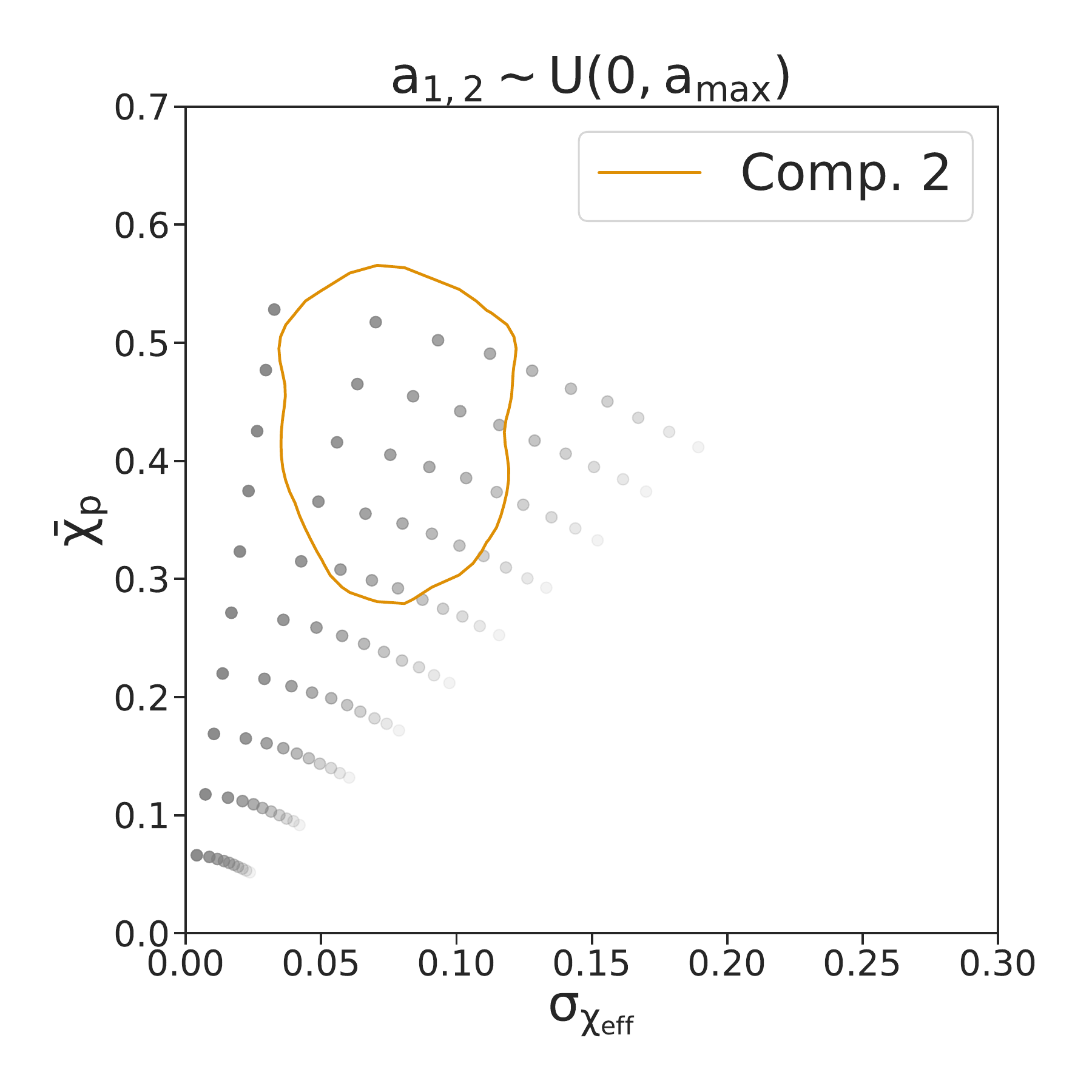}
\includegraphics[width=0.48\textwidth]{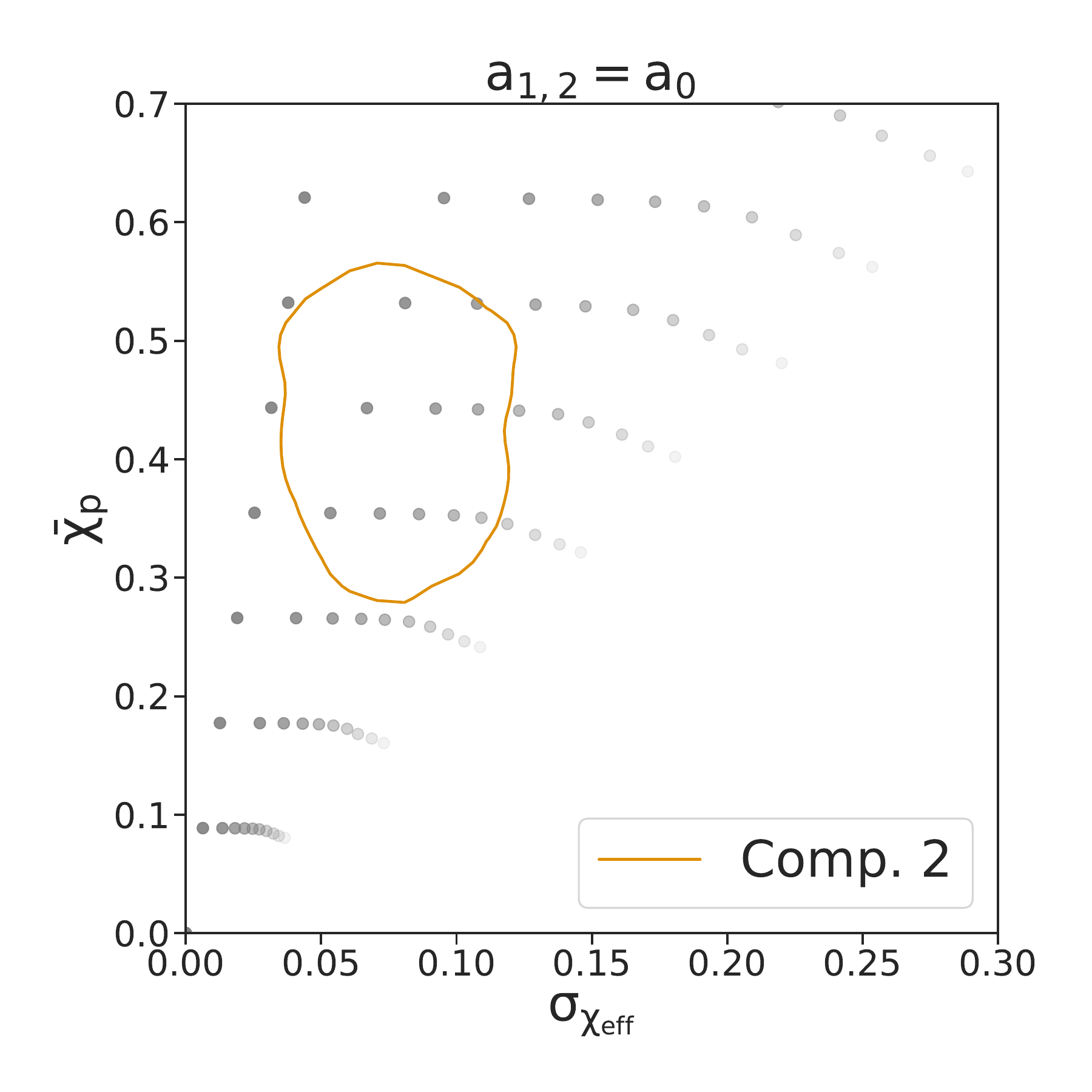}
\caption{\label{fig:sig-vs-chip} The variation of $\bar{\chi_p}$ with $\sigma_{\chi_{eff}}$ for a mixture of isotropic and in-plane populations with $\sigma_{\cos \theta}=0.1$. The opacity represents the fraction of in-plane systems and the contour represents the 90\% credible interval of these effective spin metrics for component 2 as inferred from GWTC-4.} 
\end{figure*}

We then compare, in Figure~\ref{fig:sig-vs-chip}, the inferred values of $\sigma_{\chi_{eff}}$ and $\bar{\chi}_p$ for component 2 with those derived from a mixture of isotropic and preferentially inplane orientations at $\sigma_{\cos\theta}=0.1$. We show that these metrics, while capable of constraining the fraction of excess in-plane systems, cannot yield informative conclusions on the same for component 2, given the measurement uncertainties of GWTC-4. The toy model population uses $q\sim U(0.8,1)$ even though the exact same trends are recovered for cases A and C.


\bibliography{sample701}{}

@article{Fishbach:2018edt,
    author = "Fishbach, Maya and Holz, Daniel E. and Farr, Will M.",
    title = "{Does the Black Hole Merger Rate Evolve with Redshift?}",
    eprint = "1805.10270",
    archivePrefix = "arXiv",
    primaryClass = "astro-ph.HE",
    doi = "10.3847/2041-8213/aad800",
    journal = "Astrophys. J. Lett.",
    volume = "863",
    number = "2",
    pages = "L41",
    year = "2018"
}

@article{Sridhar:2025kvi,
    author = "Sridhar, Omkar and Ray, Anarya and Kalogera, Vicky",
    title = "{Characterizing Binary Black Hole Subpopulations in GWTC-4 with Binned Gaussian Processes: On the Origins of the $35M_{\odot}$ Peak}",
    eprint = "2511.22093",
    archivePrefix = "arXiv",
    primaryClass = "astro-ph.HE",
    reportNumber = "LIGO-P2500712",
    month = "11",
    year = "2025",
    journal = "",
}

@article{Banagiri:2025dmy,
    author = "Banagiri, Sharan and Thrane, Eric and Lasky, Paul D.",
    title = "{Evidence for Three Subpopulations of Merging Binary Black Holes at Different Primary Masses}",
    eprint = "2509.15646",
    archivePrefix = "arXiv",
    primaryClass = "astro-ph.HE",
    month = "9",
    year = "2025",
    journal = "",
}

@article{Ray:2024hos,
    author = "Ray, Anarya and Maga{\~n}a Hernandez, Ignacio and Breivik, Katelyn and Creighton, Jolien",
    title = "{Searching for Binary Black Hole Subpopulations in Gravitational-wave Data Using Binned Gaussian Processes}",
    eprint = "2404.03166",
    archivePrefix = "arXiv",
    primaryClass = "astro-ph.HE",
    reportNumber = "LIGO-P2400115",
    doi = "10.3847/1538-4357/adf22a",
    journal = "Astrophys. J.",
    volume = "991",
    number = "1",
    pages = "17",
    year = "2025"
}

@article{Thrane:2018qnx,
    author = "Thrane, Eric and Talbot, Colm",
    title = "{An introduction to Bayesian inference in gravitational-wave astronomy: parameter estimation, model selection, and hierarchical models}",
    eprint = "1809.02293",
    archivePrefix = "arXiv",
    primaryClass = "astro-ph.IM",
    doi = "10.1017/pasa.2019.2",
    journal = "Publ. Astron. Soc. Austral.",
    volume = "36",
    pages = "e010",
    year = "2019",
    note = "[Erratum: Publ.Astron.Soc.Austral. 37, e036 (2020)]"
}

@inproceedings{popgw2,
  doi = {10.1063/1.1835214},
  url = {https://doi.org/10.1063/1.1835214},
  year = {2004},
  publisher = {{AIP}},
  author = {Thomas J. Loredo},
  title = {Accounting for Source Uncertainties in Analyses of Astronomical Survey Data},
  booktitle = {{AIP} Conference Proceedings}
}

@article{pop-vitale,
author = "Vitale, Salvatore and Gerosa, Davide and Farr, Will M. and Taylor, Stephen R.",
    title = "{Inferring the properties of a population of compact binaries in presence of selection effects}",
    eprint = "2007.05579",
    archivePrefix = "arXiv",
    primaryClass = "astro-ph.IM",
    doi = "10.1007/978-981-15-4702-7\_45-1",
    month = "7",
    year = "2020",
    journal = "",
}

@article{popgw3,
  title = {Reconstructing phenomenological distributions of compact binaries via gravitational wave observations},
  author = {Wysocki, Daniel and Lange, Jacob and O'Shaughnessy, Richard},
  journal = {Phys. Rev. D},
  volume = {100},
  issue = {4},
  pages = {043012},
  numpages = {21},
  year = {2019},
  month = {Aug},
  publisher = {American Physical Society},
  doi = {10.1103/PhysRevD.100.043012},
  url = {https://link.aps.org/doi/10.1103/PhysRevD.100.043012}
}

@article{Mandel:2018mve,
    author = "Mandel, Ilya and Farr, Will M. and Gair, Jonathan R.",
    title = "{Extracting distribution parameters from multiple uncertain observations with selection biases}",
    eprint = "1809.02063",
    archivePrefix = "arXiv",
    primaryClass = "physics.data-an",
    doi = "10.1093/mnras/stz896",
    journal = "Mon. Not. Roy. Astron. Soc.",
    volume = "486",
    number = "1",
    pages = "1086--1093",
    year = "2019"
}

@article{Pdet1-Farr,
doi = {10.3847/2515-5172/ab1d5f},
url = {https://dx.doi.org/10.3847/2515-5172/ab1d5f},
year = {2019},
month = {may},
publisher = {The American Astronomical Society},
volume = {3},
number = {5},
pages = {66},
author = {Will M. Farr},
title = {Accuracy Requirements for Empirically Measured Selection Functions},
journal = {Research Notes of the AAS}
}

@misc{Pdet2-essick,
      title={Precision Requirements for Monte Carlo Sums within Hierarchical Bayesian Inference}, 
      author={Reed Essick and Will Farr},
      year={2022},
      eprint={2204.00461},
      archivePrefix={arXiv},
      primaryClass={astro-ph.IM}
}

@article{Talbot2025,
  author = {Colm Talbot and Amanda Farah and Shanika Galaudage and Jacob Golomb and Hui Tong},
  title = {GWPopulation: Hardware agnostic population inference for compact binaries and beyond},
  journal = {Journal of Open Source Software},
  doi = {10.21105/joss.07753},
  url = {https://doi.org/10.21105/joss.07753},
  year = {2025},
  publisher = {The Open Journal},
  volume = {10},
  number = {109},
  pages = {7753},
  archivePrefix = {arXiv},
  eprint = {2409.14143},
  primaryClass = {astro-ph.IM},
}

@article{Antonini:2024het,
    author = "Antonini, Fabio and Romero-Shaw, Isobel M. and Callister, Thomas",
    title = "{Star Cluster Population of High Mass Black Hole Mergers in Gravitational Wave Data}",
    eprint = "2406.19044",
    archivePrefix = "arXiv",
    primaryClass = "astro-ph.HE",
    doi = "10.1103/PhysRevLett.134.011401",
    journal = "Phys. Rev. Lett.",
    volume = "134",
    number = "1",
    pages = "011401",
    year = "2025"
}

@article{Wysocki2019,
  author  = {Wysocki, Daniel and Lange, Jacob and O'Shaughnessy, Richard},
  title   = {Exploring Short Gamma-Ray Burst Progenitors with Gravitational Waves},
  journal = {Physical Review D},
  volume  = {100},
  number  = {4},
  pages   = {043012},
  year    = {2019},
  publisher = {American Physical Society},
  doi     = {10.1103/PhysRevD.100.043012}
}

@article{Sadiq:2021fin,
    author = "Sadiq, Jam and Dent, Thomas and Wysocki, Daniel",
    title = "{Flexible and fast estimation of binary merger population distributions with an adaptive kernel density estimator}",
    eprint = "2112.12659",
    archivePrefix = "arXiv",
    primaryClass = "gr-qc",
    doi = "10.1103/PhysRevD.105.123014",
    journal = "Phys. Rev. D",
    volume = "105",
    number = "12",
    pages = "123014",
    year = "2022"
}

@article{Talbot:2023pex,
    author = "Talbot, Colm and Golomb, Jacob",
    title = "{Growing pains: understanding the impact of likelihood uncertainty on hierarchical Bayesian inference for gravitational-wave astronomy}",
    eprint = "2304.06138",
    archivePrefix = "arXiv",
    primaryClass = "astro-ph.IM",
    doi = "10.1093/mnras/stad2968",
    journal = "Mon. Not. Roy. Astron. Soc.",
    volume = "526",
    number = "3",
    pages = "3495--3503",
    year = "2023"
}

@article{Gerosa:2021mno,
    author = "Gerosa, Davide and Fishbach, Maya",
    title = "{Hierarchical mergers of stellar-mass black holes and their gravitational-wave signatures}",
    eprint = "2105.03439",
    archivePrefix = "arXiv",
    primaryClass = "astro-ph.HE",
    doi = "10.1038/s41550-021-01398-w",
    journal = "Nature Astron.",
    volume = "5",
    number = "8",
    pages = "749--760",
    year = "2021"
}

@article{Skilling:2006gxv,
    author = "Skilling, John",
    title = "{Nested sampling for general Bayesian computation}",
    doi = "10.1214/06-BA127",
    journal = "Bayesian Analysis",
    volume = "1",
    number = "4",
    pages = "833--859",
    year = "2006"
}

@ARTICLE{2019PhRvD.100d3030T,
  author = {{Talbot}, Colm and {Smith}, Rory and {Thrane}, Eric and {Poole}, Gregory B.},
  title = "{Parallelized inference for gravitational-wave astronomy}",
  journal = {\prd},
  year = 2019,
  month = aug,
  volume = {100},
  number = {4},
  eid = {043030},
  pages = {043030},
  doi = {10.1103/PhysRevD.100.043030},
  archivePrefix = {arXiv},
  eprint = {1904.02863},
  primaryClass = {astro-ph.IM},
}

@article{Farah:2026jlc,
    author = "Farah, Amanda M. and Vijaykumar, Aditya and Fishbach, Maya",
    title = "{The steep redshift evolution of the hierarchical merger rate may cause the $z$-$\chi_{\rm eff}$ correlation}",
    eprint = "2601.03456",
    archivePrefix = "arXiv",
    primaryClass = "astro-ph.HE",
    month = "1",
    year = "2026",
    journal = "",
}

@article{Ray:2025xti,
    author = "Ray, Anarya and Kalogera, Vicky",
    title = "{Reexamining Evidence of a Pair-instability Mass Gap in the Binary Black Hole Population}",
    eprint = "2510.18867",
    archivePrefix = "arXiv",
    primaryClass = "astro-ph.HE",
    doi = "10.3847/2041-8213/ae374d",
    journal = "Astrophys. J. Lett.",
    volume = "998",
    number = "1",
    pages = "L20",
    year = "2026"
}

@article{Antonini:2025ilj,
    author = "Antonini, Fabio and Romero-Shaw, Isobel and Callister, Thomas and Dosopoulou, Fani and Chattopadhyay, Debatri and Gieles, Mark and Mapelli, Michela",
    title = "{Gravitational waves reveal the pair-instability mass gap and constrain nuclear burning in massive stars}",
    eprint = "2509.04637",
    archivePrefix = "arXiv",
    primaryClass = "astro-ph.HE",
    month = "9",
    year = "2025",
    journal = "",
}

@article{Tong:2025wpz,
    author = "Tong, Hui and others",
    title = "{Evidence of the pair instability gap in the distribution of black hole masses}",
    eprint = "2509.04151",
    archivePrefix = "arXiv",
    primaryClass = "astro-ph.HE",
    month = "9",
    year = "2025",
    journal = "",
}

@article{Vijaykumar:2026zjy,
    author = "Vijaykumar, Aditya and Farah, Amanda M. and Fishbach, Maya",
    title = "{The maximum mass ratio of hierarchical mergers may cause the $q$-$\chi_{\rm eff}$ correlation}",
    eprint = "2601.03457",
    archivePrefix = "arXiv",
    primaryClass = "astro-ph.HE",
    month = "1",
    year = "2026",
    journal = "",
}

@article{Farag:2022jcc,
    author = "Farag, Ebraheem and Renzo, Mathieu and Farmer, Robert and Chidester, Morgan T. and Timmes, F. X.",
    title = "{Resolving the Peak of the Black Hole Mass Spectrum}",
    eprint = "2208.09624",
    archivePrefix = "arXiv",
    primaryClass = "astro-ph.HE",
    doi = "10.3847/1538-4357/ac8b83",
    journal = "Astrophys. J.",
    volume = "937",
    number = "2",
    pages = "112",
    year = "2022"
}

@article{Ye:2024ypm,
    author = "Ye, Claire S. and Fishbach, Maya",
    title = "{The Redshift Evolution of the Binary Black Hole Mass Distribution from Dense Star Clusters}",
    eprint = "2402.12444",
    archivePrefix = "arXiv",
    primaryClass = "astro-ph.HE",
    doi = "10.3847/1538-4357/ad3ba8",
    journal = "Astrophys. J.",
    volume = "967",
    number = "1",
    pages = "62",
    year = "2024"
}

@article{Mai:2025jmk,
    author = "Mai, Aidan and Kremer, Kyle and K{\i}ro{\u{g}}lu, Fulya",
    title = "{Shadows of the Colossus: Hierarchical Black Hole Mergers in a 10-million-body Globular Cluster Simulation}",
    eprint = "2510.21916",
    archivePrefix = "arXiv",
    primaryClass = "astro-ph.GA",
    doi = "10.3847/1538-4357/ae2de5",
    journal = "Astrophys. J.",
    volume = "998",
    number = "1",
    pages = "138",
    year = "2026"
}

@article{Zevin:2020gbd,
    author = "Zevin, Michael and Bavera, Simone S. and Berry, Christopher P. L. and Kalogera, Vicky and Fragos, Tassos and Marchant, Pablo and Rodriguez, Carl L. and Antonini, Fabio and Holz, Daniel E. and Pankow, Chris",
    title = "{One Channel to Rule Them All? Constraining the Origins of Binary Black Holes Using Multiple Formation Pathways}",
    eprint = "2011.10057",
    archivePrefix = "arXiv",
    primaryClass = "astro-ph.HE",
    doi = "10.3847/1538-4357/abe40e",
    journal = "Astrophys. J.",
    volume = "910",
    number = "2",
    pages = "152",
    year = "2021"
}

@article{Colloms:2025hib,
    author = "Colloms, Storm and Berry, Christopher P. L. and Veitch, John and Zevin, Michael",
    title = "{Exploring the Evolution of Gravitational-wave Emitters with Efficient Emulation: Constraining the Origins of Binary Black Holes Using Normalizing Flows}",
    eprint = "2503.03819",
    archivePrefix = "arXiv",
    primaryClass = "astro-ph.HE",
    doi = "10.3847/1538-4357/ade546",
    journal = "Astrophys. J.",
    volume = "988",
    number = "2",
    pages = "189",
    year = "2025"
}

@article{Cheng:2023ddt,
    author = "Cheng, April Qiu and Zevin, Michael and Vitale, Salvatore",
    title = "{What You Don{\textquoteright}t Know Can Hurt You: Use and Abuse of Astrophysical Models in Gravitational-wave Population Analyses}",
    eprint = "2307.03129",
    archivePrefix = "arXiv",
    primaryClass = "astro-ph.HE",
    reportNumber = "LIGO DCC: P2300200",
    doi = "10.3847/1538-4357/aced98",
    journal = "Astrophys. J.",
    volume = "955",
    number = "2",
    pages = "127",
    year = "2023"
}

@article{Bavera:2020uch,
    author = "Bavera, Simone S. and others",
    title = "{The impact of mass-transfer physics on the observable properties of field binary black hole populations}",
    eprint = "2010.16333",
    archivePrefix = "arXiv",
    primaryClass = "astro-ph.HE",
    doi = "10.1051/0004-6361/202039804",
    journal = "Astron. Astrophys.",
    volume = "647",
    pages = "A153",
    year = "2021"
}

@article{Fishbach:2019ckx,
    author = "Fishbach, Maya and Farr, Will M. and Holz, Daniel E.",
    title = "{The Most Massive Binary Black Hole Detections and the Identification of Population Outliers}",
    eprint = "1911.05882",
    archivePrefix = "arXiv",
    primaryClass = "astro-ph.HE",
    doi = "10.3847/2041-8213/ab77c9",
    journal = "Astrophys. J. Lett.",
    volume = "891",
    number = "2",
    pages = "L31",
    year = "2020"
}

@article{Callister:2023tgi,
    author = "Callister, Thomas A. and Farr, Will M.",
    title = "{Parameter-Free Tour of the Binary Black Hole Population}",
    eprint = "2302.07289",
    archivePrefix = "arXiv",
    primaryClass = "astro-ph.HE",
    doi = "10.1103/PhysRevX.14.021005",
    journal = "Phys. Rev. X",
    volume = "14",
    number = "2",
    pages = "021005",
    year = "2024"
}

@article{Miller:2024sui,
    author = "Miller, Simona J. and Ko, Zoe and Callister, Tom and Chatziioannou, Katerina",
    title = "{Gravitational waves carry information beyond effective spin parameters but it is hard to extract}",
    eprint = "2401.05613",
    archivePrefix = "arXiv",
    primaryClass = "gr-qc",
    reportNumber = "LIGO-P2300453",
    doi = "10.1103/PhysRevD.109.104036",
    journal = "Phys. Rev. D",
    volume = "109",
    number = "10",
    pages = "104036",
    year = "2024"
}

@article{LIGOScientific:2025snk,
    author = "Abac, A. G. and others",
    collaboration = "LIGO Scientific, VIRGO, KAGRA",
    title = "{Open Data from LIGO, Virgo, and KAGRA through the First Part of the Fourth Observing Run}",
    eprint = "2508.18079",
    archivePrefix = "arXiv",
    primaryClass = "gr-qc",
    reportNumber = "LIGO-P2500167",
    month = "8",
    year = "2025",
    journal = "",
}

@article{KAGRA:2023pio,
    author = "Abbott, R. and others",
    collaboration = "KAGRA, VIRGO, LIGO Scientific",
    title = "{Open Data from the Third Observing Run of LIGO, Virgo, KAGRA, and GEO}",
    eprint = "2302.03676",
    archivePrefix = "arXiv",
    primaryClass = "gr-qc",
    reportNumber = "LIGO-P2200316",
    doi = "10.3847/1538-4365/acdc9f",
    journal = "Astrophys. J. Suppl.",
    volume = "267",
    number = "2",
    pages = "29",
    year = "2023"
}

@article{LIGOScientific:2019lzm,
    author = "Abbott, Rich and others",
    collaboration = "LIGO Scientific, Virgo",
    title = "{Open data from the first and second observing runs of Advanced LIGO and Advanced Virgo}",
    eprint = "1912.11716",
    archivePrefix = "arXiv",
    primaryClass = "gr-qc",
    reportNumber = "LIGO-P1900206",
    doi = "10.1016/j.softx.2021.100658",
    journal = "SoftwareX",
    volume = "13",
    pages = "100658",
    year = "2021"
}

@article{Zevin:2022wrw,
    author = "Zevin, Michael and Bavera, Simone S.",
    title = "{Suspicious Siblings: The Distribution of Mass and Spin across Component Black Holes in Isolated Binary Evolution}",
    eprint = "2203.02515",
    archivePrefix = "arXiv",
    primaryClass = "astro-ph.HE",
    doi = "10.3847/1538-4357/ac6f5d",
    journal = "Astrophys. J.",
    volume = "933",
    number = "1",
    pages = "86",
    year = "2022"
}

@ARTICLE{1999NuPhA.656....3A,
       author = {{Angulo}, C. and {Arnould}, M. and {Rayet}, M. and {Descouvemont}, P. and {Baye}, D. and {Leclercq-Willain}, C. and {Coc}, A. and {Barhoumi}, S. and {Aguer}, P. and {Rolfs}, C. and {Kunz}, R. and {Hammer}, J.~W. and {Mayer}, A. and {Paradellis}, T. and {Kossionides}, S. and {Chronidou}, C. and {Spyrou}, K. and {degl'Innocenti}, S. and {Fiorentini}, G. and {Ricci}, B. and {Zavatarelli}, S. and {Providencia}, C. and {Wolters}, H. and {Soares}, J. and {Grama}, C. and {Rahighi}, J. and {Shotter}, A. and {Lamehi Rachti}, M.},
        title = "{A compilation of charged-particle induced thermonuclear reaction rates}",
      journal = {nphysa},
         year = 1999,
        month = aug,
       volume = {656},
       number = {1},
        pages = {3-183},
          doi = {10.1016/S0375-9474(99)00030-5},
       adsurl = {https://ui.adsabs.harvard.edu/abs/1999NuPhA.656....3A}
}

@article{LIGOScientific:2025pvj,
    author = "Abac, A. G. and others",
    collaboration = "LIGO Scientific, VIRGO, KAGRA",
    title = "{GWTC-4.0: Population Properties of Merging Compact Binaries}",
    eprint = "2508.18083",
    archivePrefix = "arXiv",
    primaryClass = "astro-ph.HE",
    reportNumber = "LIGO-P2400004",
    month = "8",
    year = "2025",
    journal = "",
}

@article{Madau:2014bja,
    author = "Madau, Piero and Dickinson, Mark",
    title = "{Cosmic Star Formation History}",
    eprint = "1403.0007",
    archivePrefix = "arXiv",
    primaryClass = "astro-ph.CO",
    doi = "10.1146/annurev-astro-081811-125615",
    journal = "Ann. Rev. Astron. Astrophys.",
    volume = "52",
    pages = "415--486",
    year = "2014"
}

@article{Heinzel:2024hva,
    author = "Heinzel, Jack and Mould, Matthew and Vitale, Salvatore",
    title = "{Nonparametric analysis of correlations in the binary black hole population with LIGO-Virgo-KAGRA data}",
    eprint = "2406.16844",
    archivePrefix = "arXiv",
    primaryClass = "astro-ph.HE",
    doi = "10.1103/PhysRevD.111.L061305",
    journal = "Phys. Rev. D",
    volume = "111",
    number = "6",
    pages = "L061305",
    year = "2025"
}

@article{Ray:2023upk,
    author = "Ray, Anarya and Maga{\~n}a Hernandez, Ignacio and Mohite, Siddharth and Creighton, Jolien and Kapadia, Shasvath",
    title = "{Nonparametric Inference of the Population of Compact Binaries from Gravitational-wave Observations Using Binned Gaussian Processes}",
    eprint = "2304.08046",
    archivePrefix = "arXiv",
    primaryClass = "gr-qc",
    reportNumber = "LIGO-P2300098",
    doi = "10.3847/1538-4357/acf452",
    journal = "Astrophys. J.",
    volume = "957",
    number = "1",
    pages = "37",
    year = "2023"
}

@article{LIGOScientific:2025slb,
    author = "Abac, A. G. and others",
    collaboration = "LIGO Scientific, VIRGO, KAGRA",
    title = "{GWTC-4.0: Updating the Gravitational-Wave Transient Catalog with Observations from the First Part of the Fourth LIGO-Virgo-KAGRA Observing Run}",
    eprint = "2508.18082",
    archivePrefix = "arXiv",
    primaryClass = "gr-qc",
    reportNumber = "LIGO-P2400386",
    month = "8",
    year = "2025",
    journal = "",
}

@article{LIGOScientific:2014pky,
    author = "Aasi, J. and others",
    collaboration = "LIGO Scientific",
    title = "{Advanced LIGO}",
    eprint = "1411.4547",
    archivePrefix = "arXiv",
    primaryClass = "gr-qc",
    doi = "10.1088/0264-9381/32/7/074001",
    journal = "Class. Quant. Grav.",
    volume = "32",
    pages = "074001",
    year = "2015"
}

@article{VIRGO:2014yos,
    author = "Acernese, F. and others",
    collaboration = "Virgo",
    title = "{Advanced Virgo: a second-generation interferometric gravitational wave detector}",
    eprint = "1408.3978",
    archivePrefix = "arXiv",
    primaryClass = "gr-qc",
    doi = "10.1088/0264-9381/32/2/024001",
    journal = "Class. Quant. Grav.",
    volume = "32",
    number = "2",
    pages = "024001",
    year = "2015"
}

@article{KAGRA:2020agh,
    author = "Akutsu, T. and others",
    collaboration = "KAGRA",
    title = "{Overview of KAGRA: Calibration, detector characterization, physical environmental monitors, and the geophysics interferometer}",
    eprint = "2009.09305",
    archivePrefix = "arXiv",
    primaryClass = "gr-qc",
    doi = "10.1093/ptep/ptab018",
    journal = "PTEP",
    volume = "2021",
    number = "5",
    pages = "05A102",
    year = "2021"
}

@article{Wang:2025nhf,
    author = "Wang, Yuan-Zhu and Li, Yin-Jie and Gao, Shi-Jie and Tang, Shao-Peng and Fan, Yi-Zhong",
    title = "{A new group of low-spin $50-70M_\odot$ Black Holes and the high pair-instability mass cutoff}",
    eprint = "2510.22698",
    archivePrefix = "arXiv",
    primaryClass = "astro-ph.HE",
    month = "10",
    year = "2025",
    journal = "",
}

@article{Mandel:2021smh,
    author = "Mandel, Ilya and Broekgaarden, Floor S.",
    title = "{Rates of compact object coalescences}",
    eprint = "2107.14239",
    archivePrefix = "arXiv",
    primaryClass = "astro-ph.HE",
    doi = "10.1007/s41114-021-00034-3",
    journal = "Living Rev. Rel.",
    volume = "25",
    number = "1",
    pages = "1",
    year = "2022"
}

@article{Breivik:2025edm,
    author = "Breivik, Katelyn",
    title = "{Population Synthesis of Gravitational Wave Sources}",
    eprint = "2502.03523",
    archivePrefix = "arXiv",
    primaryClass = "astro-ph.HE",
    doi = "10.1016/B978-0-443-21439-4.00115-2",
    month = "2",
    year = "2025",
    journal = "",
}

@article{Mandel:2018hfr,
    author = "Mandel, Ilya and Farmer, Alison",
    title = "{Merging stellar-mass binary black holes}",
    eprint = "1806.05820",
    archivePrefix = "arXiv",
    primaryClass = "astro-ph.HE",
    doi = "10.1016/j.physrep.2022.01.003",
    journal = "Phys. Rept.",
    volume = "955",
    pages = "1--24",
    year = "2022"
}

@inbook{Mapelli:2021taw,
    author = "Mapelli, Michela",
    title = "{Formation Channels of Single and Binary Stellar-Mass Black Holes}",
    eprint = "2106.00699",
    archivePrefix = "arXiv",
    primaryClass = "astro-ph.HE",
    doi = "10.1007/978-981-15-4702-7_16-1",
    year = "2021",
    journal = "",
}

@article{Fuller:2019ckz,
    author = "Fuller, Jim and Piro, Anthony L. and Jermyn, Adam S.",
    title = "{Slowing the spins of stellar cores}",
    eprint = "1902.08227",
    archivePrefix = "arXiv",
    doi = "10.1093/mnras/stz514",
    journal = "Mon. Not. Roy. Astron. Soc.",
    volume = "485",
    number = "3",
    pages = "3661--3680",
    year = "2019"
}

@article{Ma:2019cpr,
    author = "Ma, Linhao and Fuller, Jim",
    title = "{Angular momentum transport in massive stars and natal neutron star rotation rates}",
    eprint = "1907.03713",
    archivePrefix = "arXiv",
    primaryClass = "astro-ph.SR",
    doi = "10.1093/mnras/stz2009",
    journal = "Mon. Not. Roy. Astron. Soc.",
    volume = "488",
    number = "3",
    pages = "4338--4355",
    year = "2019"
}

@article{Fuller:2019sxi,
    author = "Fuller, Jim and Ma, Linhao",
    title = "{Most Black Holes are Born Very Slowly Rotating}",
    eprint = "1907.03714",
    archivePrefix = "arXiv",
    primaryClass = "astro-ph.SR",
    doi = "10.3847/2041-8213/ab339b",
    journal = "Astrophys. J. Lett.",
    volume = "881",
    number = "1",
    pages = "L1",
    year = "2019"
}

@article{Woosley:2007qp,
    author = "Woosley, Stanford Earl and Blinnikov, S. and Heger, Alexander",
    title = "{Pulsational pair instability as an explanation for the most luminous supernovae}",
    eprint = "0710.3314",
    archivePrefix = "arXiv",
    primaryClass = "astro-ph",
    reportNumber = "LAUR-07-3795",
    doi = "10.1038/nature06333",
    journal = "Nature",
    volume = "450",
    pages = "390",
    year = "2007"
}

@article{Heger:2001cd,
    author = "Heger, A. and Woosley, S. E.",
    title = "{The nucleosynthetic signature of population III}",
    eprint = "astro-ph/0107037",
    archivePrefix = "arXiv",
    doi = "10.1086/338487",
    journal = "Astrophys. J.",
    volume = "567",
    pages = "532--543",
    year = "2002"
}

@article{Belczynski:2016jno,
    author = "Belczynski, K. and others",
    title = "{The Effect of Pair-Instability Mass Loss on Black Hole Mergers}",
    eprint = "1607.03116",
    archivePrefix = "arXiv",
    primaryClass = "astro-ph.HE",
    doi = "10.1051/0004-6361/201628980",
    journal = "Astron. Astrophys.",
    volume = "594",
    pages = "A97",
    year = "2016"
}

@article{Farmer:2020xne,
    author = "Farmer, Robert and Renzo, Mathieu and de Mink, Selma and Fishbach, Maya and Justham, Stephen",
    title = "{Constraints from gravitational wave detections of binary black hole mergers on the $^{12}\rm{C}\left(\alpha,\gamma\right)^{16}\!\rm{O}$ rate}",
    eprint = "2006.06678",
    archivePrefix = "arXiv",
    primaryClass = "astro-ph.HE",
    doi = "10.3847/2041-8213/abbadd",
    journal = "Astrophys. J. Lett.",
    volume = "902",
    number = "2",
    pages = "L36",
    year = "2020"
}

@article{vanSon:2020zbk,
    author = "van Son, L. A. C. and de Mink, S. E. and Broekgaarden, F. S. and Renzo, M. and Justham, S. and Laplace, E. and Moran-Fraile, J. and Hendriks, D. D. and Farmer, R.",
    title = "{Polluting the pair-instability mass gap for binary black holes through super-Eddington accretion in isolated binaries}",
    eprint = "2004.05187",
    archivePrefix = "arXiv",
    primaryClass = "astro-ph.HE",
    doi = "10.3847/1538-4357/ab9809",
    journal = "Astrophys. J.",
    volume = "897",
    number = "1",
    pages = "100",
    year = "2020"
}

@article{Ziegler:2020klg,
    author = "Ziegler, Joshua and Freese, Katherine",
    title = "{Filling the black hole mass gap: Avoiding pair instability in massive stars through addition of nonnuclear energy}",
    eprint = "2010.00254",
    archivePrefix = "arXiv",
    primaryClass = "astro-ph.HE",
    reportNumber = "UTTG-12-2020, NORDITA 2020-088",
    doi = "10.1103/PhysRevD.104.043015",
    journal = "Phys. Rev. D",
    volume = "104",
    number = "4",
    pages = "043015",
    year = "2021"
}

@article{Spera:2017fyx,
    author = "Spera, Mario and Mapelli, Michela",
    title = "{Very massive stars, pair-instability supernovae and intermediate-mass black holes with the SEVN code}",
    eprint = "1706.06109",
    archivePrefix = "arXiv",
    primaryClass = "astro-ph.SR",
    doi = "10.1093/mnras/stx1576",
    journal = "Mon. Not. Roy. Astron. Soc.",
    volume = "470",
    number = "4",
    pages = "4739--4749",
    year = "2017"
}

@article{Hendriks:2023yrw,
    author = "Hendriks, D. D. and van Son, L. A. C. and Renzo, M. and Izzard, R. G. and Farmer, R.",
    title = "{Pulsational pair-instability supernovae in gravitational-wave and electromagnetic transients}",
    eprint = "2309.09339",
    archivePrefix = "arXiv",
    primaryClass = "astro-ph.HE",
    doi = "10.1093/mnras/stad2857",
    journal = "Mon. Not. Roy. Astron. Soc.",
    volume = "526",
    number = "3",
    pages = "4130--4147",
    year = "2023"
}

@article{Farmer:2019jed,
    author = "Farmer, R. and Renzo, M. and de Mink, S. E. and Marchant, P. and Justham, S.",
    title = "{Mind the gap: The location of the lower edge of the pair instability supernovae black hole mass gap}",
    eprint = "1910.12874",
    archivePrefix = "arXiv",
    primaryClass = "astro-ph.SR",
    doi = "10.3847/1538-4357/ab518b",
    month = "10",
    year = "2019",
    journal = "",
}

@article{Berti:2008af,
    author = "Berti, Emanuele and Volonteri, Marta",
    title = "{Cosmological black hole spin evolution by mergers and accretion}",
    eprint = "0802.0025",
    archivePrefix = "arXiv",
    primaryClass = "astro-ph",
    doi = "10.1086/590379",
    journal = "Astrophys. J.",
    volume = "684",
    pages = "822--828",
    year = "2008"
}

@article{Hofmann:2016yih,
    author = "Hofmann, Fabian and Barausse, Enrico and Rezzolla, Luciano",
    title = "{The final spin from binary black holes in quasi-circular orbits}",
    eprint = "1605.01938",
    archivePrefix = "arXiv",
    primaryClass = "gr-qc",
    doi = "10.3847/2041-8205/825/2/L19",
    journal = "Astrophys. J. Lett.",
    volume = "825",
    number = "2",
    pages = "L19",
    year = "2016"
}

@article{Rodriguez:2019huv,
    author = "Rodriguez, Carl L. and Zevin, Michael and Amaro-Seoane, Pau and Chatterjee, Sourav and Kremer, Kyle and Rasio, Frederic A. and Ye, Claire S.",
    title = "{Black holes: The next generation{\textemdash}repeated mergers in dense star clusters and their gravitational-wave properties}",
    eprint = "1906.10260",
    archivePrefix = "arXiv",
    primaryClass = "astro-ph.HE",
    doi = "10.1103/PhysRevD.100.043027",
    journal = "Phys. Rev. D",
    volume = "100",
    number = "4",
    pages = "043027",
    year = "2019"
}

@article{Borchers:2025sid,
    author = "Borchers, Angela and Ye, Claire S. and Fishbach, Maya",
    title = "{Gravitational-wave Kicks Impact the Spins of Black Holes from Hierarchical Mergers}",
    eprint = "2503.21278",
    archivePrefix = "arXiv",
    primaryClass = "astro-ph.HE",
    doi = "10.3847/1538-4357/addec6",
    journal = "Astrophys. J.",
    volume = "987",
    number = "2",
    year = "2025"
}

@article{PortegiesZwart:1999nm,
    author = "Portegies Zwart, Simon F. and McMillan, Stephen",
    title = "{Black hole mergers in the universe}",
    eprint = "astro-ph/9910061",
    archivePrefix = "arXiv",
    doi = "10.1086/312422",
    journal = "Astrophys. J. Lett.",
    volume = "528",
    pages = "L17",
    year = "2000"
}

@article{Fitchett:1983qzq,
    author = "Fitchett, M. J.",
    title = "{The influence of gravitational wave momentum losses on the centre of mass motion of a Newtonian binary system}",
    doi = "10.1093/mnras/203.4.1049",
    journal = "Mon. Not. Roy. Astron. Soc.",
    volume = "203",
    number = "4",
    pages = "1049--1062",
    year = "1983"
}

@article{Favata:2004wz,
    author = "Favata, Marc and Hughes, Scott A. and Holz, Daniel E.",
    title = "{How black holes get their kicks: Gravitational radiation recoil revisited}",
    eprint = "astro-ph/0402056",
    archivePrefix = "arXiv",
    doi = "10.1086/421552",
    journal = "Astrophys. J. Lett.",
    volume = "607",
    pages = "L5--L8",
    year = "2004"
}

@article{Gonzalez:2006md,
    author = "Gonzalez, Jose A. and Sperhake, Ulrich and Bruegmann, Bernd and Hannam, Mark and Husa, Sascha",
    title = "{Total recoil: The Maximum kick from nonspinning black-hole binary inspiral}",
    eprint = "gr-qc/0610154",
    archivePrefix = "arXiv",
    doi = "10.1103/PhysRevLett.98.091101",
    journal = "Phys. Rev. Lett.",
    volume = "98",
    pages = "091101",
    year = "2007"
}

@article{Lousto:2009ka,
    author = "Lousto, Carlos O. and Nakano, Hiroyuki and Zlochower, Yosef and Campanelli, Manuela",
    title = "{Statistical studies of Spinning Black-Hole Binaries}",
    eprint = "0910.3197",
    archivePrefix = "arXiv",
    primaryClass = "gr-qc",
    doi = "10.1103/PhysRevD.81.084023",
    journal = "Phys. Rev. D",
    volume = "81",
    pages = "084023",
    year = "2010",
    note = "[Erratum: Phys.Rev.D 82, 129902 (2010)]"
}

@article{Gerosa:2018qay,
    author = "Gerosa, Davide and H{\'e}bert, Fran{\c{c}}ois and Stein, Leo C.",
    title = "{Black-hole kicks from numerical-relativity surrogate models}",
    eprint = "1802.04276",
    archivePrefix = "arXiv",
    primaryClass = "gr-qc",
    doi = "10.1103/PhysRevD.97.104049",
    journal = "Phys. Rev. D",
    volume = "97",
    number = "10",
    pages = "104049",
    year = "2018"
}

@article{Mahapatra:2021hme,
    author = "Mahapatra, Parthapratim and Gupta, Anuradha and Favata, Marc and Arun, K. G. and Sathyaprakash, B. S.",
    title = "{Remnant Black Hole Kicks and Implications for Hierarchical Mergers}",
    eprint = "2106.07179",
    archivePrefix = "arXiv",
    primaryClass = "astro-ph.HE",
    doi = "10.3847/2041-8213/ac20db",
    journal = "Astrophys. J. Lett.",
    volume = "918",
    number = "2",
    pages = "L31",
    year = "2021"
}

@article{Fishbach:2017dwv,
    author = "Fishbach, Maya and Holz, Daniel E. and Farr, Ben",
    title = "{Are LIGO's Black Holes Made From Smaller Black Holes?}",
    eprint = "1703.06869",
    archivePrefix = "arXiv",
    primaryClass = "astro-ph.HE",
    doi = "10.3847/2041-8213/aa7045",
    journal = "Astrophys. J. Lett.",
    volume = "840",
    number = "2",
    pages = "L24",
    year = "2017"
}

@article{Gerosa:2017kvu,
    author = "Gerosa, Davide and Berti, Emanuele",
    title = "{Are merging black holes born from stellar collapse or previous mergers?}",
    eprint = "1703.06223",
    archivePrefix = "arXiv",
    primaryClass = "gr-qc",
    doi = "10.1103/PhysRevD.95.124046",
    journal = "Phys. Rev. D",
    volume = "95",
    number = "12",
    pages = "124046",
    year = "2017"
}

@article{Kimball:2020opk,
    author = "Kimball, Chase and Talbot, Colm and L. Berry, Christopher P. and Carney, Matthew and Zevin, Michael and Thrane, Eric and Kalogera, Vicky",
    title = "{Black Hole Genealogy: Identifying Hierarchical Mergers with Gravitational Waves}",
    eprint = "2005.00023",
    archivePrefix = "arXiv",
    primaryClass = "astro-ph.HE",
    reportNumber = "DCC P2000131",
    doi = "10.3847/1538-4357/aba518",
    journal = "Astrophys. J.",
    volume = "900",
    number = "2",
    pages = "177",
    year = "2020"
}

@article{Sadiq:2023zee,
    author = "Sadiq, Jam and Dent, Thomas and Gieles, Mark",
    title = "{Binary Vision: The Mass Distribution of Merging Binary Black Holes via Iterative Density Estimation}",
    eprint = "2307.12092",
    archivePrefix = "arXiv",
    primaryClass = "astro-ph.HE",
    doi = "10.3847/1538-4357/ad0ce6",
    journal = "Astrophys. J.",
    volume = "960",
    number = "1",
    pages = "65",
    year = "2024"
}

@article{Afroz:2024fzp,
    author = "Afroz, Samsuzzaman and Mukherjee, Suvodip",
    title = "{Phase space of binary black holes from gravitational wave observations to unveil its formation history}",
    eprint = "2411.07304",
    archivePrefix = "arXiv",
    primaryClass = "astro-ph.HE",
    doi = "10.1103/7zc2-g9vq",
    journal = "Phys. Rev. D",
    volume = "112",
    number = "2",
    pages = "023531",
    year = "2025"
}

@article{Afroz:2025ikg,
    author = "Afroz, Samsuzzaman and Mukherjee, Suvodip",
    title = "{Binary Black Hole Phase Space Discovers the Signature of Pair Instability Supernovae Mass Gap}",
    eprint = "2509.09123",
    archivePrefix = "arXiv",
    primaryClass = "astro-ph.HE",
    month = "9",
    year = "2025",
    journal = "",
}

@article{Mould:2022ccw,
    author = "Mould, Matthew and Gerosa, Davide and Taylor, Stephen R.",
    title = "{Deep learning and Bayesian inference of gravitational-wave populations: Hierarchical black-hole mergers}",
    eprint = "2203.03651",
    archivePrefix = "arXiv",
    primaryClass = "astro-ph.HE",
    doi = "10.1103/PhysRevD.106.103013",
    journal = "Phys. Rev. D",
    volume = "106",
    number = "10",
    pages = "103013",
    year = "2022"
}

@article{Kimball:2020qyd,
    author = "Kimball, Chase and others",
    title = "{Evidence for Hierarchical Black Hole Mergers in the Second LIGO{\textendash}Virgo Gravitational Wave Catalog}",
    eprint = "2011.05332",
    archivePrefix = "arXiv",
    primaryClass = "astro-ph.HE",
    reportNumber = "DCC P2000466",
    doi = "10.3847/2041-8213/ac0aef",
    journal = "Astrophys. J. Lett.",
    volume = "915",
    number = "2",
    pages = "L35",
    year = "2021"
}

@article{Fishbach:2022lzq,
    author = "Fishbach, Maya and Kimball, Chase and Kalogera, Vicky",
    title = "{Limits on Hierarchical Black Hole Mergers from the Most Negative {\ensuremath{\chi}} $_{eff}$ Systems}",
    eprint = "2207.02924",
    archivePrefix = "arXiv",
    primaryClass = "astro-ph.HE",
    reportNumber = "LIGO-P2200201",
    doi = "10.3847/2041-8213/ac86c4",
    journal = "Astrophys. J. Lett.",
    volume = "935",
    number = "2",
    pages = "L26",
    year = "2022"
}

@article{Vitale:2025lms,
    author = "Vitale, Salvatore and Mould, Matthew",
    collaboration = "Society of Physicists Interested in Non-aligned Spins, SPINS, (Society of Physicists Interested in Non-aligned Spins, SPINS){\textdagger}",
    title = "{Long road to alignment: Measuring black hole spin orientation with expanding gravitational-wave datasets}",
    eprint = "2505.14875",
    archivePrefix = "arXiv",
    primaryClass = "astro-ph.HE",
    reportNumber = "INT-PUB-25-016",
    doi = "10.1103/drsl-n3wz",
    journal = "Phys. Rev. D",
    volume = "112",
    number = "8",
    pages = "083015",
    year = "2025"
}

@article{Stegmann:2025zkb,
    author = "Stegmann, Jakob and Antonini, Fabio and Olejak, Aleksandra and Biscoveanu, Sylvia and Raymond, Vivien and Rinaldi, Stefano and Flanagan, Beth",
    title = "{In-plane Black-hole Spin Measurements Suggest Most Gravitational-wave Mergers Form in Triples}",
    eprint = "2512.15873",
    archivePrefix = "arXiv",
    primaryClass = "astro-ph.HE",
    month = "12",
    year = "2025",
    journal = "",
}

@article{Plunkett:2026pxt,
    author = "Plunkett, Cailin and Callister, Thomas and Zevin, Michael and Vitale, Salvatore",
    title = "{Signatures of a subpopulation of hierarchical mergers in the GWTC-4 gravitational-wave dataset}",
    eprint = "2601.07908",
    archivePrefix = "arXiv",
    primaryClass = "gr-qc",
    month = "1",
    year = "2026",
    journal = "",
}

@article{Tong:2025xir,
    author = "Tong, Hui and Callister, Thomas A. and Fishbach, Maya and Thrane, Eric and Antonini, Fabio and Stevenson, Simon and Romero-Shaw, Isobel M. and Dosopoulou, Fani",
    title = "{A subpopulation of low-mass, spinning black holes: signatures of dynamical assembly}",
    eprint = "2511.05316",
    archivePrefix = "arXiv",
    primaryClass = "astro-ph.HE",
    month = "11",
    year = "2025",
    journal = "",
}

@article{OLeary:2005vqo,
    author = "O'Leary, Ryan M. and Rasio, Frederic A. and Fregeau, John M. and Ivanova, Natalia and O'Shaughnessy, Richard W.",
    title = "{Binary mergers and growth of black holes in dense star clusters}",
    eprint = "astro-ph/0508224",
    archivePrefix = "arXiv",
    doi = "10.1086/498446",
    journal = "Astrophys. J.",
    volume = "637",
    pages = "937--951",
    year = "2006"
}

@article{Antonini:2016gqe,
    author = "Antonini, Fabio and Rasio, Frederic A.",
    title = "{Merging black hole binaries in galactic nuclei: implications for advanced-LIGO detections}",
    eprint = "1606.04889",
    archivePrefix = "arXiv",
    primaryClass = "astro-ph.HE",
    doi = "10.3847/0004-637X/831/2/187",
    journal = "Astrophys. J.",
    volume = "831",
    number = "2",
    pages = "187",
    year = "2016"
}

@article{Tagawa:2020qll,
    author = "Tagawa, Hiromichi and Kocsis, Bence and Haiman, Zoltan and Bartos, Imre and Omukai, Kazuyuki and Samsing, Johan",
    title = "{Mass-gap Mergers in Active Galactic Nuclei}",
    eprint = "2012.00011",
    archivePrefix = "arXiv",
    primaryClass = "astro-ph.HE",
    doi = "10.3847/1538-4357/abd555",
    journal = "Astrophys. J.",
    volume = "908",
    number = "2",
    pages = "194",
    year = "2021"
}

@article{Mapelli:2021syv,
    author = "Mapelli, Michela and others",
    title = "{Hierarchical black hole mergers in young, globular and nuclear star clusters: the effect of metallicity, spin and cluster properties}",
    eprint = "2103.05016",
    archivePrefix = "arXiv",
    primaryClass = "astro-ph.HE",
    doi = "10.1093/mnras/stab1334",
    journal = "Mon. Not. Roy. Astron. Soc.",
    volume = "505",
    number = "1",
    pages = "339--358",
    year = "2021"
}

@article{Antonini:2022vib,
    author = "Antonini, Fabio and Gieles, Mark and Dosopoulou, Fani and Chattopadhyay, Debatri",
    title = "{Coalescing black hole binaries from globular clusters: mass distributions and comparison to gravitational wave data from GWTC-3}",
    eprint = "2208.01081",
    archivePrefix = "arXiv",
    primaryClass = "astro-ph.HE",
    doi = "10.1093/mnras/stad972",
    journal = "Mon. Not. Roy. Astron. Soc.",
    volume = "522",
    number = "1",
    pages = "466--476",
    year = "2023"
}

@article{Torniamenti:2024uxl,
    author = "Torniamenti, Stefano and Mapelli, Michela and P{\'e}rigois, Carole and Sedda, Manuel Arca and Artale, Maria Celeste and Dall'Amico, Marco and Vaccaro, Maria Paola",
    title = "{Hierarchical binary black hole mergers in globular clusters: Mass function and evolution with redshift}",
    eprint = "2401.14837",
    archivePrefix = "arXiv",
    primaryClass = "astro-ph.HE",
    doi = "10.1051/0004-6361/202449272",
    journal = "Astron. Astrophys.",
    volume = "688",
    pages = "A148",
    year = "2024"
}

@article{Vaccaro:2025ogk,
    author = "Vaccaro, Maria Paola",
    title = "{Hierarchical Black Hole Mergers in AGN Disks: Tracing Massive Black Hole Growth Across Cosmic Time}",
    eprint = "2508.15337",
    archivePrefix = "arXiv",
    primaryClass = "astro-ph.HE",
    month = "8",
    year = "2025",
    journal = "",
}

@article{Kiroglu:2025bbp,
    author = "K{\i}ro{\u{g}}lu, Fulya and Lombardi, James C. and Kremer, Kyle and Vanderzyden, Hans D. and Rasio, Frederic A.",
    title = "{Spin{\textendash}Orbit Alignment in Merging Binary Black Holes Following Collisions with Massive Stars}",
    eprint = "2501.09068",
    archivePrefix = "arXiv",
    primaryClass = "astro-ph.HE",
    doi = "10.3847/2041-8213/adc263",
    journal = "Astrophys. J. Lett.",
    volume = "983",
    number = "1",
    pages = "L9",
    year = "2025"
}

@article{Kiroglu:2024xpc,
    author = "K{\i}ro{\u{g}}lu, Fulya and Kremer, Kyle and Biscoveanu, Sylvia and Prieto, Elena Gonz{\'a}lez and Rasio, Frederic A.",
    title = "{Black Hole Accretion and Spin-up through Stellar Collisions in Dense Star Clusters}",
    eprint = "2410.01879",
    archivePrefix = "arXiv",
    primaryClass = "astro-ph.HE",
    doi = "10.3847/1538-4357/ada26b",
    journal = "Astrophys. J.",
    volume = "979",
    number = "2",
    pages = "237",
    year = "2025"
}

@article{Kalogera:1999tq,
    author = "Kalogera, Vassiliki",
    title = "{Spin orbit misalignment in close binaries with two compact objects}",
    eprint = "astro-ph/9911417",
    archivePrefix = "arXiv",
    doi = "10.1086/309400",
    journal = "Astrophys. J.",
    volume = "541",
    pages = "319--328",
    year = "2000"
}

@article{Steinle:2022rhj,
    author = "Steinle, Nathan and Kesden, Michael",
    title = "{Signatures of spin precession and nutation in isolated black-hole binaries}",
    eprint = "2206.00391",
    archivePrefix = "arXiv",
    primaryClass = "astro-ph.HE",
    doi = "10.1103/PhysRevD.106.063028",
    journal = "Phys. Rev. D",
    volume = "106",
    number = "6",
    pages = "063028",
    year = "2022"
}

@article{Gerosa:2018wbw,
    author = "Gerosa, Davide and Berti, Emanuele and O'Shaughnessy, Richard and Belczynski, Krzysztof and Kesden, Michael and Wysocki, Daniel and Gladysz, Wojciech",
    title = "{Spin orientations of merging black holes formed from the evolution of stellar binaries}",
    eprint = "1808.02491",
    archivePrefix = "arXiv",
    primaryClass = "astro-ph.HE",
    doi = "10.1103/PhysRevD.98.084036",
    journal = "Phys. Rev. D",
    volume = "98",
    number = "8",
    pages = "084036",
    year = "2018"
}

@article{Bavera:2020inc,
    author = "Bavera, Simone S. and Fragos, Tassos and Qin, Ying and Zapartas, Emmanouil and Neijssel, Coenraad J. and Mandel, Ilya and Batta, Aldo and Gaebel, Sebastian M. and Kimball, Chase and Stevenson, Simon",
    title = "{The origin of spin in binary black holes: Predicting the distributions of the main observables of Advanced LIGO}",
    eprint = "1906.12257",
    archivePrefix = "arXiv",
    primaryClass = "astro-ph.HE",
    doi = "10.1051/0004-6361/201936204",
    journal = "Astron. Astrophys.",
    volume = "635",
    pages = "A97",
    year = "2020"
}

@article{Mapelli:2021gyv,
    author = "Mapelli, Michela and Bouffanais, Yann and Santoliquido, Filippo and Sedda, Manuel Arca and Artale, M. Celeste",
    title = "{The cosmic evolution of binary black holes in young, globular, and nuclear star clusters: rates, masses, spins, and mixing fractions}",
    eprint = "2109.06222",
    archivePrefix = "arXiv",
    primaryClass = "astro-ph.HE",
    doi = "10.1093/mnras/stac422",
    journal = "Mon. Not. Roy. Astron. Soc.",
    volume = "511",
    number = "4",
    pages = "5797--5816",
    year = "2022"
}

@article{Chattopadhyay:2023pil,
    author = "Chattopadhyay, Debatri and Stegmann, Jakob and Antonini, Fabio and Barber, Jordan and Romero-Shaw, Isobel M.",
    title = "{Double black hole mergers in nuclear star clusters: eccentricities, spins, masses, and the growth of massive seeds}",
    eprint = "2308.10884",
    archivePrefix = "arXiv",
    primaryClass = "astro-ph.HE",
    doi = "10.1093/mnras/stad3048",
    journal = "Mon. Not. Roy. Astron. Soc.",
    volume = "526",
    number = "4",
    pages = "4908--4928",
    year = "2023"
}

@article{Rodriguez_2022,
   title={Modeling Dense Star Clusters in the Milky Way and beyond with the Cluster Monte Carlo Code},
   volume={258},
   ISSN={1538-4365},
   url={http://dx.doi.org/10.3847/1538-4365/ac2edf},
   DOI={10.3847/1538-4365/ac2edf},
   number={2},
   journal={The Astrophysical Journal Supplement Series},
   publisher={American Astronomical Society},
   author={Rodriguez, Carl L. and Weatherford, Newlin C. and Coughlin, Scott C. and Amaro-Seoane, Pau and Breivik, Katelyn and Chatterjee, Sourav and Fragione, Giacomo and Kıroğlu, Fulya and Kremer, Kyle and Rui, Nicholas Z. and Ye, Claire S. and Zevin, Michael and Rasio, Frederic A.},
   year={2022},
   month=jan, pages={22} }

@article{Antonini:2017ash,
    author = "Antonini, Fabio and Toonen, Silvia and Hamers, Adrian S.",
    title = "{Binary black hole mergers from field triples: properties, rates and the impact of stellar evolution}",
    eprint = "1703.06614",
    archivePrefix = "arXiv",
    primaryClass = "astro-ph.GA",
    doi = "10.3847/1538-4357/aa6f5e",
    journal = "Astrophys. J.",
    volume = "841",
    number = "2",
    pages = "77",
    year = "2017"
}

@article{Liu:2018nrf,
    author = "Liu, Bin and Lai, Dong",
    title = "{Black Hole and Neutron Star Binary Mergers in Triple Systems: Merger Fraction and Spin{\textendash}Orbit Misalignment}",
    eprint = "1805.03202",
    archivePrefix = "arXiv",
    primaryClass = "astro-ph.HE",
    doi = "10.3847/1538-4357/aad09f",
    journal = "Astrophys. J.",
    volume = "863",
    number = "1",
    pages = "68",
    year = "2018"
}

@article{Rodriguez:2018jqu,
    author = "Rodriguez, Carl L. and Antonini, Fabio",
    title = "{A Triple Origin for the Heavy and Low-Spin Binary Black Holes Detected by LIGO/Virgo}",
    eprint = "1805.08212",
    archivePrefix = "arXiv",
    primaryClass = "astro-ph.HE",
    doi = "10.3847/1538-4357/aacea4",
    journal = "Astrophys. J.",
    volume = "863",
    number = "1",
    pages = "7",
    year = "2018"
}

@article{Briel:2022cfl,
    author = "Briel, M. M. and Stevance, H. F. and Eldridge, J. J.",
    title = "{Understanding the high-mass binary black hole population from stable mass transfer and super-Eddington accretion in bpass}",
    eprint = "2206.13842",
    archivePrefix = "arXiv",
    primaryClass = "astro-ph.HE",
    doi = "10.1093/mnras/stad399",
    journal = "Mon. Not. Roy. Astron. Soc.",
    volume = "520",
    number = "4",
    pages = "5724--5745",
    year = "2023"
}

@article{Rodriguez:2016vmx,
    author = "Rodriguez, Carl L. and Zevin, Michael and Pankow, Chris and Kalogera, Vasilliki and Rasio, Frederic A.",
    title = "{Illuminating Black Hole Binary Formation Channels with Spins in Advanced LIGO}",
    eprint = "1609.05916",
    archivePrefix = "arXiv",
    primaryClass = "astro-ph.HE",
    doi = "10.3847/2041-8205/832/1/L2",
    journal = "Astrophys. J. Lett.",
    volume = "832",
    number = "1",
    pages = "L2",
    year = "2016"
}

@article{Farr:2017uvj,
    author = "Farr, Will M. and Stevenson, Simon and Coleman Miller, M. and Mandel, Ilya and Farr, Ben and Vecchio, Alberto",
    title = "{Distinguishing Spin-Aligned and Isotropic Black Hole Populations With Gravitational Waves}",
    eprint = "1706.01385",
    archivePrefix = "arXiv",
    primaryClass = "astro-ph.HE",
    reportNumber = "LIGO-P1700067",
    doi = "10.1038/nature23453",
    journal = "Nature",
    volume = "548",
    pages = "426",
    year = "2017"
}

@article{Antonini2023,
    author = {Antonini, Fabio and Gieles, Mark and Dosopoulou, Fani and Chattopadhyay, Debatri},
    title = {Coalescing black hole binaries from globular clusters: mass distributions and comparison to gravitational wave data from GWTC-3},
    journal = {Monthly Notices of the Royal Astronomical Society},
    volume = {522},
    number = {1},
    pages = {466-476},
    year = {2023},
    month = {03},
    issn = {0035-8711},
    doi = {10.1093/mnras/stad972},
    url = {https://doi.org/10.1093/mnras/stad972},
    eprint = {https://academic.oup.com/mnras/article-pdf/522/1/466/49913333/stad972.pdf},
}

@article{Tagawa:2020dxe,
    author = "Tagawa, Hiromichi and Haiman, Zoltan and Bartos, Imre and Kocsis, Bence",
    title = "{Spin Evolution of Stellar-mass Black Hole Binaries in Active Galactic Nuclei}",
    eprint = "2004.11914",
    archivePrefix = "arXiv",
    primaryClass = "astro-ph.HE",
    doi = "10.3847/1538-4357/aba2cc",
    journal = "Astrophys. J.",
    volume = "899",
    number = "1",
    pages = "26",
    year = "2020"
}

@article{McKernan:2021nwk,
    author = "McKernan, B. and Ford, K. E. S. and Callister, T. and Farr, W. M. and O'Shaughnessy, R. and Smith, R. and Thrane, E. and Vajpeyi, A.",
    title = "{LIGO{\textendash}Virgo correlations between mass ratio and effective inspiral spin: testing the active galactic nuclei channel}",
    eprint = "2107.07551",
    archivePrefix = "arXiv",
    primaryClass = "astro-ph.HE",
    doi = "10.1093/mnras/stac1570",
    journal = "Mon. Not. Roy. Astron. Soc.",
    volume = "514",
    number = "3",
    pages = "3886--3893",
    year = "2022"
}

@article{Li:2022cul,
    author = "Li, Gongjie and Bhaskar, Hareesh Gautham and Kocsis, Bence and Lin, Douglas N. C.",
    title = "{Secular Spin{\textendash}Orbit Resonances of Black Hole Binaries in AGN Disks}",
    eprint = "2202.11739",
    archivePrefix = "arXiv",
    primaryClass = "astro-ph.HE",
    doi = "10.3847/1538-4357/acccf1",
    journal = "Astrophys. J.",
    volume = "950",
    number = "1",
    pages = "48",
    year = "2023"
}

@article{Godfrey:2023oxb,
    author = "Godfrey, Jaxen and Edelman, Bruce and Farr, Ben",
    title = "{Cosmic Cousins: Identification of a Subpopulation of Binary Black Holes Consistent with Isolated Binary Evolution}",
    eprint = "2304.01288",
    archivePrefix = "arXiv",
    primaryClass = "astro-ph.HE",
    reportNumber = "LIGO-P2300073",
    month = "4",
    year = "2023",
    journal = "",
}

@article{Mckernan:2017ssq,
    author = "Mckernan, B. and others",
    title = "{Constraining Stellar-mass Black Hole Mergers in AGN Disks Detectable with LIGO}",
    eprint = "1702.07818",
    archivePrefix = "arXiv",
    primaryClass = "astro-ph.HE",
    doi = "10.3847/1538-4357/aadae5",
    journal = "Astrophys. J.",
    volume = "866",
    number = "1",
    pages = "66",
    year = "2018"
}

@article{Santini:2023ukl,
    author = "Santini, Alessandro and Gerosa, Davide and Cotesta, Roberto and Berti, Emanuele",
    title = "{Black-hole mergers in disklike environments could explain the observed q-{\ensuremath{\chi}}eff correlation}",
    eprint = "2308.12998",
    archivePrefix = "arXiv",
    primaryClass = "astro-ph.HE",
    doi = "10.1103/PhysRevD.108.083033",
    journal = "Phys. Rev. D",
    volume = "108",
    number = "8",
    pages = "083033",
    year = "2023"
}

@article{McKernan:2023xio,
    author = "McKernan, B. and Ford, K. E. S.",
    title = "{Constraining the LVK AGN channel with black hole spins}",
    eprint = "2309.15213",
    archivePrefix = "arXiv",
    primaryClass = "astro-ph.HE",
    doi = "10.1093/mnras/stae1351",
    journal = "Mon. Not. Roy. Astron. Soc.",
    volume = "531",
    number = "3",
    pages = "3479--3485",
    year = "2024"
}

@article{McKernan:2024kpr,
    author = "McKernan, Barry and Ford, K. E. Saavik and Cook, Harrison E. and Delfavero, Vera and McPike, Emily and Nathaniel, Kaila and Postiglione, Jake and Ray, Shawn and O'Shaughnessy, Richard",
    title = "{McFACTS I: Testing the LVK AGN Channel with Monte Carlo for AGN Channel Testing and Simulation (McFACTS)}",
    eprint = "2410.16515",
    archivePrefix = "arXiv",
    primaryClass = "astro-ph.HE",
    doi = "10.3847/1538-4357/adf114",
    journal = "Astrophys. J.",
    volume = "990",
    number = "2",
    pages = "217",
    year = "2025"
}

@article{Cook:2024ajp,
    author = "Cook, Harrison E. and McKernan, Barry and Ford, K. E. Saavik and Delfavero, Vera and Nathaniel, Kaila and Postiglione, Jake and Ray, Shawn and McPike, Emily J. and O'Shaughnessy, Richard",
    title = "{McFACTS. II. Mass Ratio{\textendash}Effective Spin Relationship of Black Hole Mergers in the Active Galactic Nucleus Channel}",
    eprint = "2411.10590",
    archivePrefix = "arXiv",
    primaryClass = "astro-ph.HE",
    doi = "10.3847/1538-4357/adfd56",
    journal = "Astrophys. J.",
    volume = "993",
    number = "2",
    pages = "163",
    year = "2025"
}

@article{Fabj:2025vza,
    author = "Fabj, Gaia and Tiede, Christopher and Rowan, Connar and Pessah, Martin and Samsing, Johan",
    title = "{Spin-Orbit Misalignments of Eccentric Black Hole Mergers in AGN Disks}",
    eprint = "2510.07952",
    archivePrefix = "arXiv",
    primaryClass = "astro-ph.HE",
    month = "10",
    year = "2025",
    journal = "",
}

@article{Bruel:2025sdq,
    author = "Bruel, Tristan and Lamberts, Astrid and Rodriguez, Carl L. and Feldmann, Robert and Grudic, Michael Y. and Moreno, Jorge",
    title = "{Great Balls of FIRE - IV. Contribution of massive star clusters to the astrophysical population of merging binary black holes}",
    eprint = "2503.03810",
    archivePrefix = "arXiv",
    primaryClass = "astro-ph.GA",
    doi = "10.1051/0004-6361/202554454",
    journal = "Astron. Astrophys.",
    volume = "701",
    pages = "A252",
    year = "2025"
}

@article{vanSon:2022myr,
    author = "van Son, L. A. C. and de Mink, S. E. and Renzo, M. and Justham, S. and Zapartas, E. and Breivik, K. and Callister, T. and Farr, W. M. and Conroy, C.",
    title = "{No Peaks without Valleys: The Stable Mass Transfer Channel for Gravitational-wave Sources in Light of the Neutron Star{\textendash}Black Hole Mass Gap}",
    eprint = "2209.13609",
    archivePrefix = "arXiv",
    primaryClass = "astro-ph.HE",
    doi = "10.3847/1538-4357/ac9b0a",
    journal = "Astrophys. J.",
    volume = "940",
    number = "2",
    pages = "184",
    year = "2022"
}

@dataset{ligo_scientific_collaboration_and_virgo_2025_16053484,
  author       = {LIGO Scientific Collaboration and Virgo Collaboration and KAGRA Collaboration},
  title        = {GWTC-4.0: Parameter estimation data release},
  month        = aug,
  year         = 2025,
  publisher    = {Zenodo},
  doi          = {10.5281/zenodo.16053484},
  url          = {https://doi.org/10.5281/zenodo.16053484},
}

@article{Galaudage:2026opk,
    author = "Galaudage, Shanika",
    title = "{Compactness Peaks and Subpopulations: Probing Stellar Physics and Formation Channels of Merging Binary Black Holes}",
    eprint = "2605.25994",
    archivePrefix = "arXiv",
    primaryClass = "astro-ph.HE",
    month = "5",
    year = "2026"
}

@article{Galaudage:2024meo,
    author = "Galaudage, Shanika and Lamberts, Astrid",
    title = "{Compactness peaks: An astrophysical interpretation of the mass distribution of merging binary black holes}",
    eprint = "2407.17561",
    archivePrefix = "arXiv",
    primaryClass = "astro-ph.HE",
    doi = "10.1051/0004-6361/202451654",
    journal = "Astron. Astrophys.",
    volume = "694",
    pages = "A186",
    year = "2025"
}

@dataset{ligo_scientific_collaboration_and_virgo_2021_5546663,
  author       = {LIGO Scientific Collaboration and Virgo Collaboration and KAGRA Collaboration},
  title        = {GWTC-3: Compact Binary Coalescences Observed by
                   LIGO and Virgo During the Second Part of the Third
                   Observing Run — Parameter estimation data release
                  },
  month        = nov,
  year         = 2021,
  publisher    = {Zenodo},
  doi          = {10.5281/zenodo.5546663},
  url          = {https://doi.org/10.5281/zenodo.5546663},
}

@dataset{ligo_scientific_collaboration_and_virgo_2022_6513631,
  author       = {LIGO Scientific Collaboration and  Virgo Collaboration},
  title        = {GWTC-2.1: Deep Extended Catalog of Compact Binary
                   Coalescences Observed by LIGO and Virgo During the
                   First Half of the Third Observing Run - Parameter
                   Estimation Data Release
                  },
  month        = may,
  year         = 2022,
  publisher    = {Zenodo},
  version      = {v2},
  doi          = {10.5281/zenodo.6513631},
  url          = {https://doi.org/10.5281/zenodo.6513631},
}

@dataset{ligo_scientific_collaboration_2025_16740128,
  author       = {LIGO Scientific Collaboration and
                  Virgo Collaboration and
                  KAGRA Collaboration},
  title        = {GWTC-4.0 Cumulative Search Sensitivity Estimates},
  month        = aug,
  year         = 2025,
  publisher    = {Zenodo},
  doi          = {10.5281/zenodo.16740128},
  url          = {https://doi.org/10.5281/zenodo.16740128},
}

@article{deMink:2016vkw,
    author = "de Mink, S. E. and Mandel, I.",
    title = "{The chemically homogeneous evolutionary channel for binary black hole mergers: rates and properties of gravitational-wave events detectable by advanced LIGO}",
    eprint = "1603.02291",
    archivePrefix = "arXiv",
    primaryClass = "astro-ph.HE",
    doi = "10.1093/mnras/stw1219",
    journal = "Mon. Not. Roy. Astron. Soc.",
    volume = "460",
    number = "4",
    pages = "3545--3553",
    year = "2016"
}

@article{deBoer:2017ldl,
    author = "deBoer, R. J. and others",
    title = "{The $^{12}$C({\ensuremath{\alpha}},{\ensuremath{\gamma}})$^{16}$O reaction and its implications for stellar helium burning}",
    eprint = "1709.03144",
    archivePrefix = "arXiv",
    primaryClass = "nucl-ex",
    doi = "10.1103/RevModPhys.89.035007",
    journal = "Rev. Mod. Phys.",
    volume = "89",
    number = "3",
    pages = "035007",
    year = "2017"
}

@article{Roy:2025ktr,
    author = "Roy, Soumendra Kishore and van Son, Lieke A. C. and Farr, Will M.",
    title = "{A mid-thirties crisis: dissecting the properties of gravitational wave sources near the 35 solar mass peak}",
    eprint = "2507.01086",
    archivePrefix = "arXiv",
    primaryClass = "astro-ph.HE",
    reportNumber = "LIGO document number LIGO-P2500403",
    doi = "10.1088/1361-6382/ae1921",
    journal = "Class. Quant. Grav.",
    volume = "42",
    number = "22",
    pages = "225008",
    year = "2025"
}

@article{Zevin:2022bfa,
    author = "Zevin, Michael and Holz, Daniel E.",
    title = "{Avoiding a Cluster Catastrophe: Retention Efficiency and the Binary Black Hole Mass Spectrum}",
    eprint = "2205.08549",
    archivePrefix = "arXiv",
    primaryClass = "astro-ph.HE",
    doi = "10.3847/2041-8213/ac853d",
    journal = "Astrophys. J. Lett.",
    volume = "935",
    pages = "L20",
    year = "2022"
}

@article{Tauris:2022ggv,
    author = "Tauris, Thomas M.",
    title = "{Tossing Black Hole Spin Axes}",
    eprint = "2205.02541",
    archivePrefix = "arXiv",
    primaryClass = "astro-ph.HE",
    doi = "10.3847/1538-4357/ac86c8",
    journal = "Astrophys. J.",
    volume = "938",
    pages = "66",
    year = "2022"
}

@article{Wysocki:2017isg,
    author = "Wysocki, Daniel and Gerosa, Davide and O'Shaughnessy, Richard and Belczynski, Krzysztof and Gladysz, Wojciech and Berti, Emanuele and Kesden, Michael and Holz, Daniel E.",
    title = "{Explaining LIGO{\textquoteright}s observations via isolated binary evolution with natal kicks}",
    eprint = "1709.01943",
    archivePrefix = "arXiv",
    primaryClass = "astro-ph.HE",
    reportNumber = "LIGO-P1700174",
    doi = "10.1103/PhysRevD.97.043014",
    journal = "Phys. Rev. D",
    volume = "97",
    number = "4",
    pages = "043014",
    year = "2018"
}

@article{Callister:2020vyz,
    author = "Callister, Thomas A. and Farr, Will M. and Renzo, Mathieu",
    title = "{State of the Field: Binary Black Hole Natal Kicks and Prospects for Isolated Field Formation after GWTC-2}",
    eprint = "2011.09570",
    archivePrefix = "arXiv",
    primaryClass = "astro-ph.HE",
    doi = "10.3847/1538-4357/ac1347",
    journal = "Astrophys. J.",
    volume = "920",
    number = "2",
    pages = "157",
    year = "2021"
}

@article{Fragione:2021qtg,
    author = "Fragione, Giacomo and Loeb, Abraham and Rasio, Frederic A.",
    title = "{Impact of Natal Kicks on Merger Rates and Spin{\textendash}Orbit Misalignments of Black Hole{\textendash}Neutron Star Mergers}",
    eprint = "2108.06538",
    archivePrefix = "arXiv",
    primaryClass = "astro-ph.HE",
    doi = "10.3847/2041-8213/ac225a",
    journal = "Astrophys. J. Lett.",
    volume = "918",
    number = "2",
    pages = "L38",
    year = "2021"
}

@article{Mandel:2015eta,
    author = "Mandel, Ilya",
    title = "{Estimates of black-hole natal kick velocities from observations of low-mass X-ray binaries}",
    eprint = "1510.03871",
    archivePrefix = "arXiv",
    primaryClass = "astro-ph.HE",
    doi = "10.1093/mnras/stv2733",
    journal = "Mon. Not. Roy. Astron. Soc.",
    volume = "456",
    number = "1",
    pages = "578--581",
    year = "2016"
}

@article{Mirabel:2016msh,
    author = "Mirabel, F{\'e}lix",
    title = "{The Formation of Stellar Black Holes}",
    eprint = "1609.08411",
    archivePrefix = "arXiv",
    primaryClass = "astro-ph.HE",
    doi = "10.1016/j.newar.2017.04.002",
    journal = "New Astron. Rev.",
    volume = "78",
    pages = "1--15",
    year = "2017"
}

@article{Heggie:1997gq,
    author = "Heggie, D. C.",
    title = "{Star cluster simulations}",
    eprint = "astro-ph/9711185",
    archivePrefix = "arXiv",
    month = "11",
    year = "1997",
    journal = "",
}

@article{Tiwari:2025oah,
    author = "Tiwari, Vaibhav",
    title = "{Population of Binary Black Holes Inferred from One Hundred and Fifty Gravitational Wave Signals}",
    eprint = "2510.25579",
    archivePrefix = "arXiv",
    primaryClass = "astro-ph.HE",
    month = "10",
    year = "2025",
    journal = "",
}

@article{Tiwari:2021yvr,
    author = "Tiwari, Vaibhav",
    title = "{Exploring Features in the Binary Black Hole Population}",
    eprint = "2111.13991",
    archivePrefix = "arXiv",
    primaryClass = "astro-ph.HE",
    doi = "10.3847/1538-4357/ac589a",
    journal = "Astrophys. J.",
    volume = "928",
    number = "2",
    pages = "155",
    year = "2022"
}

@article{Tiwari:2020otp,
    author = "Tiwari, Vaibhav and Fairhurst, Stephen",
    title = "{The Emergence of Structure in the Binary Black Hole Mass Distribution}",
    eprint = "2011.04502",
    archivePrefix = "arXiv",
    primaryClass = "astro-ph.HE",
    doi = "10.3847/2041-8213/abfbe7",
    journal = "Astrophys. J. Lett.",
    volume = "913",
    number = "2",
    pages = "L19",
    year = "2021"
}

@article{Olejak:2024qxr,
    author = "Olejak, Aleksandra and Klencki, Jakub and Xu, Xiao-Tian and Wang, Chen and Belczynski, Krzysztof and Lasota, Jean-Pierre",
    title = "{Unequal-mass, highly-spinning binary black hole mergers in the stable mass transfer formation channel}",
    eprint = "2404.12426",
    archivePrefix = "arXiv",
    primaryClass = "astro-ph.HE",
    doi = "10.1051/0004-6361/202450480",
    journal = "Astron. Astrophys.",
    volume = "689",
    pages = "A305",
    year = "2024"
}

@dataset{olejak:2024-12530199,
  author       = {Olejak, Aleksandra},
  title        = {Unequal-mass, highly-spinning binary black hole
                   mergers in the stable mass transfer formation
                   channel - database
                  },
  month        = jun,
  year         = 2024,
  publisher    = {Zenodo},
  doi          = {10.48550/arXiv.2404.12426},
  url          = {https://doi.org/10.48550/arXiv.2404.12426},
}

@article{Belczynski:2020bca,
    author = "Belczynski, Krzysztof",
    title = "{The most ordinary formation of the most unusual double black hole merger}",
    eprint = "2009.13526",
    archivePrefix = "arXiv",
    primaryClass = "astro-ph.HE",
    doi = "10.3847/2041-8213/abcbf1",
    journal = "Astrophys. J. Lett.",
    volume = "905",
    number = "2",
    pages = "L15",
    year = "2020"
}

@article{Belczynski:2005mr,
    author = "Belczynski, Krzystof and Kalogera, Vassiliki and Rasio, Frederic A. and Taam, Ronald E. and Zezas, Andreas and Bulik, Tomasz and Maccarone, Thomas J. and Ivanova, Natalia",
    title = "{Compact object modeling with the startrack population synthesis code}",
    eprint = "astro-ph/0511811",
    archivePrefix = "arXiv",
    doi = "10.1086/521026",
    journal = "Astrophys. J. Suppl.",
    volume = "174",
    pages = "223",
    year = "2008"
}

@article{Belczynski:2017gds,
    author = "Belczynski, K. and others",
    title = "{Evolutionary roads leading to low effective spins, high black hole masses, and O1/O2 rates for LIGO/Virgo binary black holes}",
    eprint = "1706.07053",
    archivePrefix = "arXiv",
    primaryClass = "astro-ph.HE",
    doi = "10.1051/0004-6361/201936528",
    journal = "Astron. Astrophys.",
    volume = "636",
    pages = "A104",
    year = "2020"
}

@article{Olejak:2021iux,
    author = "Olejak, Aleksandra and Belczynski, Krzysztof",
    title = "{The Implications of High BH Spins on the Origin of BH{\textendash}BH Mergers}",
    eprint = "2109.06872",
    archivePrefix = "arXiv",
    primaryClass = "astro-ph.HE",
    doi = "10.3847/2041-8213/ac2f48",
    journal = "Astrophys. J. Lett.",
    volume = "921",
    number = "1",
    pages = "L2",
    year = "2021"
}

@article{Olejak:2022zee,
    author = "Olejak, Aleksandra and Fryer, Chris L. and Belczynski, Krzysztof and Baibhav, Vishal",
    title = "{The role of supernova convection for the lower mass gap in the isolated binary formation of gravitational wave sources}",
    eprint = "2204.09061",
    archivePrefix = "arXiv",
    primaryClass = "astro-ph.HE",
    doi = "10.1093/mnras/stac2359",
    journal = "Mon. Not. Roy. Astron. Soc.",
    volume = "516",
    number = "2",
    pages = "2252--2271",
    year = "2022"
}

@article{Olejak:2021fti,
    author = "Olejak, Aleksandra and Belczynski, Krzysztof and Ivanova, Natalia",
    title = "{Impact of common envelope development criteria on the formation of LIGO/Virgo sources}",
    eprint = "2102.05649",
    archivePrefix = "arXiv",
    primaryClass = "astro-ph.HE",
    doi = "10.1051/0004-6361/202140520",
    journal = "Astron. Astrophys.",
    volume = "651",
    pages = "A100",
    year = "2021"
}

@article{Dorozsmai:2022wff,
    author = "Dorozsmai, Andris and Toonen, Silvia",
    title = "{Importance of stable mass transfer and stellar winds for the formation of gravitational wave sources}",
    eprint = "2207.08837",
    archivePrefix = "arXiv",
    primaryClass = "astro-ph.SR",
    doi = "10.1093/mnras/stae152",
    journal = "Mon. Not. Roy. Astron. Soc.",
    volume = "530",
    number = "4",
    pages = "3706--3739",
    year = "2024"
}

@article{Broekgaarden:2022nst,
    author = "Broekgaarden, Floor S. and Stevenson, Simon and Thrane, Eric",
    title = "{Signatures of Mass Ratio Reversal in Gravitational Waves from Merging Binary Black Holes}",
    eprint = "2205.01693",
    archivePrefix = "arXiv",
    primaryClass = "astro-ph.HE",
    doi = "10.3847/1538-4357/ac8879",
    journal = "Astrophys. J.",
    volume = "938",
    number = "1",
    pages = "45",
    year = "2022"
}
\bibliographystyle{aasjournalv7}

\end{document}